\documentclass[a4paper,fleqn,usenatbib]{mnras}

\usepackage{newtxtext,newtxmath}
\usepackage[T1]{fontenc}
\usepackage{ae,aecompl}
\usepackage{times}
\usepackage{rotating}
\usepackage{enumitem}
\usepackage{times}
\usepackage{amsmath}
\usepackage{amssymb}
\usepackage{textcomp}
\usepackage{verbatim}
\usepackage{esdiff}

\newcommand{\msun}{{\,\rm M_\odot}}

\newcommand{\kms}{\,{\rm km}\,{\rm s}^{-1}}
\newcommand{\cm}{\,{\rm cm}}

\newcommand{\Gyr}{\,{\rm Gyr}}

\newcommand{\nm}{\,{\rm nm}}
\newcommand{\Myr}{\,{\rm Myr}}

\newcommand{\kpc}{\,{\rm kpc}}
\newcommand{\pkpc}{\,{\rm pkpc}}
\newcommand{\Mpc}{\,{\rm Mpc}}

\newcommand{\mum}{\,\mu{\rm m}}
\newcommand{\mmag}{\,{\rm mag}}

\def\aap{A\&A}
\def\apj{ApJ}

\def\apjl{ApJ}
\def\mnras{MNRAS}
\def\araa{ARA\&A}
\def\aj{AJ}

\def\nat{Nature}

\def\apjs{ApJS}

\usepackage{graphicx}	% Including figure files
\usepackage{amsmath}	% Advanced maths commands
\usepackage{amssymb}	% Extra maths symbols
\usepackage{subfig}
\makeatletter

\newcommand{\Rmnum}[1]{\expandafter\@slowromancap\romannumeral #1@}

\renewcommand\paragraph{\@startsection{paragraph}{4}{\z@}{3.25ex\@plus1ex\@minus.2ex}{-1em}{\normalfont\it\normalsize}}

\makeatother
\title[IllustrisTNG high redshift luminosity functions]
{High redshift {\it \textbf{JWST}} predictions from IllustrisTNG: Dust modelling and galaxy luminosity functions}

      \author[M. Vogelsberger et al.] {\parbox{18.5cm}{
          Mark Vogelsberger$^1$\thanks{email:mvogelsb@mit.edu}, 
          Dylan Nelson$^2$,
          Annalisa Pillepich$^3$,
          Xuejian Shen$^1$,
          Federico Marinacci$^{5,4,1}$,
          Volker Springel$^2$,
          R\"udiger Pakmor$^2$,
          Sandro Tacchella$^4$,
          Rainer Weinberger$^4$,
          Paul Torrey$^{6,1}$,
          Lars Hernquist$^4$
        }\vspace{0.3cm}\\
        $^1$ Department of Physics, Kavli Institute for Astrophysics and Space Research, Massachusetts Institute of Technology, Cambridge, MA 02139, USA\\
        $^2$ Max-Planck-Institut f\"{u}r Astrophysik, Karl-Schwarzschild-Str. 1, 85741 Garching, Germany\\
        $^3$ Max-Planck-Institut f\"{u}r Astronomie, K\"{o}nigstuhl 17, 69117 Heidelberg, Germany\\        
         $^4$ Harvard-Smithsonian Center for Astrophysics, 60 Garden Street, Cambridge, MA, 02138, USA\\
         $^5$Department of Physics \& Astronomy, University of Bologna, via Gobetti 93/2, 40129 Bologna, Italy\\
         $^6$Department of Astronomy, University of Florida, 316 Bryant Space Sciences 
  Center, Gainesville, FL 32611 USA\\
}

\begin{document}
\label{firstpage}
\pagerange{\pageref{firstpage}--\pageref{lastpage}}
\maketitle

% Abstract of the paper
\begin{abstract}
The {\it James Webb Space Telescope} ({\it JWST}) promises to revolutionise our understanding of the early Universe, and contrasting its upcoming observations with predictions of the $\Lambda$CDM model requires detailed theoretical forecasts. Here, we exploit the large dynamic range of the IllustrisTNG simulation suite, TNG50, TNG100, and TNG300, to derive multi-band galaxy luminosity functions from $z=2$ to $z=10$. We put particular emphasis on the exploration of different dust attenuation models to determine galaxy luminosity functions for the rest-frame ultraviolet (UV), and apparent wide NIRCam bands. Our most detailed dust model is based on continuum Monte Carlo radiative transfer calculations employing observationally calibrated dust properties. This calibration results in constraints on the redshift evolution of the dust attenuation normalisation and dust-to-metal ratios yielding a stronger redshift evolution of the attenuation normalisation compared to most previous theoretical studies. Overall we find good agreement between the rest-frame UV luminosity functions and observational data for all redshifts, also beyond the regimes used for the dust model calibrations. Furthermore, we also recover the observed high redshift ($z=4-6$) UV luminosity versus stellar mass relation, the H$\alpha$ versus star formation rate relation, and the H$\alpha$ luminosity function at $z=2$. 
The bright end ($M_{\rm UV}>-19.5$) cumulative galaxy number densities are consistent with observational data.
For the F200W NIRCam band, we predict that {\it JWST} will detect $\sim 80$ ($\sim 200$) galaxies with a signal-to-noise ratio of $10$ ($5$) within the NIRCam field of view, $2.2\times2.2 \,{\rm arcmin}^{2}$, for a total exposure time of $10^5{\rm s}$ in the redshift range $z=8 \pm 0.5$. These numbers drop to $\sim 10$ ($\sim 40$) for an exposure time of $10^4{\rm s}$. 
\end{abstract}


%%%%%%%%%%%%%%%%%%%%%%%%%%%%%%%%%%%%%%%%%%%%%%%%%%

%%%%%%%%%%%%%%%%% BODY OF PAPER %%%%%%%%%%%%%%%%%%

\section{Introduction}
\label{sec:Section1}

The concordance paradigm of structure formation, the $\Lambda$CDM model~\citep[e.g.,][]{Planck2016}, provides testable predictions for the growth of dark matter halos and is the basis of our current theory for galaxy formation \citep[][]{White1978, Blumenthal1984}. Confronting these predictions with observations at multiple epochs of cosmic evolution is crucial to confirm or falsify this theoretical framework. One of the most basic quantities of the galaxy population is the galaxy luminosity function that measures the comoving number density of galaxies as a function of their luminosities at different redshifts. 

{\it JWST} will open a new window into the high redshift Universe to study faint and distant galaxies during the early phases of cosmic evolution. Specifically, {\it JWST} will quantify the galaxy population and galaxy luminosity functions at higher redshifts than ever before. Most importantly it will also dramatically increase the statistical sample sizes of high redshift galaxies. 
Two {\it JWST} instruments, the Near InfraRed Camera (NIRCam) and the Mid InfraRed Instrument (MIRI), are designed to obtain broadband photometry over the wavelength range from $0.7$ to $25.5\mum$ with unprecedented sensitivity and angular resolution. 
This wavelength coverage allows {\it JWST} to probe the rest-frame UV, optical, and near-infrared (IR) spectral energy distributions (SEDs)
of high-redshift galaxies, which will lead to novel insights into the evolution of the earliest galaxy populations. These observations are crucial to study the physics that shapes the galaxy population in the early Universe. Furthermore, quantifying the high redshift galaxy population is also important to understand the sources
of reionisation in the early Universe. 

These upcoming {\it JWST} observations will extend the successes of previous studies to explore the high redshift Universe. For example, the Lyman-break technique~\citep[e.g.,][]{Steidel1993, Steidel1996} allowed the detection of galaxies at $z \sim 3$. These early limits were pushed further with the Advanced Camera on the Hubble Space Telescope ({\it HST}), which extended the Lyman-break technique to $z \sim 6$~\citep[][]{Bouwens2003, Stanway2003}. The Wide-Field Camera 3 with near-IR filters then further increased the number of galaxies that could be identified at $z \sim 7$~\citep[e.g.,][]{Bouwens2010, Wilkins2010,Oesch2010b}, enlarging the sample sizes of galaxies at these redshifts substantially, with a few
examples at $z \sim 10$. Specifically, {\it HST} has detected about $2000$ galaxy candidates at high redshifts ($z \sim 4 - 10$) from both blank and gravitationally lensed fields~\citep[e.g.,][]{Koekemoer2013, Lotz2017}. Brighter objects have also been discovered with ground-based facilities, such as the United Kingdom Infra-Red Telescope (UKIRT) and the Visible and Infrared Survey Telescope for Astronomy (VISTA) ~\citep[][]{McLure2009, Bowler2015}. These combined observations have provided constraints on the space density of relatively
bright galaxy populations up to $z \sim 10$~\citep[e.g.,][]{McLure2009, McLure2013, Castellano2010, vanderBurg2010, Oesch2013, Oesch2014, Oesch2018, Schenker2013, Tilvi2013, Bowler2014, Bowler2015, Bouwens2014b, Bouwens2015, Bouwens2016, Bouwens2017, Schmidt2014, Atek2015a, McLeod2015, McLeod2016, Finkelstein2015, Livermore2017, Ishigaki2018}. 

\begin{table*}
\begin{tabular}{llccrrrccccc}
\hline
{\bf IllustrisTNG Simulation} & run &   {volume side length} & $N_{\rm gas}$  & $N_{\rm dm}$ & $m_{\rm b}$ & $m_{\rm dm}$ & $\epsilon_{\rm dm,stars}$ &  $\epsilon_{\rm gas}^{\rm min}$ \\
 & &   $[h^{-1}{\rm Mpc}]$  & & &  $[h^{-1}{\rm M}_\odot]$ & $[h^{-1}{\rm M}_\odot]$ &  $[h^{-1}{\rm kpc}]$ & $[h^{-1}{\rm kpc}]$\\
\hline
\hline
{\bf TNG300}   &  TNG300(-1)  & 205  & $2500^3$  & $2500^3$   & $7.4\times 10^6$   & $4.0\times 10^7$   & 1.0 & 0.25\\
%               &  TNG300-2  & 205  & $1250^3$  & $1250^3$     & $5.9\times 10^7$   & $3.2\times 10^8$   & 2.0 & 0.5 \\
%               &  TNG300-3  & 205  & $625^3$   & $625^3$      & $4.8\times 10^8$   & $2.5\times 10^9$   & 4.0 & 1.0 \\
\hline
{\bf TNG100}   &  TNG100(-1)  & 75   & $1820^3$  & $1820^3$   & $9.4\times 10^5$   & $5.1\times 10^6$   & 0.5 & 0.125 \\
%               &  TNG100-2  & 75   & $910^3$   & $910^3$      & $7.6\times 10^6$   & $4.0\times 10^7$   & 1.0 & 0.25\\
%               &  TNG100-3  & 75   & $455^3$   & $455^3$      & $6.0\times 10^7$   & $3.2\times 10^8$   & 2.0 & 0.5\\
\hline
{\bf TNG50}    &  TNG50(-1)   & 35   & $2160^3$  & $2160^3$   & $5.7\times 10^4$   & $3.1\times 10^5$   & 0.2 & 0.05 \\
%               &  TNG50-2   & 35   & $1080^3$  & $1080^3$     & $4.6\times 10^5$   & $2.5\times 10^6$   & 0.4 & 0.1\\
%               &  TNG50-3   & 35   & $540^3$   & $540^3$      & $3.7\times 10^6$   & $2.0\times 10^7$   & 0.8 & 0.2 \\
%               &  TNG50-4   & 35   & $270^3$   & $270^3$      & $2.9\times 10^7$   & $1.6\times 10^8$   & 1.6 & 0.4\\               
\hline
\end{tabular}
\caption{{\bf IllustrisTNG simulation suite.} The table presents the basic numerical parameters of all IllustrisTNG simulations studied in this paper: simulation volume side length, number of gas cells ($N_{\rm gas}$), number of dark matter particles ($N_{\rm dm}$), baryon mass resolution ($m_{\rm b}$), dark matter mass resolution ($m_{\rm DM}$), Plummer-equivalent maximum physical softening length of dark matter and stellar particles ($\epsilon_{\rm dm,stars}$) and the minimal comoving cell softening length $\epsilon_{\rm gas}^{\rm min}$. In the following we will identify TNG50 with TNG50-1, TNG100 with TNG100-1 and TNG300 with TNG300-1. 
\label{tab:tabsims}}
\end{table*}

Despite these successes, it still remains difficult to observationally quantify evolutionary details of the high redshift galaxy luminosity functions. 
For example, probing the high redshift faint-end galaxy population and luminosity functions with existing observational facilities is challenging. 
Most of the currently existing constraints rely on fields that are lensed by massive foreground galaxy clusters with uncertain magnification corrections~\citep[e.g.,][]{Kawamata2016, Bouwens2017, Priewe2017}. It is therefore difficult to derive robust results for the evolution of the faint-end of the galaxy luminosity function.  {\it JWST} will probe the luminosity function in unlensed fields much deeper, and importantly also provide  more robust redshift measurements for high redshift candidates that are
currently selected by the Lyman-break technique. This will lead to a better understanding of the faint-end population of the galaxy luminosity function and its evolution with redshift with important consequences for the source of reionisation.

Given the prospects of {\it JWST} to quantify the galaxy population of the high redshift Universe, theoretical predictions are required to provide templates for comparisons with upcoming {\it JWST} observational data. 
Various hydrodynamical simulations have therefore been employed to explore the high redshift Universe and to derive {\it JWST} predictions, for example: the First Billion Years simulation suites~\citep[e.g.,][]{Paardekooper2013}, the BlueTides simulation~\citep[e.g.,][]{Wilkins2016, Wilkins2017}, the Renaissance simulations suite~\citep[e.g.,][]{Xu2016, Barrow2017}, the FIRE-2 simulations~\citep[e.g.,][]{ma2018b}, the Sphinx simulation~\citep[e.g.,][]{Rosdahl2018} and others~\citep[e.g.,][]{Dayal2013, Shimizu2014}.
In addition, semi-analytic models of galaxy formation~\citep[e.g.,][]{Cowley2018, Yung2018}, and empirical models~\citep[e.g.,][]{tacchella2018} have also been studied to derive {\it JWST} predictions and high redshift luminosity functions. We note that most of the studied hydrodynamical simulations have only been evolved down to relatively high redshifts due to the demanding computational requirements. It is therefore often unclear how reliable some of the high redshift predictions are, given that some of these models could not be tested towards lower redshifts, where  observational data for comparisons is available. These uncertainties then directly propagate into the derived {\it JWST} predictions for the high redshift Universe. We therefore stress that reliable high redshift galaxy population forecasts require a galaxy formation model that has been calibrated towards lower redshifts. Such predictions based on well-tested hydrodynamical cosmological simulations and galaxy formation models are currently largely missing.

The goal of this paper is to fill this gap by analysing the recently finished IllustrisTNG simulation suite to predict the high redshift galaxy population as it will be observed by {\it JWST}. The focus of this paper is a study of high redshift galaxy luminosity functions to derive galaxy number densities as a function of galaxy luminosities at different epochs ranging from $z=2$ to $z=10$. We maximise the dynamic range of our analysis by combining all three IllustrisTNG simulations, TNG50, TNG100 and TNG300. Dust plays a crucial role in shaping the bright end of the galaxy luminosity function, and we therefore explore multiple dust models in our study. One of our dust models is based on full Monte Carlo dust radiative transfer calculations where the dust-to-metal ratio  is calibrated against current constraints on the rest-frame UV luminosity functions at $z=2-10$. This dust calibration procedure leads to novel constraints on the dust-to-metal ratios, and dust attenuation normalisations. Based on the luminosity functions for various {\it JWST} NIRCam bands, we can also derive predictions for the expected number of detected galaxies given specific NIRCam configurations.

The structure of this paper is as follows. In Section~\ref{sec:Section2} we briefly describe the IllustrisTNG simulation suite and state the numerical parameters of TNG50, TNG100 and TNG300 in detail. Our dust attenuation models are then described in Section~\ref{sec:Section3}. The main results are presented in Section~\ref{sec:Section4}, where we discuss various galaxy luminosity functions and {\it JWST} survey forecasts. Specifically, we discuss rest-frame UV luminosity functions, and then explore {\it JWST} luminosity functions for all NIRCam wide filter bands. Towards the end of Section~\ref{sec:Section4} we also briefly discuss the predicted mass-to-light ratio relations and H$\alpha$ luminosity star formation rate relation that provide another important verification of our simulation predictions and radiative transfer based dust modelling in particular. Our summary and conclusions are presented in Section~\ref{sec:Section5}.

\section{The Illustris~TNG Simulation Suite}
\label{sec:Section2}

Our analysis is based on the IllustrisTNG simulation suite~\citep[][]{Marinacci2018, Naiman2018, Nelson2018, Pillepich2018b, Springel2018}, including the newest addition of TNG50 ~\citep[][]{Nelson2019, Pillepich2019}, which consists of three primary simulations: TNG50, TNG100 and TNG300. IllustrisTNG is the follow-up project of the Illustris simulations~\citep{Vogelsberger2014a, Vogelsberger2014b, Genel2014, Nelson2015, Sijacki2015}. The IllustrisTNG simulation suite employs the following cosmological parameters~\citep[][]{Planck2016}: $\Omega_{\rm m} = 0.3089$, $\Omega_{\rm b} = 0.0486$, $\Omega_{\Lambda} = 0.6911$, $H_0 = 100\,h\,\kms \Mpc^{-1} = 67.74\,\kms \Mpc^{-1}$, $\sigma_{8} = 0.8159$, and $n_{\rm s} = 0.9667$. The three major simulations cover three different periodic, uniformly sampled volumes,
roughly ${\sim 50^3}, 100^3, 300^3\,{\rm Mpc}^3$ for TNG50, TNG100 and TNG300, respectively. The numerical parameters of the different simulations are summarised in Table~\ref{tab:tabsims}.
All simulations were carried out with the moving-mesh code {\sc Arepo}
\citep{Springel2010,Pakmor2016} combined with the IllustrisTNG galaxy formation model~\citep[][]{Weinberger2017, Pillepich2018a} which is an updated version of the Illustris galaxy formation
model~\citep{Vogelsberger2013, Torrey2014}. We note that TNG50, TNG100, and TNG300 differ in their highest numerical resolution as listed in Table~\ref{tab:tabsims}. In the following we will identify TNG50 with TNG50-1, TNG100 with TNG100-1 and TNG300 with TNG300-1.

\section{Galaxy luminosities and dust models}
\label{sec:Section3}

The central goal of our paper is to provide high redshift galaxy luminosity functions. We therefore have to assign luminosities and multiple band magnitudes to each galaxy of all different simulations of IllustrisTNG. We will describe this assignment process in this section. In this work we define a galaxy as being either a central or satellite galaxy
as identified by the {\sc Subfind} algorithm~\citep[][]{Springel2001, Dolag2009}. For the following analysis, we impose a stellar mass cut for galaxies. We only consider galaxies with a stellar mass larger than $100$ times the baryonic mass resolution, $100 \times m_{\rm b}$, within twice the stellar half mass radius.  Galaxies resolved with a lower number of resolution elements will not be considered, since we assume that their structure is not reliably modelled. Furthermore, galaxy luminosities, magnitudes, stellar masses and star formation rates are calculated within a fixed physical aperture of $30\pkpc$ and based on gravitationally bound particles and cells.

\subsection{Dust-free galaxy emission}
\label{sec:intrinsic}

We first describe our method to derive dust-free or intrinsic magnitudes, i.e. assuming that no
dust is absorbing or scattering photons within the galaxy. Consequently these dust-free magnitudes also do not include dust emission. We note that dust emission also has no impact on the derived magnitudes for the given redshift and
filter ranges studied in this paper.
To calculate dust-free
magnitudes we treat each stellar particle in the simulation as a simple stellar
population (SSP) using a stellar population synthesis method. Here we employ the Flexible Stellar Population Synthesis ({\sc Fsps}) code~\citep{conroy2009,Conroy2010} with MIST
isochrones~\citep{paxton2011,paxton2013,paxton2015, choi2016,dotter2016} and
the MILES stellar library~\citep{MILES1,MILES2}, assuming a Chabrier initial mass
function~\citep{chabrier2003} consistent with the IllustrisTNG galaxy formation model. To accelerate the luminosity and magnitude calculations, we first generate
a two-dimensional magnitude grid with one axis containing initial stellar
metallicity values and the other axis containing stellar age values. The
initial metallicity, $Z_{\rm i}$, axis contains $13$ values uniformly
logarithmically spaced from $10^{-4.35}$ to $10^{-1.35}$. The stellar ages, $t_{\rm ssp}$,
axis contains $107$ values uniformly logarithmically spaced from $10^{-4}$ to
$10^{1.3}\Gyr$. For each point on this grid, we construct the corresponding SSP
and calculate the rest-frame SED normalised to an initial
stellar mass of $1\msun$. We then use the {\sc Sedpy}\footnote{\url{https://github.com/bd-j/sedpy}} code to convolve the
resulting fluxes, either absolute or apparent, with transmission curves of arbitrary bands and calculate
rest-frame magnitudes. We note that all the magnitudes in this work are AB magnitudes, i.e. assuming a reference flux of $3631\;{\rm Jy}$. To extract rest-frame UV magnitudes,
$M_{\rm UV}$, we convolve the fluxes with a tophat filter of width $10\nm$
centred on $150\nm$; i.e. $M_{\rm UV} = M_{\rm 1500}$. This definition of UV magnitude is consistent with most other observational and theoretical studies~\citep[e.g.,][]{Duncan2014,Bowler2014,Bowler2015,Finkelstein2015,liu2016,Cullen2017}.

To calculate apparent dust-free magnitudes, we redshift the rest-frame SED,
$L_\nu(\nu_{\rm em})$, and calculate the apparent flux at a given frequency as:
\begin{equation}
f_\nu(\nu_{\rm obs})=\dfrac{(1+z)\,L_\nu(\nu_{\rm em})}{4\pi \, d_{\rm L}^{2}}\,e^{-\tau_{\rm IGM}^{\rm eff}(\nu_{\rm obs})},
\end{equation}
where $\nu_{\rm obs}=\nu_{\rm em}/(1+z)$, $z$ is the redshift of the source,
$d_{\rm L}$ is the luminosity distance to the source and $\tau_{\rm IGM}^{\rm
eff}(\nu)$ is the effective optical depth of the clumpy intergalactic medium (IGM).
This optical depth, $\tau_{\rm IGM}^{\rm eff}$, is calculated based on the IGM
absorption model of \cite{Madau1995} and \cite{madau1996}. We note that we do not consider IGM
absorption corrections for rest-frame magnitudes since this
absorption occurs at $\lesssim 120\nm$ (Lyman-$\alpha$), which is shorter than
the wavelength ranges of all bands studied in our work. Based on this apparent flux we then derive band magnitudes through a convolution as described above. 

{H\small\Rmnum{2}} regions surrounding young stellar populations reprocess the
Lyman continuum photons into nebular continuum and line emissions, which
represents an important contribution to galaxy SEDs.
This
emission depends on both the spectra of ionising photons and the properties of
the surrounding medium, including geometry, gas density, chemical content and
the covering fraction of Lyman continuum photons. Here we adopt the nebular
emission model of~\cite{byler2017}, which is based on photoionization
calculations using the~{\sc Cloudy} code~\citep{ferland2013}. The model assumes
that the fraction of escaping Lyman continuum photons is zero and nebular
emission is purely determined by the gas phase metallicity and the ionisation
parameter. Following~\citet{Wilkins2016} and~\cite{byler2017}, we choose the gas
phase metallicity to be the same as the initial metallicity of the stellar particle, which
is inherited from the gas cell from which the stellar particle has been created. The ionisation parameter, as defined in
\citet{byler2017}, encodes the intensity of the ionising source
and the geometry of the gas cloud. We choose the ionisation parameter to be
$0.01$ as suggested in~\citet{byler2018}. The time span that a SSP is surrounded by
its birth cloud is limited, so nebular emission is only active for SSPs with
$t_{\rm ssp}<t_{\rm esc}=10\Myr$. Stars
older than $t_{\rm esc}$ can escape the birth clouds.

Based on the magnitude grid, we assign rest-frame and apparent band
magnitudes to stellar particles of galaxies.  For each
stellar particle we employ a bilinear interpolation of the grid values to
calculate the band magnitudes corresponding to the $Z_{\rm i}$ and
$t_{\rm ssp}$ values of the stellar particle. In a last step, we sum up the luminosity contribution from all stellar particles in a galaxy within the physical aperture of $30\pkpc$. We mass-weight each particle
by its initial mass $m_{\rm i}$ to derive the dust-free luminosities and
the corresponding band magnitudes of the whole galaxy. This model is essentially the same as the dust-free model A of \citet{Nelson2018}.

\subsection{Dust attenuation models}

Dust attenuation can significantly alter the SED of a galaxy, and correspondingly its luminosity and magnitude in certain bands.
Consequently, dust attenuation also changes the galaxy luminosity function,
especially towards the bright end, where large amounts of dust are
available within galaxies. It is therefore necessary to model the effects of
dust attenuation to turn the dust-free band magnitudes of the previous section into dust-corrected band magnitudes. In the following we explore three different dust
attenuation models of increasing complexity:
\begin{itemize}[leftmargin=*]
\item Model A -- empirical dust model
\item Model B -- resolved dust optical depth model 
\item Model C -- resolved dust radiative transfer model
\end{itemize}
For Model A, we employ observationally derived empirical scaling relations to
link dust-free rest-frame UV magnitudes with observed dust-attenuated
rest-frame UV magnitudes. This model is solely based on the dust-free rest-frame
UV magnitudes, and this approach does not require any further input from the simulations. For Model B, we consider in addition also the simulated gas
distribution and spatial configuration of stars within each simulated galaxy to
derive dust column densities to attenuate the light of stellar particles without explicit radiative transfer. This
second model therefore takes into account additional information from the
simulated galaxies to derive dust-attenuated luminosities and magnitudes. For Model C, we also
consider the local gas distribution in galaxies and perform full dust Monte Carlo radiative
transfer calculations of photons propagating through the dusty interstellar medium of simulated galaxies. Most
theoretical high redshift luminosity function studies have employed Model A~\citep[e.g.,][]{tacchella2013,mason2015,liu2016,tacchella2018} or
Model B~\citep[e.g.,][]{kitzbichler2007,somerville2012,clay2015,trayford2015,Wilkins2017,Yung2018} to correct for the attenuation of dust. Some works have also explored Monte Carlo radiative transfer calculations~\citep[e.g.,][]{trayford2017}, however not at high redshift employing a calibrated dust model. Here we go beyond those
approaches by also performing full dust radiative transfer for galaxies in TNG50,
TNG100, and TNG300 using a calibrated dust model from $z=2$ to $z=10$ to derive dust-attenuated galaxy luminosity functions. In the following we describe the
details of each dust model.

\subsubsection{Model A -- empirical dust model}
Instead of directly modelling dust attenuation based on the local gas and dust properties
around stars in simulated galaxies, we can employ observationally motivated empirical scaling
relations to derive dust-attenuated rest-frame UV magnitudes of the simulated
galaxies. This is the basic idea of our dust Model A. We note that this approach is limited to UV magnitudes since the empirical scaling relations have only been established for this band. It is therefore not possible to derive dust-attenuated luminosities or magnitudes for other bands. Predictions for those bands will be derived based on dust Models B and C as discussed below.

To implement this dust Model A, we first assume that the observed, and therefore dust-attenuated, UV spectrum
of a galaxy can be characterised by a simple power law $f_{\lambda}\sim
\lambda^{\beta}$, where $\beta$ is the slope of the UV continuum. The
attenuation of UV light and the corresponding reprocessed light into IR
wavelengths correlates with this slope. Many works have studied this  ${{\rm
IRX} = (L_{\rm IR}/L_{\rm UV})}$ versus $\beta$ relation in detail finding
that~\citep[e.g.,][]{meurer1999,siana2009,casey2014}:
\begin{equation}
\rm{IRX}=B\,[10^{0.4\,(C_{0}+C_{1}\,\beta)}-1],
\label{eq:irxbeta}
\end{equation}
where $B$, $C_{0}$ and $C_{1}$ are constants. Since ${\rm{IRX}=B[10^{0.4\,A_{\rm
UV}}-1]}$, the attenuation of the galaxy UV light, $A_{\rm UV}$, depends then
linearly on $\beta$: $A_{\rm UV}=C_{0}+C_{1}\beta$. For the constants,
\cite{meurer1999} found ${C_{0}=4.43,\ C_{1}=1.99}$ for the originally proposed relation.
For the Small Magellanic Cloud ${\rm IRX}-\beta$ relation~\citep{Lequeux1982,Prevot1984,Bouchet1985,Bouwens2016}, these constants are ${C_{0}=2.45,\ C_{1}=1.1}$. More recent measurements find
${C_{0}=3.36,\ C_{1}=2.04}$~\citep{casey2014}. Here, we adopt the original
Meurer relation for our Model A. 

Using the $A_{\rm UV}-\beta$ relation we can calculate the dust attenuation
strength based on the $\beta$ slope. The slope, $\beta$, itself furthermore
correlates with the observed dust-attenuated rest-frame UV magnitude. This
empirical relation can, for example, be parametrised
as~\citep[e.g,][]{Bouwens2012,Bouwens2014a}:
\begin{equation}
\langle \beta \rangle (M_{\rm UV}^{\rm dust})=\diff{\beta}{M_{\rm UV}^{\rm dust}}(z)\,[M_{\rm UV}^{\rm dust}-M_0]+\beta_{M_0}(z),
\label{eq:betamuv}
\end{equation}
where $M_{\rm UV}^{\rm dust}$ is the rest-frame UV magnitude observed after
dust attenuation.  Some observationally derived values for
$\beta_{M_0}(z)$, ${\rm d}\beta/{\rm d}M_{\rm UV}^{\rm dust}(z)$ and $M_0$ are
summarised in Table~\ref{tab:modelA_parameter}. 

\begin{table*}
\centering
\begin{tabular}
{ p{0.05\textwidth}|p{0.04\textwidth}|p{0.13\textwidth}|p{0.09\textwidth}|p{0.05\textwidth}|p{0.07\textwidth}|p{0.09\textwidth}|p{0.04\textwidth}|p{0.13\textwidth}|p{0.09\textwidth}}
\hline 
{\bf redshift} & $M_0$$^{\rm a}$ & $\beta_{M_0}$$^{\rm a}$ & ${\rm d}\beta/{\rm d}M_{\rm UV}$$^{\rm a}$ & 
$M_0$$^{\rm b}$ & $\beta_{M_0}$$^{\rm b}$ & ${\rm d}\beta/{\rm d}M_{\rm UV}$$^{\rm b}$ &
$M_0$$^{\rm c}$ & $\beta_{M_0}$$^{\rm c}$ & ${\rm d}\beta/{\rm d}M_{\rm UV}$$^{\rm c}$
\\
$\mathbf{z}$ & $[{\rm mag}]$ & & $[{\rm mag}^{-1}]$ & 
$[{\rm mag}]$ & & $[{\rm mag}^{-1}]$ &
$[{\rm mag}]$ & & $[{\rm mag}^{-1}]$\\
\hline
\hline
{\bf 4} &-19.5 & ${-1.85}\pm0.01\pm0.06$ & ${-0.11}\pm0.01$ & -20.31 & $-1.86^{+0.03}_{-0.02}$ & $+0.01\pm0.03$ & -19.5 &$-2.00\pm0.02\pm0.10$ & $-0.11\pm0.01$\\
{\bf 5} &-19.5 & ${-1.91}\pm0.02\pm0.06$ & ${-0.14}\pm0.02$ & -19.94 & $-1.97^{+0.07}_{-0.04}$ & $+0.00\pm0.06$ & -19.5 &$-2.08\pm0.03\pm0.10$ & $-0.16\pm0.03$ \\
{\bf 6} &-19.5 & ${-2.00}\pm0.05\pm0.08$ & ${-0.20}\pm0.04$ & -19.54 & $-2.02^{+0.13}_{-0.13}$ & $-0.10\pm0.07$ & -19.5 &$-2.20\pm0.05\pm0.14$ & $-0.15\pm0.04$\\
{\bf 7} &-19.5 & ${-2.05}\pm0.09\pm0.13$ & ${-0.20}\pm0.07$ & -19.39 & $-2.42^{+0.31}_{-0.20}$ & $-0.20\pm0.11$ & -19.5 & $-2.27\pm0.07\pm0.28$ &  $-0.21\pm0.07$\\
{\bf 8} &-19.5 & ${-2.13}\pm0.44\pm0.27$ & ${-0.15}$ (fixed)& -19.35 & $-2.03^{+0.46}_{-0.38}$ & $-0.03\pm0.26$\\
\hline
\end{tabular}
\raggedright
\\
$^{\rm a}$ \citet{Bouwens2014a},
$^{\rm b}$ \citet{finkelstein2012} ($M_{\rm 0}$ is calculated based on $0.5L^{\ast}$, where the luminosity bin is $0.25L^{\ast}-0.75L^{\ast}$ and $L^{\ast}$ is taken from the reference),
$^{\rm c}$ \citet{Bouwens2012}
\caption{{\bf Observationally derived parameters for the $\mathbf{{M_{\rm UV}^{\rm dust}-\langle \beta
\rangle}}$ relation.} We also state random and systematic errors, respectively. The quoted values differ quite significantly between the different observational studies.}
\label{tab:modelA_parameter}
\end{table*}

To derive the mean attenuation at a given $M_{\rm UV}^{\rm dust}$ and
ultimately link $M_{\rm UV}^{\rm dust}$ with  $M_{\rm UV}^{\rm dust-free}$, the
dust-free rest-frame UV magnitude, we briefly introduce two commonly employed
approaches:  {(i)} taking averages of the inferred
dust-attenuated and dust-free luminosities or {(ii)} taking averages of the
inferred dust-attenuated and dust-free magnitudes. These two approaches are used to derive dust-attenuated rest-frame
UV magnitudes based on the dust-free rest-frame UV magnitudes, or for the
reverse operation. For each approach, the mean dust attenuation can furthermore be derived in two ways either analytically or numerically. We note that these various methods of establishing the link between $M_{\rm UV}^{\rm dust}$ and $M_{\rm UV}^{\rm dust-free}$ are not
fully equivalent once scatter in $\beta$ at a given $M_{\rm UV}^{\rm dust}$ is considered. However, this scatter only causes minor differences in the actual mapping between $M_{\rm UV}^{\rm dust}$ and $M_{\rm UV}^{\rm
dust-free}$ as we will demonstrate below.\\

\noindent {\it Analytical approach:} We first discuss an analytic approach to link
dust-attenuated and dust-free magnitudes.\\

\noindent {\it (i) Luminosity based average:}
Here we derive an analytic relation between the dust-free rest-frame UV
magnitude, $M_{\rm UV}^{\rm dust-free}$, and the dust-corrected rest-frame UV
magnitude, $M_{\rm UV}^{\rm dust}$, based on luminosity averages. This method has often been adopted in previous works~\citep[e.g.,][]{tacchella2013,mason2015,tacchella2018}.
Most of these studies also commonly assume that there is some scatter in $\beta$ for a given
$M_{\rm UV}^{\rm dust}$. This scatter is assumed to follow a Gaussian distribution at each $M_{\rm
UV}^{\rm dust}$ with a constant dispersion
$\sigma_{\beta}=0.34$~\citep[e.g.,][]{smit2012,tacchella2013,tacchella2018}. We
can then calculate the average attenuation as a function of $\langle \beta
\rangle$. To derive this attenuation, we start from the ${\rm IRX}-\beta$ relation and
derive $\langle \rm{IRX} \rangle$ for the Gaussian $\beta$ distribution.
$\rm{IRX}$ correlates linearly with the ratio between dust-free UV
luminosity and dust-attenuated UV luminosity as:
\begin{equation}
1+\rm{IRX}/B=10^{0.4A_{\rm UV}}=L_{\rm UV}^{\rm dust-free}/L_{\rm UV}^{\rm dust}.
\end{equation}
We can then derive the dust-free UV luminosity for a given dust-attenuated luminosity as:
\begin{equation}
\ln{\left(\dfrac{ L_{\rm UV}^{\rm dust-free} }{L_{\rm UV}^{\rm dust}}\right)}\!=\!\ln{(1+\rm{IRX}/B)}\!=\! 0.4\,\ln{10}\,(C_{0}+C_{1}\beta) .
\end{equation}
The $L_{\rm UV}^{\rm dust-free}/L_{\rm UV}^{\rm dust}$ values follow a log-normal distribution since $\beta$ has a Gaussian distribution.
%The mean value, $\mu$, of $\ln{\big(L_{\rm UV}^{\rm dust-free}/L_{\rm UV}^{\rm dust}\big)}$ is $0.4\,\ln{10}\,(C_{0}+C_{1}\langle \beta \rangle)$ with a standard deviation, $\sigma$, of $0.4\,\ln{10}\,C_{1}\sigma_{\beta}$. This results in:
%\begin{equation}
%\dfrac{\langle L_{\rm UV}^{\rm dust-free}\rangle}{L_{\rm UV}^{\rm dust}} \! = \!
%\!=&\exp\big[0.4\,\ln{10}\,(C_{0}\!+\!C_{1}\langle \beta 
%\rangle)\!+\!(0.4\,\ln{10}\,C_{1}\sigma_{\beta})^{2}/2\big].
%\end{equation}
Since this equation is evaluated at a given $L_{\rm UV}^{\rm dust}$, we have ${\langle L_{\rm UV}^{\rm dust-free}/L_{\rm UV}^{\rm dust}\rangle=\langle L_{\rm UV}^{\rm dust-free} \rangle/L_{\rm UV}^{\rm dust}}$. Then the mean attenuation at a given $L_{\rm UV}^{\rm dust}$ or $M_{\rm UV}^{\rm dust}$ is given by:
\begin{align}
\label{eq:auvbeta}
\langle A_{\rm UV} \rangle (M_{\rm UV}^{\rm dust}) =& -2.5\log\left(\langle L_{\rm UV}^{\rm dust-free} \rangle/L_{\rm UV}^{\rm dust}\right)\\
=&\,C_{0}+0.2\,\ln{10}\,C_{1}^{2}\,\sigma_{\beta}^{2}+C_{1}\langle \beta \rangle (M_{\rm UV}^{\rm dust}).\nonumber
\end{align}
We can now use this relation of averages to link $M_{\rm UV}^{\rm dust-free}$
with $M_{\rm UV}^{\rm dust}$ based on ${M_{\rm UV}^{\rm dust}=M_{\rm UV}^{\rm dust-free}+\langle A_{\rm UV} \rangle (M_{\rm UV}^{\rm dust})}$, where we set
negative $\langle A_{\rm UV} \rangle (M_{\rm UV}^{\rm dust})$ values to zero to avoid
an unphysical attenuation.  Replacing $\langle A_{\rm UV} \rangle (M_{\rm UV}^{\rm
dust})$ with Equation~\ref{eq:auvbeta}, and then substituting $\langle \beta
\rangle$ with Equation~\ref{eq:betamuv} finally yields:
\begin{align}
M_{\rm UV}^{\rm dust} =
\begin{cases}
\frac{1}{1-C_{1}\diff{\beta}{M_{\rm UV}^{\rm dust}}(z)}\Big[ & M_{\rm UV}^{\rm dust-free}+C_{0}+C_{1}\beta_{M_{0}}(z)- \\ &C_{1}\diff{\beta}{M_{\rm UV}^{\rm dust}}(z)M_{0}
+0.2\,\ln{10}\,C_{1}^{2}\,\sigma_{\beta}^{2}  \Big], \\ & \rm{for}\,\langle A_{\rm UV} \rangle (M_{\rm UV}^{\rm dust})>0 \\ \\
M_{\rm UV}^{\rm dust-free},&\,\rm{for}\,\langle A_{\rm UV} \rangle (M_{\rm UV}^{\rm dust})\leq0.
\end{cases}
\end{align}
This relation links $M_{\rm UV}^{\rm dust-free}$ with $M_{\rm UV}^{\rm dust}$ and
allows us to apply dust attenuation corrections to the dust-free galaxy
rest-frame UV magnitudes calculated in Section~\ref{sec:intrinsic}. \\

\noindent {\it (ii) Magnitude based average:}
Instead of calculating average luminosities, we can also directly calculate
average dust-free rest-frame UV magnitudes for a given dust-attenuated
rest-frame UV magnitude. Taking magnitudes instead of luminosity averages
then leads to:
\begin{equation}
\langle A_{\rm UV} \rangle (M_{\rm UV}^{\rm dust}) =C_{0}+C_{1}\langle \beta \rangle (M_{\rm UV}^{\rm dust}).
\label{eq:auvbeta2}
\end{equation}
Replacing $\langle A_{\rm UV} \rangle (M_{\rm UV}^{\rm dust})$ in ${M_{\rm UV}^{\rm dust}=M_{\rm UV}^{\rm dust-free}+\langle A_{\rm UV} \rangle (M_{\rm UV}^{\rm dust})}$ with
Equation~\ref{eq:auvbeta2}, and then replacing $\langle \beta \rangle$ with
Equation~\ref{eq:betamuv} we find:
\begin{align}
M_{\rm UV}^{\rm dust} =
\begin{cases}
\frac{1}{1-C_{1}\diff{\beta}{M_{\rm UV}^{\rm dust}}(z)}\Big[ & M_{\rm UV}^{\rm dust-free}+C_{0}+ \\ & C_{1}\beta_{M_{0}}(z)- C_{1}\diff{\beta}{M_{\rm UV}^{\rm dust}}(z)M_{0}
\Big], \\ & \rm{for}\,\langle A_{\rm UV} \rangle (M_{\rm UV}^{\rm dust})>0 \\ \\
M_{\rm UV}^{\rm dust-free},&\,\rm{for}\,\langle A_{\rm UV} \rangle (M_{\rm UV}^{\rm dust})\leq0.
\end{cases}
\end{align}
Again this relation links $M_{\rm UV}^{\rm dust-free}$ with $M_{\rm UV}^{\rm dust}$
and allows us to apply dust attenuation corrections to the dust-free
rest-frame UV magnitudes of galaxies derived in Section~\ref{sec:intrinsic}. However, the $\sigma_\beta$ term does not appear in
this relation since the average has been performed over magnitudes rather than
luminosities. We will demonstrate below that this does not affect the resulting dust correction in any significant way.\\

\begin{figure}
\includegraphics[width=0.51\textwidth]{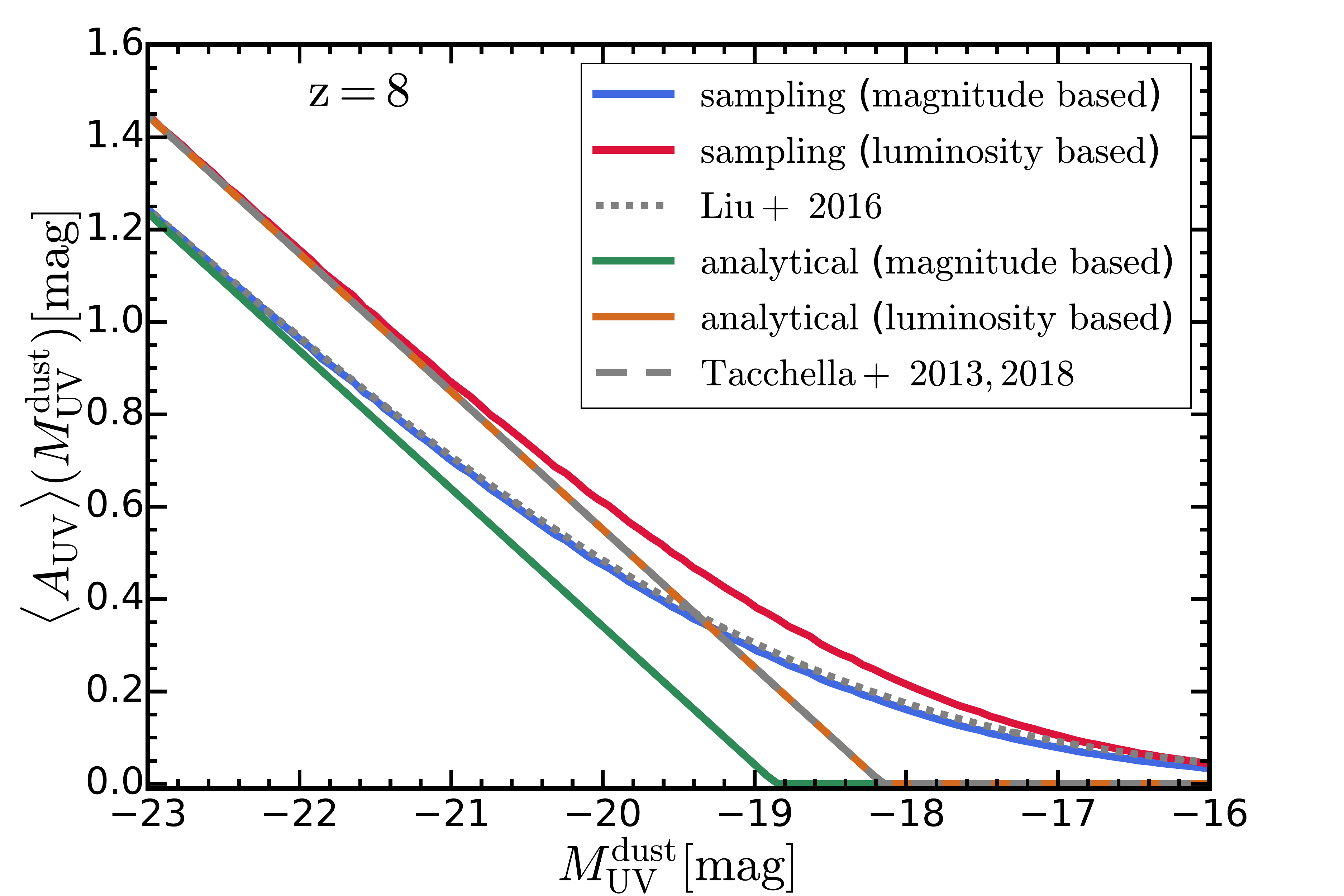}
\includegraphics[width=0.51\textwidth]{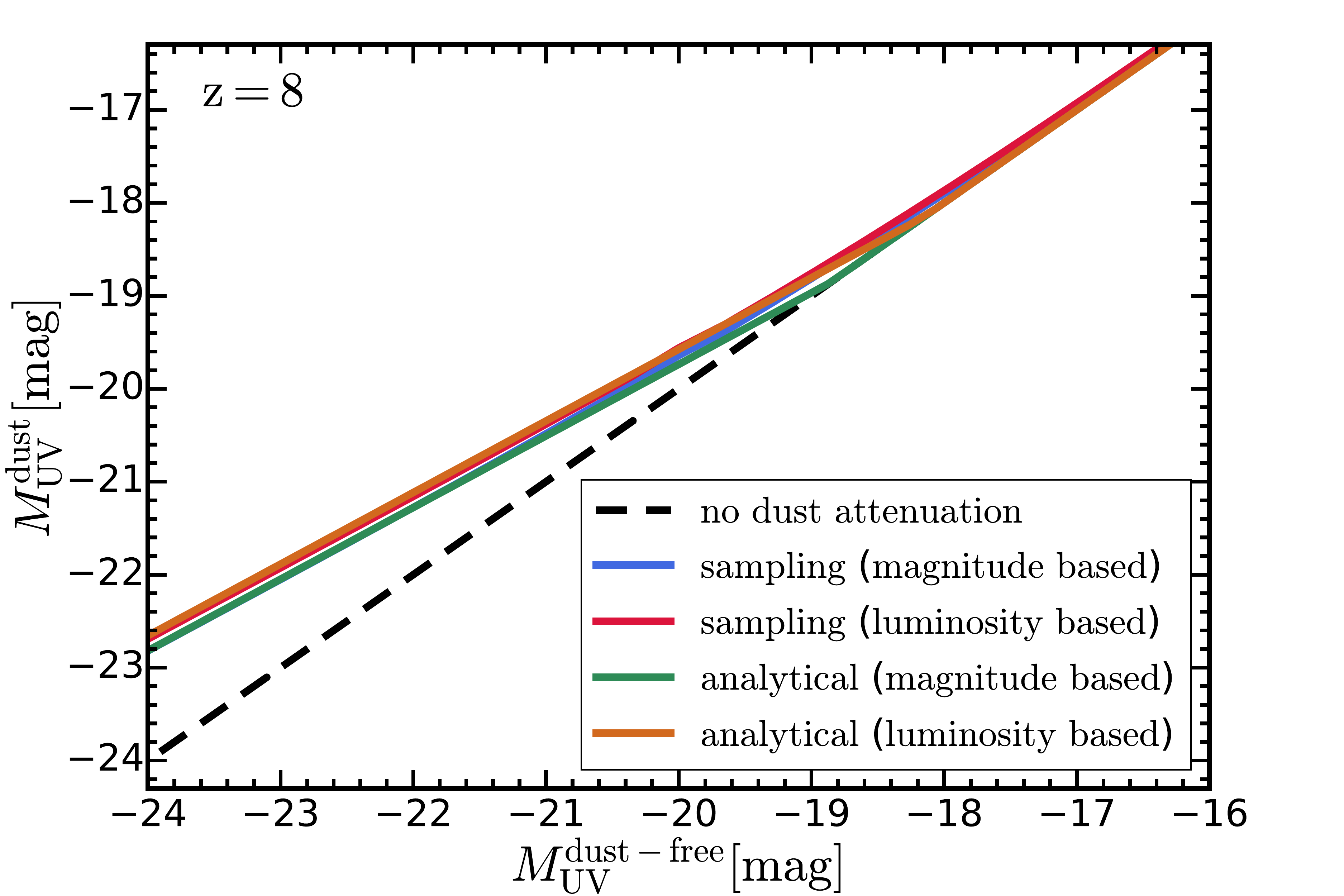}
\caption{{\bf Dust attenuation relations of dust Model A.} {\it Top panel:} Derived ${\langle A_{\rm UV} \rangle (M_{\rm UV}^{\rm dust}) - M_{\rm UV}^{\rm dust}}$ relations at $z=8$ for the different approaches described in the text. For the parameters of the $M_{\rm UV}^{\rm dust}-\langle \beta\rangle$ relation we employ the values of \citet{Bouwens2014a}. The various methods lead to visible differences for the derived attenuation relations. {\it Bottom panel:} Derived $M_{\rm UV}^{\rm dust-free}$ to $M_{\rm UV}^{\rm dust}$ mapping at $z=8$ for the different approaches. Despite the differences in the ${\langle A_{\rm UV} \rangle (M_{\rm UV}^{\rm dust}) - M_{\rm UV}^{\rm dust}}$ relations, we find that the mappings between dust-free rest-frame UV magnitudes and dust-attenuated magnitudes differ only slightly for the different approaches. 
We also compare the relations to results from~\protect\cite{tacchella2013,liu2016,tacchella2018}. Despite the small differences between the various mappings all approaches lead to the same effective dust attenuation correction of dust-free magnitudes.}
\label{fig:AuvMuv}
\end{figure}

\noindent{\it Numerical sampling approach:} We next discuss a numerical approach to link dust-attenuated and dust-free magnitudes that is based on a numerical sampling of the underlying distributions. This method can also be applied to luminosities or magnitudes directly. For this approach the attenuation of numerical samples with negative $A_{\rm UV}$ will be set to zero, which leads to a rise in $\langle A_{\rm UV} \rangle$. This is the main difference compared to the analytical approach. This method has, for example, been adopted in \citet{smit2012,liu2016}.

\noindent {\it (i) Luminosity based average:}
We first select $200$ linearly spaced $M_{\rm UV}^{\rm dust}$ values ranging
from $-27\mmag$ to $-13\mmag$.  At each $M_{\rm UV}^{\rm dust}$ or correspondingly $L_{\rm UV}^{\rm
dust}$, we sample $10^5$ $\beta$ values drawn from a normal distribution
centred around $\langle \beta \rangle (M_{\rm UV}^{\rm dust})$ with
$\sigma_{\beta}=0.34$. We then calculate the $L_{\rm UV}^{\rm dust-free}$ value for each
individual sample using the ${\rm IRX}-\beta$ relation. If the resulting $L_{\rm
UV}^{\rm dust-free}$ is smaller than $L_{\rm UV}^{\rm dust}$, we set $L_{\rm UV}^{\rm
dust-free}=L_{\rm UV}^{\rm dust}$; i.e. we set the attenuation to zero for these negative cases. Finally, we combine all the samples and
divide them based on their $M_{\rm UV}^{\rm dust-free}$ values into $30$ linearly spaced
bins ranging from $-25\mmag$ to $-15\mmag$. In each bin, the mean $L_{\rm UV}^{\rm dust}$
of the samples within each bin is then calculated. The $M_{\rm UV}^{\rm dust}$ value
corresponding to this mean $L_{\rm UV}^{\rm dust}$ is then derived. This then finally results in an average mapping from $M_{\rm UV}^{\rm dust-free}$ to $M_{\rm
UV}^{\rm dust}$. \\

\noindent {\it (ii) Magnitude based average:}
Similar to the luminosity average approach, we first select $200$ linearly spaced
$M_{\rm UV}^{\rm dust}$ values ranging from $-27\mmag$ to $-13\mmag$. At each $M_{\rm
UV}^{\rm dust}$, we sample $10^5$ $\beta$ values with a normal distribution
centred at $\langle \beta \rangle (M_{\rm UV}^{\rm dust})$ with
$\sigma_{\beta}=0.34$. We then calculate the $A_{\rm UV}$ value for each individual
sample with the relation $A_{\rm UV}=C_{0}+C_{1}\beta$. Negative $A_{\rm UV}$
values are set to zero. We then calculate $M_{\rm UV}^{\rm dust-free}$ for
each individual sample with $M_{\rm UV}^{\rm dust-free}=M_{\rm UV}^{\rm dust}-A_{\rm
UV}$. Finally, we combine all the samples and split them based on their
$M_{\rm UV}^{\rm dust-free}$ values into $30$ linearly spaced bins ranging from $-25\mmag$ to
$-15\mmag$. In each bin, the mean $M_{\rm UV}^{\rm dust}$ of the samples within the
bin is then calculated. This then results in an averaged mapping from
$M_{\rm UV}^{\rm dust-free}$ to $M_{\rm UV}^{\rm dust}$.\\

In Figure~\ref{fig:AuvMuv}, we compare the  ${\langle A_{\rm UV} \rangle (M_{\rm
UV}^{\rm dust}) - M_{\rm UV}^{\rm dust}}$ relations (top panel) and the
resulting mappings from $M_{\rm UV}^{\rm dust-free}$ to $M_{\rm UV}^{\rm dust}$
(bottom panel) derived by the different approaches evaluated at $z=8$. For the parameters of the $M_{\rm UV}^{\rm dust}-\langle \beta\rangle$ relation we employ the values of \citet{Bouwens2014a}. The dust attenuation curves differ between the analytical and numerical sampling
approach towards the faint end of the galaxy population. For the numerical sampling
approach, samples with negative $A_{\rm UV}$ are set to zero
attenuation which leads to a rise in $\langle A_{\rm UV} \rangle$, especially
towards the faint end where $A_{\rm UV}$ can easily drop below zero. The magnitude based and luminosity based averages also lead to differences mainly at the bright end. Luminosity based averages lead to stronger attenuation due to the extra term caused by the assumed scatter in $\beta$. Although these different approaches and methods result in differences in the derived  ${\langle A_{\rm UV}
\rangle (M_{\rm UV}^{\rm dust}) - M_{\rm UV}^{\rm dust}}$ relations, the
differences in the final magnitude mappings are negligible as demonstrated in the lower panel of Figure~\ref{fig:AuvMuv}. In the following we
will therefore adopt the analytic approach with luminosity based averages to correct our
dust-free magnitudes for dust in Model A. 

The advantage of this empirical dust model is that it follows from
observational scaling relations, and only depends on the dust-free galaxy magnitudes as predicted by the simulation. At the same time however, this is also a weakness. Specifically, both the observed ${M_{\rm UV}^{\rm dust}-\langle \beta
\rangle}$ relation and the observed ${\rm IRX}-\beta$ relation still contain significant
uncertainties. \cite{Bouwens2014a} provides currently the most complete observational dataset to derive the ${M_{\rm UV}^{\rm dust}-\langle \beta
\rangle}$ relation with more than $4000$ galaxies at $z=4-8$. However, there
have been several other studies of the ${M_{\rm UV}^{\rm dust}-\langle \beta
\rangle}$ relation~\citep[][]{Wilkins2011, finkelstein2012,
Bouwens2012, dunlop2013, rogers2013}. Besides different sample selections, also
the measurement of the $\beta$ slopes differ between these studies, which causes
significant differences of up to $\Delta\beta=0.5$.
After correcting for different systematic biases, \cite{Bouwens2014a} demonstrated
that these different observational studies are roughly in agreement
($\Delta\beta<0.3$) with each other. Furthermore, the underlying ${\rm IRX}-\beta$
relation is also uncertain~\citep{meurer1999}. Specifically, no consensus has
yet been reached regarding the detailed ${\rm IRX}-\beta$ relation at $z\gtrsim3$. Several
studies~\citep[e.g.,][]{reddy2010, reddy2012, nordon2013, McLure2013,
 koprowski2018, mclure2018, reddy2018} have explored the ${\rm IRX}-\beta$ relationship
at high redshifts. For typical but little or modestly obscured systems, the
results are broadly consistent with the~\cite{meurer1999} ${\rm IRX}-\beta$ relation, but deviate from it for more IR luminous, highly obscured galaxies. However,
there is still a debate on whether they support a gray attenuation
curve~\citep[e.g.,][]{calzetti2000, mclure2018, koprowski2018} or are more
Small Magellanic Cloud-like~\citep[e.g.,][]{capak2015,reddy2018}.  All these caveats have to be taken
into account when using Model A. Given these observational inconsistencies, we will in the following treat the two parameters of the ${M_{\rm UV}^{\rm dust}-\langle \beta
\rangle}$ relation, the intercept $\beta_{M_0}$ and the slope  ${\rm d}\beta/{\rm d}M_{\rm UV}$, as adjustable parameters that can be derived through a calibration procedure described below. We have also tested explicitly  that the observationally derived ${M_{\rm UV}^{\rm dust}-\langle \beta
\rangle}$ relations lead to IllustrisTNG dust-attenuated galaxy luminosity functions that are inconsistent with observed rest-frame UV luminosity functions. This also motivates a calibration based approach for dust Model A.

\subsubsection{Model B -- resolved dust optical depth model}

For our second dust model we explicitly consider the resolved gas properties
around every stellar particle to derive the dust attenuation based on an
estimate of the dust optical depth of the surrounding gas. We note that this
model does not include dust emission, which mainly affects the rest-frame SED
of galaxies in far-IR and therefore has no influence on rest-frame UV or other
band magnitudes that we study in this paper. In addition to the resolved
component, this second dust model also includes an unresolved dust component for dust
attenuation. This unresolved component is required due to the limited
numerical resolution of our simulations. Below we describe the implementation of both
dust components in detail. This method is similar to the fiducial dust-model C of \citet{Nelson2018} which was used to compare IllustrisTNG colour distributions versus SDSS at $z\sim0$.

{\it Resolved dust:} To calculate the resolved dust component, we first map the gas distribution around each galaxy
onto a cubic grid. This cube has a side length of $60\pkpc$ centred on
the most bound particle of each subhalo hosting the galaxy. This physical side length is sufficiently
large to cover most of the gas that contributes to dust attenuation at all
redshifts. In the following we observe each galaxy along the $z$ axis. The cubic
grid has a pixel size of $1\pkpc$ in the $x$-$y$ plane and samples the
$z$ direction with a spacing of $0.1\pkpc$.  We verified that the
resulting luminosities and magnitudes do not change if we, for example, increase the $x-y$ plane pixel size
to $2\pkpc$. We assume that dust is only forming in cold phases of the interstellar medium (ISM) and gas
cells therefore need to be either star-forming or cold enough with temperature
$<8000\,{\rm K}$ to contribute to the dust grid values. The hydrogen mass and gas metallicity within each grid cell are then calculated based on this criterion. For each stellar particle we then find its $z$-location on the grid and calculate accumulated cold gas values of all the grid cells in front of the stellar particle along the line-of-sight direction. Specifically, we calculate the hydrogen column density $N_{\rm H}$ and hydrogen mass-weighted gas metallicity $Z_{\rm g}$. Finally, we compute the V band ($547.7\nm$) resolved optical depth for this stellar particle as~\citep{guiderdoni1987,Devriendt1999,Devriendt2000,DeLucia2007,kitzbichler2007,Fontanot2009,guo2009,clay2015,Nelson2018}:
\begin{equation}
\tau_{\rm V}^{\rm res} \, = \, \tau_{\rm dust}(z) \,\, \left(\frac{Z_{\rm g}}{Z_{\odot}}\right )^\gamma \,\, \left( \frac{N_{\rm H}}{N_{\rm H,0}} \right)\,,
\label{eq:tau_res_V}
\end{equation}
where ${\gamma=1}$, ${Z_{\odot}=0.0127}$ and ${N_{\rm H,0}=2.1\times
10^{21}\cm^{-2}}$ are normalisation values for gas metallicity and hydrogen
column density, respectively. In~\cite{guiderdoni1987}, $\gamma$ is typically wavelength dependent
with two regimes interpolating extinction curves between the solar neighbourhood and
the Magellanic clouds. Here, we use this relation to derive the V band optical
depth in systems with different metallicity. \citet{guiderdoni1987} and some later works set $\gamma=1.6$ for
$\lambda>200\nm$. However, in \citet{somerville2012, Yung2018}, the optical
depth for the V band is assumed to linearly depend on the metal mass in the galaxy
disc, which is equivalent to $\gamma=1$. \cite{Wilkins2017} also assumed that the dust optical depth depends linearly
on metal column density. \citet{ma2017}, following the approach in
\citet{Hopkins2005}, also scaled dust attenuation linearly with metallicity.
We have tested explicitly that different $\gamma$ values do not lead
to significant differences in the resulting luminosity functions. We therefore adopt
$\gamma=1$ to derive the V band optical depth in the following. This then also implies that the redshift dependent
scale factor ${ \tau_{\rm dust}(z)}$ scales like the average dust-to-metal ratio. We note that there are also
some variations in the literature in terms of the employed gas density and metallicity values entering
the optical depth relation above.
For example, \cite{Nelson2018} used the neutral hydrogen density and neutral hydrogen mass weighted metallicity instead of hydrogen density and gas metallicity in Equation~\ref{eq:tau_res_V}. However, they studied only the low redshift galaxy population.  Other studies~\citep[e.g.,][]{guiderdoni1987,Devriendt1999,Devriendt2000,clay2015}, referred to $N_{\rm H}$ as hydrogen column density, while \citet{guo2009} referred to it as neutral hydrogen column density. \cite{kitzbichler2007} referred to the term used in \citet{Devriendt1999} as neutral hydrogen column density, while they referred to their $N_{\rm H}$ as hydrogen column density. As stated above, we employ here the hydrogen column density for calculating the V band optical depth.

\begin{figure}
\centering
\includegraphics[width=0.51\textwidth]{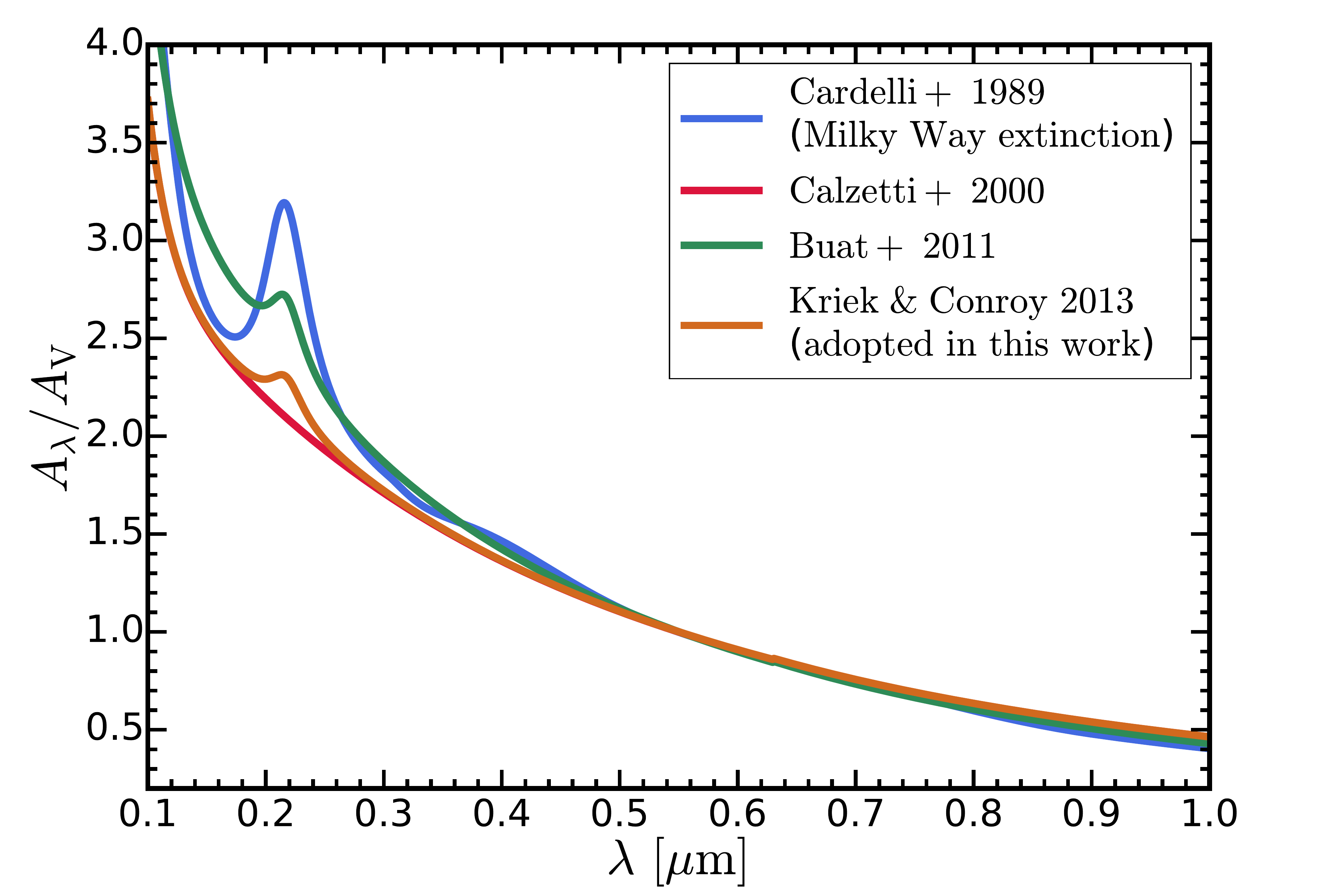}
\caption{{\bf Different dust attenuation and extinction curves.} The Milky Way extinction curve~\citep{Cardelli1989} shows a strong UV feature. The Calzetti attenuation curve~\citep{calzetti2000} is based on observations of local starburst galaxies. The attenuation curve we adopt for our work~\citep{kriek2013} adds a parametrised slope correction and UV feature on top of the Calzetti attenuation curve~\citep{noll2009}. We set the slope correction $\delta$ to zero while keeping the UV feature. Based on observations, \citet{Buat2011} derived the sample averaged parameters for this attenuation curve. They followed the same attenuation parametrisation as \citet{kriek2013,noll2009} but determined the parameters through average values of $30$ galaxies at in the redshift range $1<z<2$.}
\label{fig:extinction_curves}
\end{figure}

Next, we have to translate the V band optical depth to an actual dust attenuation. The attenuation depends on the geometry of dust and stars.
In the following, we employ the dust geometry model described in
\citet{calzetti1994}, where the ionised gas and dust are co-spatial and
uniformly mixed. We note that we calculate here dust attenuation and not
extinction. Extinction only considers the removal of photons from the line of
sight, both absorption and scattering. The light comes from a background point
source and the dust is entirely foreground to the source. Extinction therefore
is insensitive to the detailed spatial distribution of the foreground dust.
However, attenuation refers to the situation where the radiation sources are
distributed within the dust with a range of depths. Both radiation sources and
dust have an extended and complex spatial distribution. The photons will not
only be removed from a given line of sight, they can also be scattered into the line of
sight from other points in the extended source. To calculate the dust
attenuation, the radiative transfer equations have to be solved to derive the
relation between observed luminosity and intrinsic luminosity. Using the
solution of the radiative transfer equation for a homogeneous mixture of stars and dust,
the V band attenuation can be written as~\citep{calzetti1994}:
\begin{equation}
A_{\rm V}^{\rm res}\,=\,-2.5\log\left(\frac{L_{\rm V}^{\rm dust}}{L_{\rm V}^{\rm dust-free}}\right)\,=\,-2.5\log\left(\frac{1-e^{-\tau_{\rm V}^{\rm res}}}{\tau_{\rm V}^{\rm res}}\right).
\end{equation}
Besides specifying the V band optical depth and attenuation, we also need to assume a certain
functional form for the attenuation curve to calculate the dust attenuation at
other wavelengths. This curve depends on details of the underlying dust
distribution, like density, grain size distribution and chemical composition.
In the absence of detailed knowledge of these quantities, we have to assume a
certain attenuation curve. The \cite{calzetti2000} attenuation curve is commonly adopted for the
study of high redshift galaxies. This curve is shown in
Figure~\ref{fig:extinction_curves} along with other extinction and attenuation
curves. The Calzetti relation was first proposed for local starburst galaxies and can be described
as~\citep{calzetti2000}:
\begin{equation}
A^{\rm res}(\lambda) \, = \, A_{\rm V}^{\rm res} \, \left ( \frac{k^\prime(\lambda)}{4.05} \right )\,,
\end{equation}
where $k^{'}(\lambda)$ is the attenuation curve normalised to $A_{\rm V}=4.05$: 
\begin{equation}
k^\prime(\lambda) = 4.05 + 2.659 \, 
\begin{cases} 
\left(-1.857+\frac{1.040}{\lambda}\right), & \\ \quad {\rm for} \quad 0.63\mum<\lambda<2.20\mum \\ \\
\left(-2.156+\frac{1.509}{\lambda}-\frac{0.198}{\lambda^2}+\frac{0.011}{\lambda^3}\right),& \\ \quad {\rm for} \quad 0.12\mum<\lambda<0.63\mum.
\end{cases}
\end{equation}
Attenuation outside of these wavelength ranges can be derived through
extrapolation. We note that the Calzetti attenuation law does not
include the $217.5\nm$ UV bump feature in dust attenuation, which is non-negligible for high redshift
galaxies~\citep[e.g.,][]{kriek2013, scoville2015}. In addition, the slope of
the attenuation curve is observed to vary among different galaxies or
sight lines, which imprints non-universal features in dust
attenuation~\citep[e.g.,][]{noll2009, kriek2013, salmon2016}. Therefore
\cite{kriek2013} proposed a modified version of the Calzetti law to account for
the UV bump feature and the change in slope:
\begin{equation}
A^{\rm res}(\lambda) \, = \, \frac{A_{\rm V}^{\rm res}}{4.05} \, \big[k^\prime(\lambda)+D(\lambda)\big] \, \left(\frac{\lambda}{\lambda_{\rm V}}\right)^{\delta},
\end{equation}
where $\delta$ is the correction of the steepness of the attenuation curve and
$D(\lambda)$ is a Lorentzian-like Drude profile to parameterise the UV bump
feature defined as:
\begin{equation}
D(\lambda)=\frac{E_{\rm b}(\lambda \, \Delta\lambda)^2}{(\lambda^2-\lambda_{0}^2)^2+(\lambda \, \Delta\lambda)^2},
\end{equation}
where $E_{\rm b}$ is the parameter defining the strength of the UV bump,
${\lambda_{0}=217.5\nm}$ is the central wavelength of the UV bump, and
$\Delta\lambda=35\nm$ is the FWHM of the bump~\citep{Seaton1979,noll2009}.
\cite{kriek2013} found a relation between $E_{\rm b}$ and $\delta$:
\begin{equation}
E_{\rm b}=(0.85\pm 0.09)-(1.9\pm0.4) \, \delta.
\end{equation} 
Therefore the shape of the attenuation curve is purely characterised by
$\delta$. In \citet{Buat2011}, $30$ galaxies at $1<z<2$ were studied and they
derived an averaged attenuation curve with $\delta=-0.13$, $E_{\rm b}=1.26$ and
$\Delta\lambda=35.6\nm$. For simplicity, we assume $\delta=0$ in this work,
which implies no correction to the slope of the Calzetti attenuation curve while
introducing the UV bump feature.

We then finally get $A^{\rm res}(\lambda)$ for each stellar particle in the
galaxy as:
\begin{equation}
A^{\rm res}(\lambda) \, = \, \frac{A_{\rm V}^{\rm res}}{4.05} \, \big[k^\prime(\lambda)+D(\lambda)\big].
\end{equation}
To calculate the dust attenuated magnitudes of an arbitrary band, we determine the
attenuation at the effective wavelength of the band, $A(\lambda_{\rm eff})$, to
represent the attenuation in the entire wavelength range covered by a given filter: 
\begin{equation}
M^{\rm dust}=M^{\rm dust-free}+A^{\rm res}(\lambda_{\rm eff}),
\label{eq:m_dust_band}
\end{equation}
where $M^{\rm dust}$ and $M^{\rm dust-free}$ are the dust-attenuated and dust-free
magnitudes of an arbitrary band, respectively. We have verified that Equation~\ref{eq:m_dust_band} is
a good approximation of the detailed procedure applying the attenuation to each wavelength and then convolving with the filter transmission function. For example, for a SSP with initial stellar mass of $10^{10}\msun$ and solar metallicity and applying the Calzetti attenuation curve with $A_{V}=0.2$, we find that the rest-frame magnitude differences between these two approaches are $\leq 0.01$ in all SDSS optical bands and the GALEX FUV band.

{\it Unresolved dust:} Apart from the resolved dust attenuation of the ISM,
there are also dust components which cannot be resolved by our simulations due to
the limited numerical resolution. One major component that cannot be resolved
by our simulations are the birth clouds surrounding young stars, which can
cause significant dust attenuation. We also have to account for this dust
attenuation contribution in addition to the resolved dust component. 
It is common to include this contribution using the power law extinction
model presented in~\cite{charlot2000}. This model specifically considers the
attenuation contribution from birth clouds around young stars. Many
semi-analytic and hydrodynamical works have adopted this model to extend their
resolved ISM attenuation model to smaller scales to add the attenuation
contribution from birth clouds. The original~\citet{charlot2000} model includes
the attenuation by finite lifetime birth clouds surrounding young stellar
populations as well as the ambient diffuse ISM. According to this model the V
band optical depth can then be described as a combination of these two
contributions:
\begin{equation}
\tau_{\rm V} =  \, 
\begin{cases} 
(\tau_1+\tau_2) , & \quad {\rm for} \quad t_{\rm ssp}\leq t_{\rm esc} \\ \\
\tau_2 , & \quad {\rm for} \quad t_{\rm ssp}>t_{\rm esc},
\end{cases}
\label{eq:bc_att}
\end{equation}
where $\tau_{1}$ is the optical depth of unresolved birth clouds around young
stars, $\tau_{2}$ is the optical depth of the ambient ISM gas and $t_{\rm esc}$
is an age criterion defining young stars associated with birth clouds. Stars
older than $t_{\rm esc}$ can escape the birth clouds and their light is not
subject to attenuation due to the birth cloud environment anymore.

Our dust model already includes the resolved ISM component as described above. However, it does not
include an attenuation contribution from the birth cloud environment. We
therefore have to add this unresolved birth cloud component. This approach has
also been commonly employed in other studies~\citep[e.g.,][]{guo2009,
somerville2012, clay2015}. We note that sometimes only the
resolved ISM contribution to dust is considered neglecting the birth cloud
attenuation~\citep[][]{kitzbichler2007, Yung2018}. For example, \cite{Yung2018}
states that the addition of attenuation due to birth clouds leads to an
overprediction of attenuation at high redshifts. Some models also include a
resolved ISM component along with the birth cloud and ambient ISM
combination~\citep[][]{Nelson2018}.

To apply the model for the birth cloud dust attenuation contribution, we have to know the
optical depths values entering Equation~\ref{eq:bc_att}.  At $z=0$, commonly
adopted values are: $\tau_1 =0.7,\ \tau_2 =0.3,\ t_{\rm
esc}=10\Myr$~\citep[e.g.,][]{trayford2015,byler2017,Nelson2018}. At higher
redshifts some works have assumed certain redshift dependencies of these
parameters. For example, \citet{somerville2012} assumed a $(1+z)^{-1}$
($1/[(1+z)\,z]$ for $z>1$) redshift dependence for the optical depth $\tau_1$ and
$(1+z)^{-1}$ for the optical depth $\tau_2$. These redshift dependencies were
chosen to match the observed $L_{\rm IR}/L_{\rm bol}$ relation and the bright
end of the observed UV luminosity functions at $z=0.5-5$. They also assume a
redshift dependence for the age criteria for young stellar populations, $t_{\rm
esc}$, while most other studies set  this to a constant value of $10\Myr$ at all redshifts.
\citet{guo2009}, on the other hand, assumed that the optical depth of young stellar populations is
three times that of the old stellar population, which is suggested by
\citet{charlot2000}. They then chose $(1+z)^{-0.4}$ for the redshift dependence of
the optical depths. \citet{trayford2015} adopted the same ratio between the optical
depth of ISM and birth cloud. \citet{Cullen2017} found in their best-fit model
that the dust optical depth of old stellar populations is $0.3$ times that of
young stellar population. They also found $t_{\rm esc}$ to be around $12\Myr$.
\citet{DeLucia2007} and \citet{clay2015} assumed that the additional optical depth
of dust in birth clouds scales linearly with the resolved dust optical depth.
The linear scale factor is assumed to have some random variation around the median
value of $2.33$ for different galaxies. They have employed a $(1+z)^{-1}$ redshift scaling for the optical depths values.

After reviewing existing unresolved dust models, we now describe our specific implementation of unresolved dust. First, we discard the
attenuation of diffuse ISM from the~\citet{charlot2000} model since this
contribution is already included in our resolved ISM component. This leaves us
with an additional attenuation due to the birth cloud environment that we have
to implement for the unresolved dust component. In the absence of a detailed
physical model for the dust abundance in birth clouds, we assume that the additional
optical depth of birth clouds is two times~\citep{charlot2000,Kong2004,DeLucia2007,guo2009,Henriques2015,clay2015} the mean optical
depth of resolved dust averaged over all the stellar particles in the galaxy:
\begin{equation}
\tau_{\rm V}^{\rm unres} =  \, 
\begin{cases} 
2 \, \langle \tau_{\rm V}^{\rm res}\rangle , & \quad {\rm for} \quad t_{\rm ssp}\leq t_{\rm esc} \\ \\
0 , & \quad {\rm for} \quad t_{\rm ssp}>t_{\rm esc},
\end{cases}
\end{equation}
where $t_{\rm esc}=10\Myr$, and $\langle \tau_{\rm V}^{\rm res}\rangle$ is the
average optical depth of the resolved dust component. This mean value is derived by a simple
average of $\tau_{\rm V}^{\rm res}$ over all stellar particles that contribute to
the stellar light of the galaxy. Our complete dust model therefore has only one free redshift-scaling parameter, which is the
redshift-dependent normalisation of the optical depth $\tau_{\rm dust}(z)$. 

Under a simple uniform dust screen model the solution of the radiative transfer equation can then be used to derive the V band attenuation based on this unresolved optical depth~\citet{charlot2000}:
\begin{equation}
A_{\rm V}^{\rm unres}\!=\!-2.5\log\left(\frac{L_{\rm dust}}{L_{\rm dust-free}}\right)\!=\!-2.5\log\left(e^{-\tau_{\rm V}^{\rm unres}}\right).
\end{equation}
Here the dust screen lies between the observer and source and the source is treated as point source. We note that the equation above simply evaluates to ${A_{\rm V}^{\rm unres}\!=\!1.086\,\tau_{\rm V}^{\rm unres}}$. We adopt a power law attenuation curve for unresolved dust with power law index $-0.7$, the best-fit value in \citet{charlot2000}, such that the attenuation at
other wavelengths is given by:
\begin{equation}
A^{\rm unres}(\lambda)=A_{\rm V}^{\rm unres} \left (\frac{\lambda}{\lambda_{\rm V}}\right)^{-0.7}.
\end{equation}
Similar to the resolved dust model, we then have:
\begin{equation}
M^{\rm dust}=M^{\rm dust-free}+A^{\rm unres}(\lambda_{\rm eff}).
\end{equation}
This then fully specifies the unresolved dust component.

{\it Combined resolved and unresolved dust attenuation:} Given the dust-free $M^{\rm dust-free}$ magnitude of a galaxy, we combine the attenuation from
the resolved and unresolved dust component to calculate the final observed
dust-corrected band magnitude as:
\begin{equation}
M^{\rm dust}=M^{\rm dust-free}+A^{\rm unres}(\lambda_{\rm eff})+A^{\rm res}(\lambda_{\rm eff}),
\end{equation}
where $\lambda_{\rm eff}$ is the effective wavelength of the band filter. This
combined dust attenuation fully specifies our dust Model B.

\subsubsection{Model C -- dust radiative transfer}
For the final dust model we consider the distribution of stars and dust in the
simulated galaxies to perform full Monte Carlo continuum dust radiative transfer
calculations using a modified version of the publicly available {\sc Skirt}
code~\citep{Baes2011,camps2013,camps2015,saftly2014}. {\sc Skirt} traces dust scattering
and absorption for randomly sampled photon packets until they reach a detector. This results
in integrated fluxes for each simulated galaxy. Based on these fluxes we then calculate the rest-frame magnitudes for each band through a convolution with the corresponding {\sc sedpy} transmission functions. In the following we
describe the different components and steps that are required for the radiative transfer
calculation. This method of using {\sc Skirt} post-processing radiative transfer is similar to that recently developed in \citet{Rodriguez2018} to compare IllustrisTNG and PanSTARRS optical galaxy morphologies.

{\it Stellar SEDs:} Stellar particle SEDs are modelled with {\sc Fsps} to be
consistent with the SEDs employed in our other two dust models. To this end
we have modified the {\sc Skirt} code in the following way. We first use {\sc Fsps} to
generate a series of dust-free SED templates as described in the previous
sections. These SED templates have $5994$ wavelength points covering a
wavelength range from $9.1\nm$ to $1\cm$ for the adopted MILES stellar library.
Then we modify the ``FSPSVariableSEDFamily'' class in {\sc Skirt} to implement
this family of template SEDs.  Based on the initial mass, metallicity and age of a given stellar
particle, {\sc Skirt} then generates the SED of the particle through
interpolation over those SED templates. Doppler shifts of spectra due to
the motion of stellar particles are not included here. 

\begin{figure}
\centering
\includegraphics[width=0.5\textwidth]{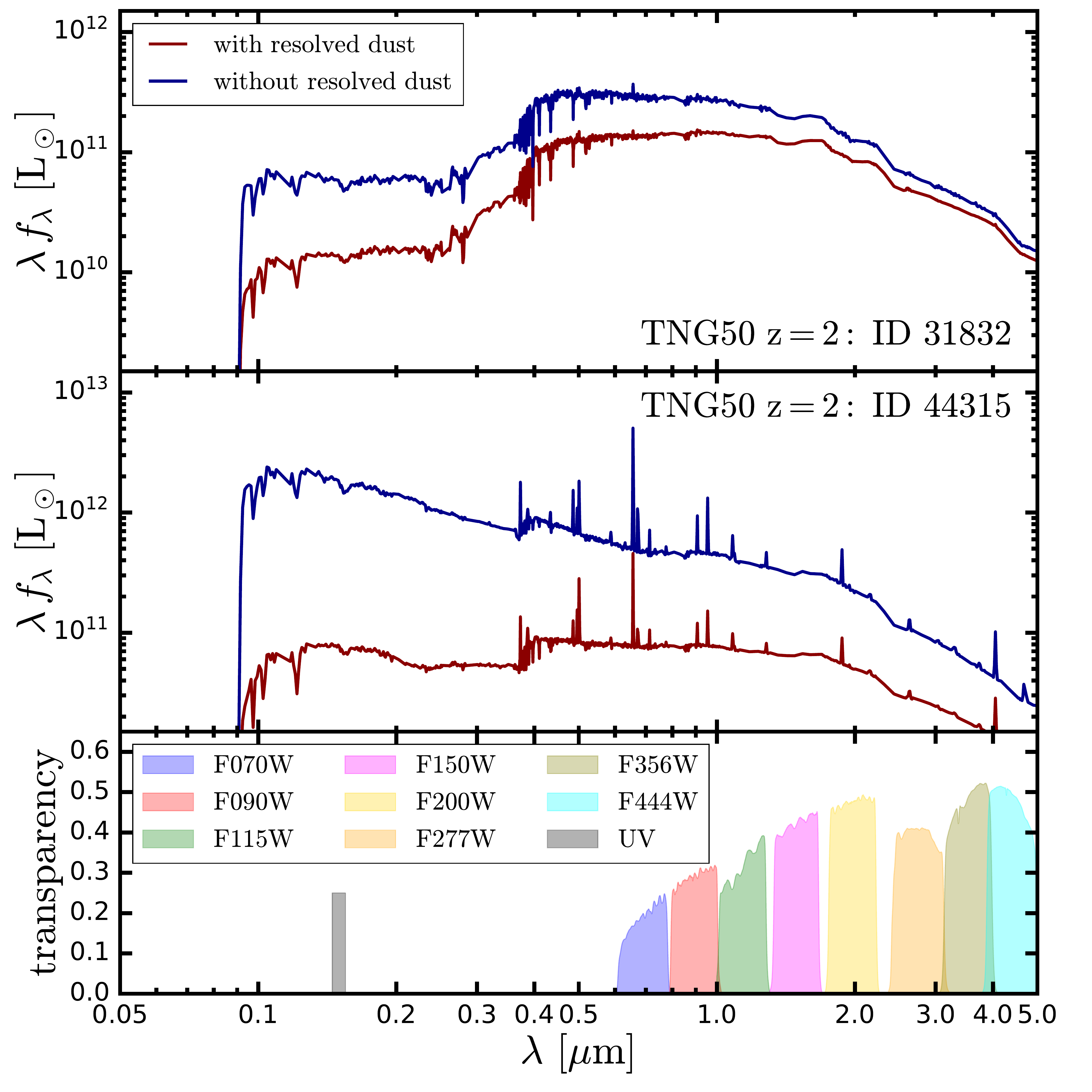}
\caption{{\bf Spectral energy distribution for two example galaxies from TNG50 at $\mathbf{z=2}$.} We show the intrinsic spectral energy distribution (blue) and the dust attenuated distribution (red) using {\sc Skirt} assuming a dust-to-metal ratio of 0.9, which represents the best-fit value of our dust parameter calibration procedure described in the text. The bottom panel shows the transmission functions of the different band filters employed in this work: UV band filter, and the eight wide {\it JWST} NIRCam filters ($\lambda/\Delta\lambda \sim 4-5$).}
\label{fig:modelC_SED}
\end{figure}

{\it Wavelength grid construction:} {\sc Skirt} employs a wavelength grid that
stores the wavelength values for emitted photons. At each of those wavelengths
{\sc Skirt} then emits a prescribed number of photon packets. The output flux
is also sampled by this wavelength grid.  For the purpose of our high
redshift luminosity function predictions between $z=2$ and $z=10$, it is
sufficient to have a wavelength grid between $0.05\mu m$ and $5\mu m$. This
covers all relevant {\it JWST} NIRCam  wide filters at all redshifts, and the rest-frame UV
band.  The SEDs with nebular emission have very narrow and sharp emission
line features. We have therefore designed our wavelength grid in  a specific way to account for this. We
start from the wavelength grid of the input {\sc Fsps} SED templates. This
high resolution grid has $5164$ sampled points between wavelengths of
$0.05\mum$ and $5\mum$. In order to efficiently run {\sc Skirt}, we need to
reduce the number of wavelength points without losing important spectral
features like emission lines. We therefore first select every $12^{\rm th}$ grid point on this grid
resulting in a $12$ times lower resolution grid. In a next step, we select
additional $11$ points around each emission line to ensure that those are not
lost during the grid down-sampling process. The exact wavelengths of these emission lines are
taken from the {\sc Fsps} datafile. Finally, in order to make accurate
predictions for UV magnitudes, we keep all the wavelength grid points between $140\nm$ to $180\nm$, which includes the coverage of our UV
filter from the original fine grid.  This procedure results in a wavelength
grid with $1168$ grid points covering wavelengths from $0.05\mu m$ to $5\mu m$, which suits our need
to efficiently and accurately calculate galaxy fluxes and photometric
properties.

{\it Source distribution:} Next we have to provide {\sc Skirt} with smoothing
length values for the stellar particles representing the radiation sources. This is
required to calculate internally a smoothed photon source distribution function.
We employ a K-D tree algorithm to calculate the smoothing length enclosing $64\pm1$ nearest stellar particles for all stellar particles within the galaxy. Given the spatial location and smoothing length values of stellar particles,
{\sc Skirt} then creates a photon source distribution and emissivity profile through the
entire space by interpolating over these kernels. For each wavelength on the
wavelength grid, $N_{\rm p}$ photon packets are then launched isotropically, for
which the launch positions are determined by the luminosity distribution given
by the smoothed stellar particle distribution at that wavelength. The photon
packets then propagate through the resolved ISM and interact with the dust
cells randomly before they are finally collected by the photon detector. We set $N_{\rm p}$
equal to the number of all the stellar particles bound to a galaxy. This number is by construction 
larger than the number of stellar particles within the $30\pkpc$ aperture. In addition we
apply $N_{\rm p}=10^{5}$ as an upper and $N_{\rm p}=10^{2}$ as a lower limit for the number of generated photon packets. We have tested the convergence of radiative transfer calculations on selected galaxies in TNG50 with $M_{\ast}$ ranging from $10^{7}$ to $10^{11}{\rm M}_{\odot}$. Doubling the number of photon packets leads only to minor, $<0.01\mmag$, changes in all rest-frame UV and optical bands.

\begin{figure*}
\centering
\raisebox{0.2\textwidth}{\rotatebox[origin=t]{90}{\bf\large without resolved dust}}
\includegraphics[width=0.44\textwidth]{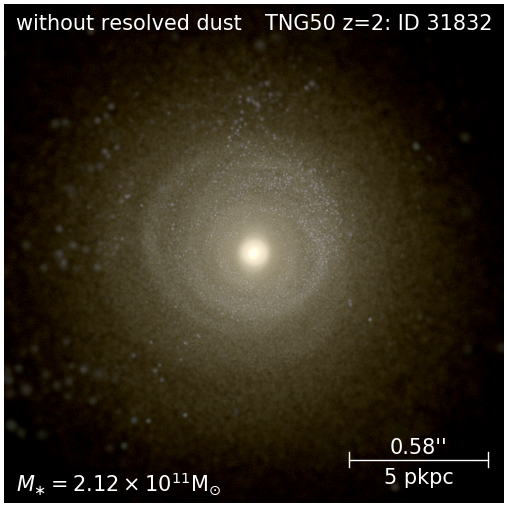}
\includegraphics[width=0.44\textwidth]{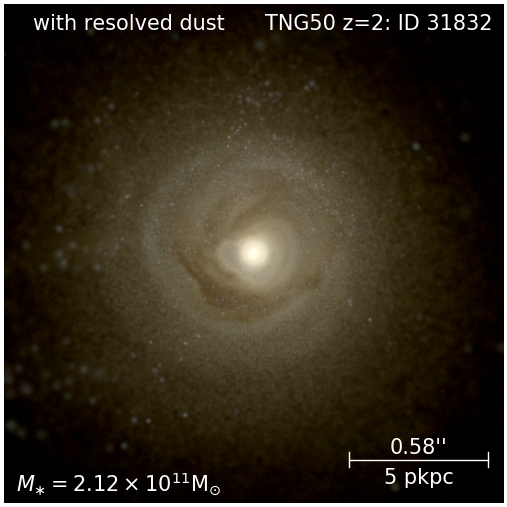}
\raisebox{0.2\textwidth}{\rotatebox[origin=t]{90}{\bf\large with resolved dust (dust-to-metal ratio 0.3)}}\\
\hspace{-0.08cm}\includegraphics[width=0.44\textwidth]{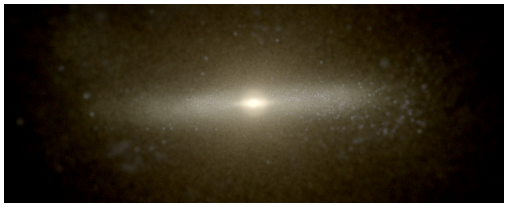}
\hspace{0.025cm}\includegraphics[width=0.44\textwidth]{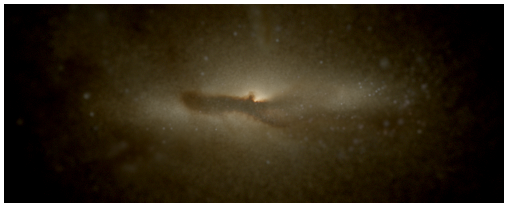}\\
\hspace{-0.15cm}\raisebox{0.2\textwidth}{\rotatebox[origin=t]{90}{\bf\large with resolved dust (dust-to-metal ratio 0.6)}}
\includegraphics[width=0.44\textwidth]{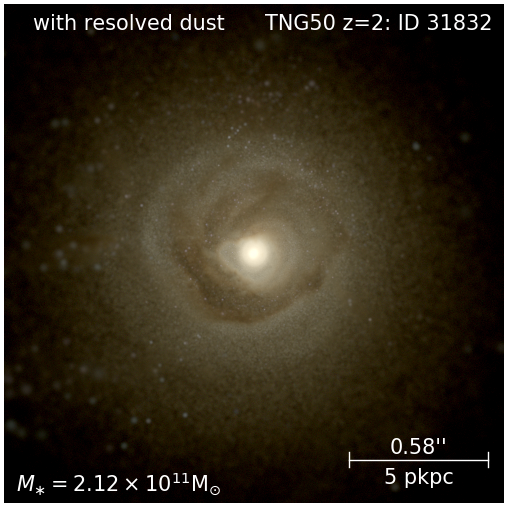}
\hspace{0.02cm}\includegraphics[width=0.44\textwidth]{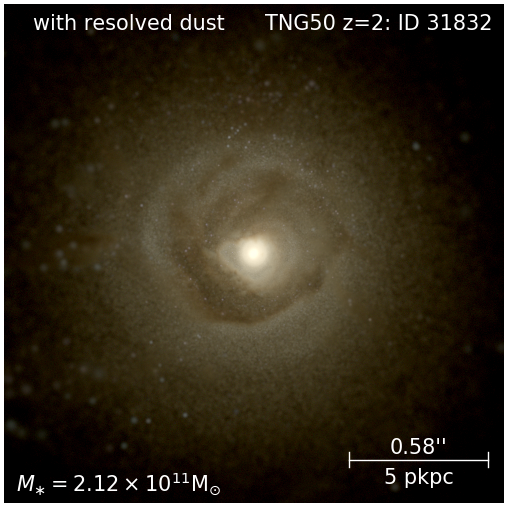}\hspace{0.04cm}\raisebox{0.2\textwidth}{\rotatebox[origin=t]{90}{\bf\large with resolved dust (dust-to-metal ratio 0.9)}}\\
\hspace{-0.02cm}\includegraphics[width=0.44\textwidth]{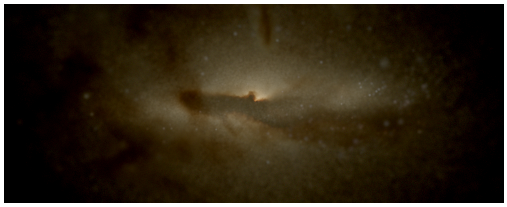}
\includegraphics[width=0.44\textwidth]{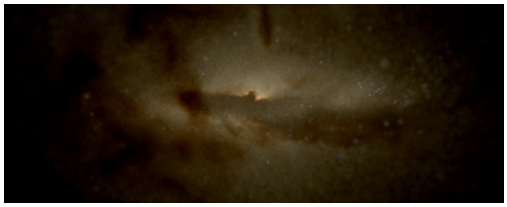}
\caption{{\bf {\it \textbf{JWST}} NIRCam face-on and edge-on images of a TNG50 galaxy at $\mathbf{z=2}$}. {\it Top left panel:} Galaxy images without resolved dust. {\it Other  panels:} Images with resolved dust for three dust-to-metal ratios: $0.3$ (upper right), $0.6$ (lower left) and $0.9$ (lower right).  The images are based on apparent F115W, F150W, F200W filter fluxes of the galaxy ID 31832. The SED of this galaxy is shown in Figure~\ref{fig:modelC_SED}.  The images cover a $18\pkpc \times 18\pkpc$ field of view with $500\times 500$ pixels. The radiative transfer employs $10^7$ photon packets per wavelength on the reduced wavelength grid with $234$ points covering $0.05\mu m$ to $5\mu m$. Our fiducial SED grid employs a finer grid with $1168$ points. The central bright bulge region is covered by dust absorption when observed in the edge-on view. }
\label{fig:image}
\end{figure*}

{\it Dust properties:} The position, gas density and metallicity of the Voronoi
gas cells from the simulation are used to calculate the distribution of dust in
galaxies. We assume a
spatially constant dust-to-metal ratio at each redshift to link the cold gas density and metallicity with the dust density:
\begin{equation}
\rho_{\rm dust}=f_{\rm z}(z)\,\,Z_{\rm g}\,\,\rho_{\rm g}\,,
\end{equation}
where $Z_{\rm g}$ and $\rho_{\rm g}$ are the metallicity and  density of
the gas cell, and $f_{\rm z}(z)$ is a redshift dependent dust-to-metal ratio, which has to be calibrated in the absence of a dust
model to predict self-consistently the dust distribution within galaxies. Following
Model B, only star-forming  gas cells or gas
cells colder than $<8000\,{\rm K}$ contribute to the ISM dust budget.
Gas cells that do not meet any of these criteria have their dust-to-metal ratio set to zero. Dust emission is self-consistently included in this model and we assume a
\citet{draine2007} dust mixture of amorphous silicate and graphitic
grains, including varying amounts of Polycyclic Aromatic Hydrocarbons (PAHs) particles. This model reproduces the average Milky Way extinction curve and is widely used~\citep[e.g.,][]{jonsson2010,Baes2011,Kimm2013,RemyRuyer2014}. The model is an update of the~\citet{Li2001} model to adjust PAHs to match early Spitzer results. The dust mass distribution is then mapped to an adaptively refined grid on which the optical depth at each wavelength is calculated. Specifically, the space is discretized by an octree
algorithm and the root cell is a cube with sidelength of $60\pkpc$ centred
on the position of the most bound particle of the subhalo. The octree is then
refined between a minimum refinement level of $3$ and a maximum refinement level
of $k$. The minimum octree cell size should be smaller than the physical spatial
resolution of the simulation, $60\pkpc/2^{k}\leqslant \epsilon_{\rm gas\,,phy}^{\rm
min}$. For TNG50, the minimum comoving gravitational softening length is
$\epsilon_{\rm gas}^{\rm min} = 0.074\kpc$ comoving, which correspondingly requires a maximum refinement level $k=14$ at $z=10$. For TNG100, $\epsilon_{\rm gas}^{\rm min}=0.185\kpc$ comoving and requires $k=12$ at $z=10$. The cell splitting criterion is set to $2\times 10^{-6}$. This sets the
maximum fraction of total dust mass that can be contained in a single dust
cell. If the dust mass fraction within a dust cell is larger than this value
and the maximum refinement level has not been reached, the cell will be
subdivided further.

{\it Magnitude Calculation:} To finally calculate galaxy magnitudes, we place a detector $1\Mpc$ away from the simulated
galaxy along the positive $z$-direction to collect emitted photons. The integrated galaxy flux is then generated by {\sc Skirt} and then convolved
with the transmission curve of the band transmission curves using the {\sc sedpy} code.
For the calculation of apparent magnitudes, the rest-frame flux is
redshifted and converted to observed spectra. In addition, we also 
correct for IGM absorption~\citep{Madau1995,madau1996} as described in Section~\ref{sec:intrinsic}.

{\it Unresolved components:} Similar to Model B, Model C also suffers from the
coarse numerical resolution of the simulation. Specifically, this limited resolution is
problematic for the modelling of the young stellar component. The treatment of
the young stellar component is affected by two limitations of the simulation:
{\it (i)} the relatively coarse sampling of star formation due to the limited mass
resolution, and {\it (ii)} the inability of resolving birth cloud absorption
associated with recent star formation. In principle, a single newly formed stellar particle or a star-forming
gas cell should represent an ensemble of giant {H\small\Rmnum{2}} regions with
stars formed inside of them with different formation times. Young and massive stars are
then surrounded by dense birth clouds. Consequently, their dust-free stellar
light emission will be reprocessed by a variety of physical processes. {H\small\Rmnum{2}}
regions surrounding stellar populations reprocess the Lyman continuum
photons into nebular continuum and line emissions. Furthermore, the unresolved
dust in {H\small\Rmnum{2}} and photodissociation regions absorbs the
UV/optical light and reprocesses it to IR.

The coarse sampling of star formation can be alleviated through a resampling technique~\citep[][]{trayford2017}. However, this introduces a few additional free parameters for the modelling. We therefore follow another approach here by exploiting the high mass resolution of TNG50, which is about a factor of 15 and 120 higher than TNG100 and TNG300, respectively. Therefore, the TNG50 results are expected to be less affected by the problem of a coarse sampling of star formation given its high mass resolution. Using TNG50 we then apply a resolution correction to TNG100 and TNG300 to match the mass resolution of TNG50. This  resolution
correction procedure, as described in Section~\ref{sec:corr}, alleviates the problem of a too coarse sampling of star formation in TNG100 and TNG300 without introducing additional free parameters.  

The second problem, i.e. the inclusion of unresolved dust attenuation associated with young
stellar populations, requires a modification of the  stellar particles SEDs. Stellar
particles with ages greater than $10\Myr$ are still modelled with {\sc Fsps}
spectra templates while stellar particles younger than $10\Myr$ are modelled with the
{\sc Mappings-\Rmnum{3}} spectra library~\citep[][]{groves2008}. These are based on  1D
photoionization and radiative transfer calculations to model the
radiative transfer of radiation from a newly formed massive stellar
cluster with a \cite{kroupa2001} initial stellar mass function, using {\sc Starburst99}~\citep{leitherer1999} spectra for the dust-free 
stellar light emission. The radiation propagates through the surrounding spherically
symmetric {H\small\Rmnum{2}} and photodissociation regions, whose
covering fractions decrease as a function of time as it is cleared away by the strong winds
from the massive cluster. We note that the use of  the {\sc Starburst99} population synthesis
model and Kroupa initial stellar mass functions~\citep{kroupa2001} causes some inconsistency in our approach, since old
particles are modelled with {\sc Fsps} and a Chabrier initial stellar mass function. However, the resulting
differences are small in the UV and optical
ranges.  \citet{gonzalez2014} found that the differences in NIR are significant between different population synthesis models due to variations in the treatment of the thermally pulsating asymptotic giant branch stars, which should in principle only be important for older stellar population. We have explicitly compared the SEDs of galaxies with young stellar population modelled by {\sc Fsps} versus {\sc Mappings-\Rmnum{3}} and found no difference in the NIR range. We are therefore confident that the inconsistency in the population synthesis models does not have a significant impact on our results for all bands and all redshifts. The final output SEDs then include the stellar light attenuated by dust in the {H\small\Rmnum{2}} and photodissociation regions,
and the corresponding dust, nebular continuum and line emissions. We note
that the {\sc Fsps} nebular emission model adopted in Model B is no longer employed here since those are already modelled through the {\sc Mappings-\Rmnum{3}} calculation. 

\begin{table*}
\centering
\begin{tabular}{ 
p{0.075\textwidth}|p{0.1\textwidth}|p{0.1\textwidth}|p{0.15\textwidth}|p{0.15\textwidth}|p{0.15\textwidth}}
\hline
{\bf redshift} &Model A&& & Model B& Model C\\
$\mathbf{z}$ & $M_0 [{\rm mag}]$ & $\beta_{M_0}$ & ${\rm d}\beta/{\rm d}M_{\rm UV} [{\rm mag}^{-1}]$ & $\tau_{\rm dust}$ & dust-to-metal ratio\\ 
\hline
\hline
{\bf 2} & -19.5 & -1.38   & -0.28   & 0.46  & 0.90\\ 
{\bf 3} & -19.5 & -1.63   & -0.26   & 0.20  & 0.41\\ 
{\bf 4} & -19.5 & -1.80   & -0.30   & 0.13  & 0.24\\  
{\bf 5} & -19.5 & -2.03   & -0.29   & 0.08  & 0.15\\ 
{\bf 6} & -19.5 & -2.32   & -0.47   & 0.06  & 0.11\\ 
{\bf 7} & -19.5 & -2.53   & -0.40   & 0.04  & 0.08\\ 
{\bf 8} & -19.5 & -2.66   & -0.34   & 0.03  & 0.06\\
\hline
\end{tabular}
\caption{{\bf Best-fit dust parameters for dust Models A, B, and C.} For Model A we quote the parameters for the ${M_{\rm UV}^{\rm dust} - \langle \beta \rangle}$ relation: $M_0$, $\beta_{M_0}$, ${\rm d}\beta/{\rm d}M_{\rm UV}$. We note that we keep $M_0=-19.5\mmag$ constant and assume an underlying \citet{meurer1999} relation for the Model A fits. For Model B we quote the optical depth normalisation: $\tau_{\rm dust}$. For Model C we quote the dust-to-metal ratios. These best-fit parameters are all redshift dependent and then uniformly applied to all galaxies at these redshifts. }
\label{tab:modelA_fit}
\end{table*}

The {\sc Mappings-\Rmnum{3}} SED of a specific stellar particle is controlled by five
parameters that we specify in the following: {\it Star Formation Rate:} The star formation rate is
required because the {\sc Mappings-\Rmnum{3}} SED templates are normalised to
$\rm{SFR}=1\msun/\rm yr$. We assume $\rm{SFR}=m_{\rm i}/10\Myr$ to maintain
conservation of stellar mass, where $m_{\rm i}$ is the initial mass of the
stellar particle.  {\it Metallicity of the birth cloud:} The metallicity of the
birth cloud is assumed to be the initial metallicity of the stellar particle,
representing the metallicity of its birth environment.  {\it Pressure of the
surrounding ISM:} This pressure can be calculated based on the density of the star-forming gas. However,
as demonstrated in \citet{groves2008}, the impact of pressure on UV to NIR spectra
is very small. We therefore simply employ a fixed typical value of $\log_{10}{(P_{0}/k_{\rm
B})/\cm^{-3}\rm{K}}=5$~\citep{groves2008,Rodriguez2018}.  {\it The compactness
of the birth cloud:} This describes the density of the {H\small\Rmnum{2}}
region and can be calculated as
$\log_{10}{C}=3/5\log_{10}{M_{\rm cl}/\msun}+2/5\log_{10}{P_{0}/k_{\rm B}/(\rm{cm}^{-3}\rm K)}$~\citep{groves2008}, where $M_{\rm cl}$ is the mass of the stellar cluster and $P_{0}$ is the
pressure of {H\small\Rmnum{2}} region. The compactness parameter mainly affects
the temperature of the dust and its emission in FIR. It has almost no impact on
optical/UV/NIR photometric properties~\citep{groves2008}. We therefore also fix this
to a typical value of $\log_{10}{C}=5$~\citep{groves2008,Rodriguez2018}.  {\it
Photodissociation regions covering fraction:} This defines the time-averaged fraction of the stellar cluster solid angle that is covered by the photodissociation regions. The  {\sc Mappings-\Rmnum{3}} spectra library includes SEDs for the case of a fully covered or uncovered {H\small\Rmnum{2}} region. The final output SED is a linear combination of these
two extreme cases controlled by the parameter $f_{\rm PDR}$. We adopt here the fiducial value
$0.2$ for $f_{\rm PDR}$~\citep[][]{groves2008,jonsson2010,Rodriguez2018}. These parameters then fully specify the  {\sc Mappings-\Rmnum{3}} SEDs and therefore our unresolved dust attenuation model for Model C.

In Fig~\ref{fig:modelC_SED}, we demonstrate the impact of dust on two example rest-frame galaxy SEDs. The galaxies are selected at $z=2$ from the TNG50 simulation, which has the highest baryonic mass resolution of all IllustrisTNG simulations. The dust-to-metal ratio is set to $0.9$ for the generation of the SEDs for these two galaxies. This dust-to-metal ratio is the best-fit value of our dust calibration procedure described below. The SED of galaxy ID 44315 (middle panel) has a much stronger UV flux than galaxy ID 31832 (top panel) and the impact of dust attenuation is also stronger. We present face-on and edge-on images of galaxy ID 31832 in Figure~\ref{fig:image}. In the top left panel we present the galaxy without resolved dust attenuation. The remaining three panels present images with resolved dust attenuation. For the upper right panel we employ a dust-to-metal ratio of $0.3$, for the lower left panel we use $0.6$, and finally for the lower right panel we employ a dust-to-metal ratio of $0.9$, which is the best-fit value of our calibration procedure at this redshift as we will describe below. We note that we include the unresolved dust component for all cases since this is intrinsic to the employed {\sc Mappings-\Rmnum{3}} spectral templates. These images are based on apparent F115W, F150W, F200W NIRCam wide filter fluxes. They cover a $18\pkpc \times 18\pkpc$ field of view with $500\times 500$ pixels. The {\sc Skirt} radiative transfer calculation employed $10^7$ photon packets per wavelength on a reduced  wavelength grid with $234$ grid points in the wavelength range from $0.05\mu m$ to $5\mu m$. We stress that this grid is still constructed to cover all nebular emission lines reliably. The inclusion and sufficient sampling of these emission lines is crucial for a correct modelling of the appearance of galaxies~\citep[e.g.,][]{Camps2016,trayford2017}. Figure~\ref{fig:image} clearly demonstrates the important role of dust. In fact, the central bright bulge region of the galaxy is nearly completely covered by dust absorption when observed from the edge-on perspective. Even for a low dust-to-metal ratio of $0.3$ the appearance of the galaxy is dramatically different from the image without resolved dust attenuation. Any analysis or predictions based on the stellar light emission of simulated galaxies, e.g. UV, optical or lines like H$\alpha$, have to take dust attenuation into account. Neglecting this effect, can significantly alter the results, interpretation and lead to wrong conclusions.

\subsection{Numerical resolution corrections}
\label{sec:corr}
The highest mass resolutions of the different IllustrisTNG simulations, TNG50, TNG100, and TNG300, differ from each other, see Table~\ref{tab:tabsims}. Combining results
of TNG50, TNG100, and TNG300 therefore requires a resolution correction for the 
luminosities, magnitudes, stellar masses and star formation rates. 
More specifically, this requires a resolution correction
for TNG100 and TNG300 to match the numerical resolution of TNG50, which provides the highest numerical resolution of all simulations of the IllustrisTNG simulation suite. Once such a numerical resolution correction has been applied, the results of the three simulations can be combined. We achieve this resolution correction through a band magnitude resolution correction factor. 
A similar procedure has been applied for stellar masses and metal
abundances of IllustrisTNG simulations~\citep[e.g.,][]{Pillepich2018b, Vogelsberger2018}. We stress that the resolution corrections differ among the different bands since the SED resolution dependence is wavelength dependent. Furthermore, these corrections also have to be derived for dust-free and dust-attenuated cases separately, and all correction factors are redshift dependent. 

In the following we describe briefly the procedure to derive the resolution corrections for
the different band magnitudes. To this end we study the resolution dependence of the three simulations TNG50, TNG100
and TNG300 at their various resolution levels.
Specifically, we explore the dependence of galaxy magnitudes on the baryonic mass resolution for the different bands, simulations, resolutions and redshifts. For each simulation, we first divide the galaxies in $20$ logarithmically spaced halo mass bins ranging from
$10^{9}\msun$ to $10^{14}\msun$. We assume that the impact of resolution
on the halo mass is relatively mild and that the halo mass is a stable quantity to
divide galaxies into groups from simulations with different mass resolutions.
We then calculate for each simulation the median band magnitude of all galaxies in a given halo mass bin for every band.
By comparing the median magnitude of TNG100 (TNG300) with the median magnitude of TNG50 for every halo mass bin, we can derive a magnitude
correction for TNG100 (TNG300) to the resolution level of TNG50 at a given
halo mass. 
The simulation volume of TNG50 is too small to provide a large number of galaxies towards larger halo masses. We therefore set the magnitude correction factor to zero once the number of galaxies in a halo mass bin drops below $10$ for TNG50. This assumption is supported by the fact that TNG100 and TNG300 have nearly identical luminosity functions and stellar mass functions towards the bright and massive end. We can therefore safely assume a vanishing correction factor in this regime. We apply this whole procedure of deriving correction factors to all redshifts of interest. The resulting magnitude correction factors for each dust model as a function of halo mass and
redshift are then stored such that we can easily correct all band magnitudes for all simulations at all redshifts. 
Resolution corrected band magnitudes can then be used to construct resolution corrected galaxy luminosity functions. We note that the described resolution correction procedure can also be applied to other quantities like galaxy stellar masses and star formation rates. Those resolution corrected values build the basis to eventually derive the combined galaxy luminosity functions at all redshifts, for different bands, and for the three different dust models. Before describing how to derive this combined luminosity function based on the resolution corrected magnitudes, we first briefly discuss the dust model calibration step to determine the free dust parameters for the three dust models.  

\begin{figure*}
\centering
\includegraphics[width=0.48\textwidth]{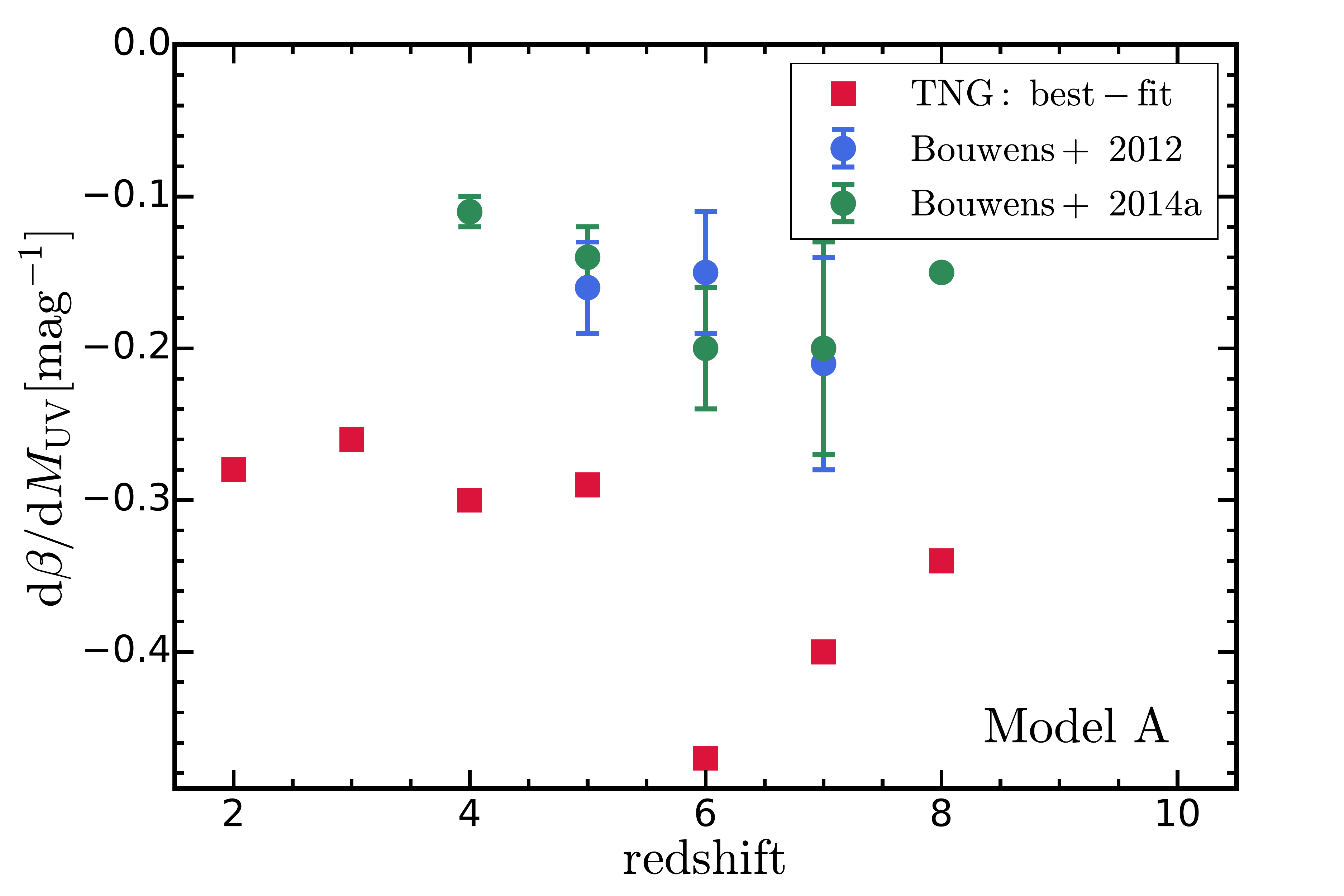}
\includegraphics[width=0.48\textwidth]{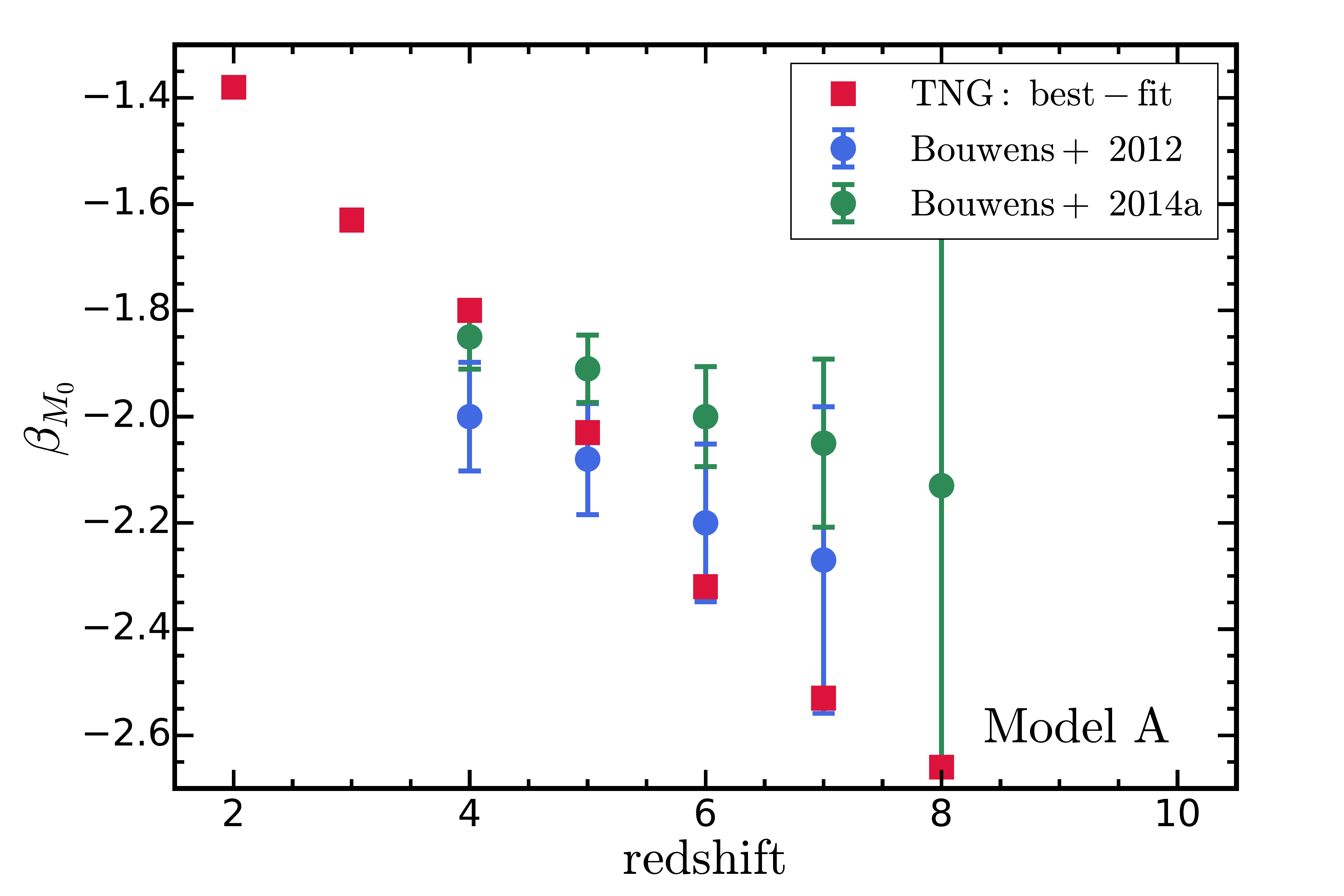}\\
\hspace{0.07cm}\includegraphics[width=0.48\textwidth]{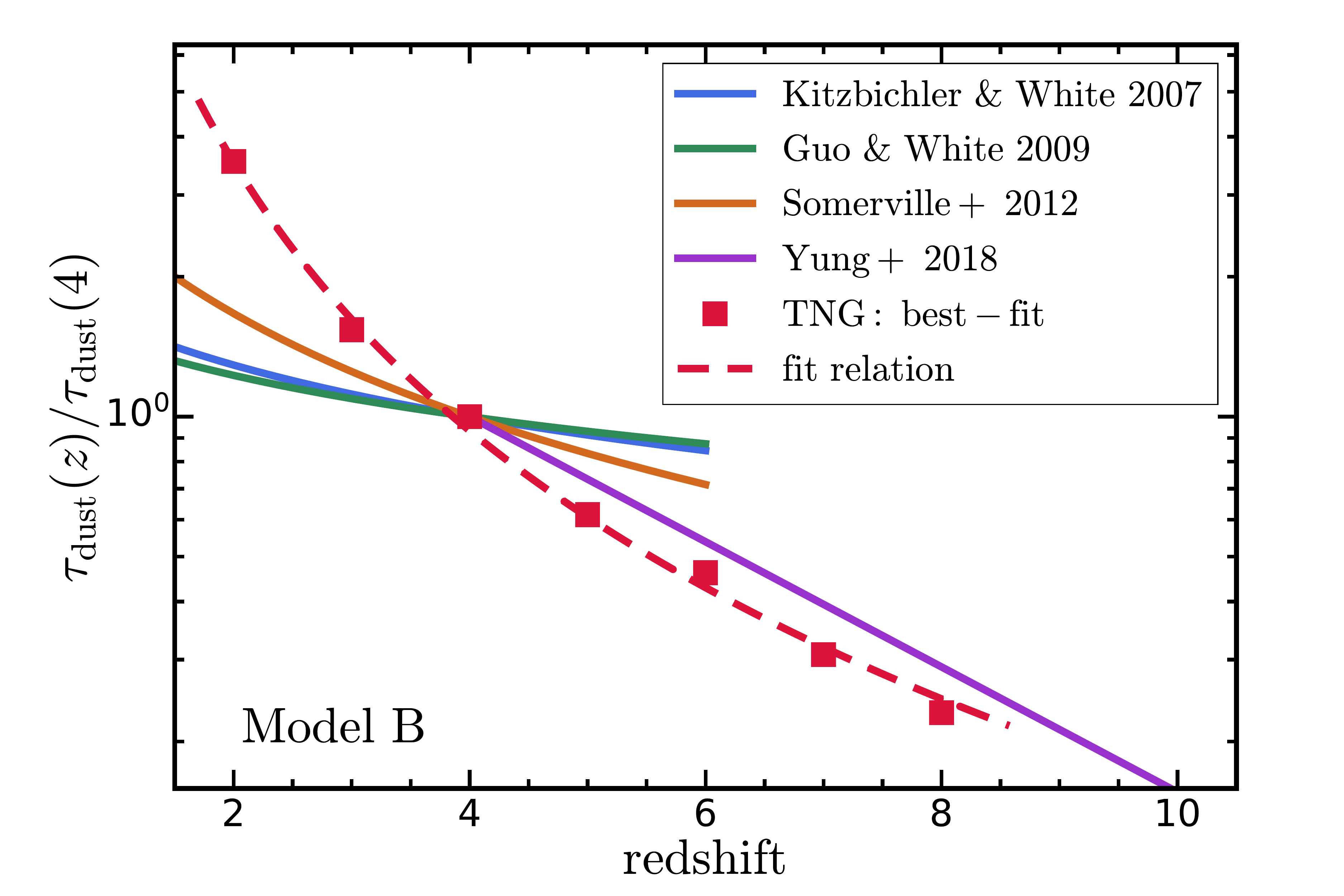}
\includegraphics[width=0.48\textwidth]{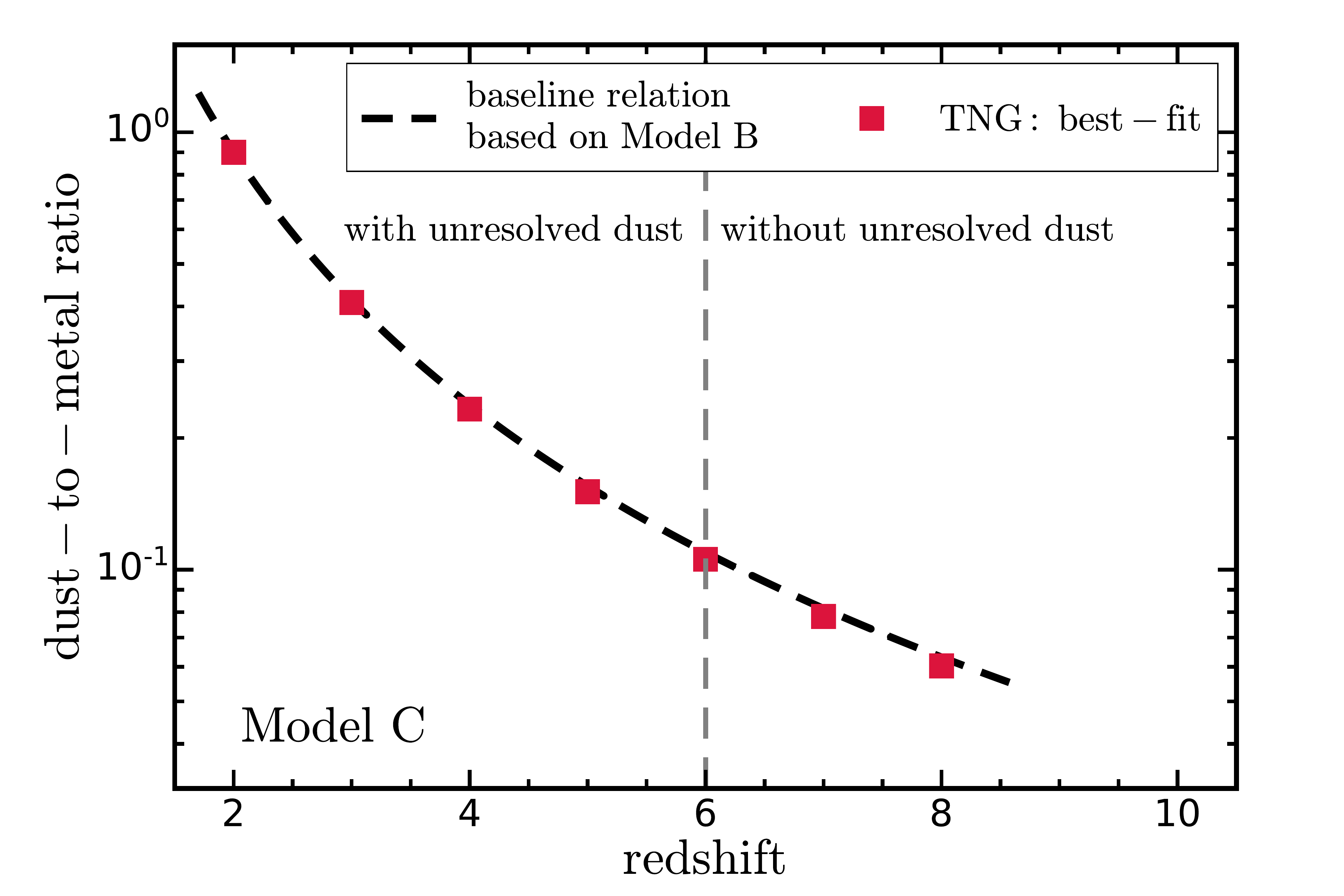}
\caption{{\bf Redshift dependence of best-fit dust parameters for dust Models A, B, and C.} {\it Upper left and upper right panels:} Best-fit values for the ${M_{\rm UV}^{\rm dust} - \langle \beta \rangle}$ relation assuming an underlying \citet{meurer1999} relation and fixed ${M_0=-19.5\mmag}$ for dust Model A. We also present the observationally inferred relations of~\protect\cite{Bouwens2012,Bouwens2014a}. These deviate from the best-fit values for the slope, ${{\rm d}\beta/{\rm d}M_{\rm UV}}$, redshift evolution but are consistent for the intercept $\beta_{M_0}$. {\it Lower left panel:} Best-fit $\tau_{\rm dust}(z)$ values of Model B. We also include the redshift evolution from other theoretical works, mostly semi-analytic models, for the evolution of the dust optical depth normalisation employing similar dust models. {\it Lower right panel:} Best-fit dust-to-metal ratio for Model C. The redshift evolution of the dust-to-metal ratio is based on the optical depth redshift evolution of Model B as baseline relation.}
\label{fig:model_parameter_tuning}
\end{figure*}

\subsection{Dust model calibration}
\label{sec:calibration}
All three dust models presented above depend on some free parameters. Specifically, Model A depends on two dust parameters; Models B and C depend each on one free dust parameter. Ultimately, these free parameters reflect the uncertain knowledge of the amount and distribution of dust in the simulated galaxies. 
The exact amount of dust and its distribution is not directly tracked in
IllustrisTNG. Only recently, some galaxy formation simulations and semi-analytical models have been able to track this dust component in detail through the direct modelling of the
relevant dust physics~\citep[][]{McKinnon2016, Aoyama2017, McKinnon2017, Popping2017,
McKinnon2018,Vogelsberger2019, Dave2019, Vijayan2019}. We therefore have to rely on the gas phase metallicity as a proxy
for the dust component in the galaxy. It is common to assume here a linear
scaling and express the amount of dust in terms of a dust-to-metal ratio. This
parameter is however non-universal, and also redshift dependent. We therefore
have to perform a dust model calibration for all dust models and all redshifts.

For the calibration process we compare the predicted dust-attenuated rest-frame UV luminosity
functions to existing observational data of rest-frame UV luminosity functions.  Our calibration is based on the observational data
compilation of \cite{Finkelstein2016} at $z=4-10$, which consists of both
ground- and space-based observations from
\cite{McLure2009,Castellano2010,vanderBurg2010,McLure2013,Oesch2013,Oesch2014,Schenker2013,Tilvi2013,Bowler2014,Bowler2015,Bouwens2015,Bouwens2016,Finkelstein2015,Schmidt2014,McLeod2015,McLeod2016}.
Additional observational constraints are taken from \citet{Alavi2014} at $z=2$,
\citet{Mehta2017} at $z=2,3$, \citet{Parsa2016} at $z=2-4$, \citet{reddy2008}
and \citet{reddy2009} at $z=3$, \citet{vanderBurg2010} at $z=3-5$,
\citet{Bouwens2017,Atek2018} at $z=6$, \citet{Atek2015b,Ouchi2009} at $z=7$ and
\citet{Oesch2018} at $z=10$.  We perform the calibration of dust parameters
based on the resolution corrected TNG100 simulation result, which represents
the best compromise of volume to sample the bright end, which is most sensitive
to dust attenuation, and numerical resolution. We note that results of the dust calibration
are sensitive to the radial cut applied for the stars contributing to the
luminosity calculation of a given galaxy. Throughout this work, we use $24$ points linearly spaced from $-16\mmag$ to $-24\mmag$ to create $23$ bins for the UV luminosity function calculation.

{\it Model A:} In principle, Model A contains no free parameters, since all the relations and parameters of this model are directly tied to observational scaling relations. However, as discussed above, there are significant observational uncertainties in the parameters of these relations. Most importantly the ${M_{\rm UV}^{\rm dust}-\langle \beta
\rangle}$ relation and its redshift evolution are still quite uncertain. 
Instead of relying on observational relations, we will therefore find the parameters for the ${M_{\rm UV}^{\rm dust}-\langle \beta
\rangle}$ relation through a best-fit to the rest-frame UV luminosity function. We note that we have explicitly verified that directly applying existing observationally inferred ${M_{\rm UV}^{\rm dust}-\langle \beta
\rangle}$ relations to our simulation results leads to predicted galaxy luminosity functions that are inconsistent with observed rest-frame UV luminosity functions. For the calibration of Model A, we set up a two dimensional grid for each redshift, with ${\rm d}\beta/{\rm d}M_{\rm
UV}$ and $\beta_{M_0}$ as the two axes.  The ${\rm d}\beta/{\rm d}M_{\rm UV}$
axis contains $81$ points ranging from $-0.8\mmag^{-1}$ to $0\mmag^{-1}$. The $\beta_{M_0}$ axis
contains $301$ points ranging from $-4$ to $-1$. For each grid point we apply the dust attenuation of Model A and compare
the resulting resolution corrected TNG100 rest-frame UV luminosity function with
our sample of observed rest-frame UV luminosity functions and calculate $\chi^2$ as:
\begin{equation}
\label{eq:chisq}
\frac{\chi^{2}}{n_{\rm eff}-\nu}=\frac{1}{n_{\rm eff}-\nu}\Bigg[\sum_{\rm i=1}^{n_{\rm eff}}\Big(\dfrac{\log{\phi}(M_{\rm ob}^{\rm i})-\log{\phi}_{\rm ob}^{\rm i}}{\sigma_{\rm ob}^{\rm i}}\Big)^{2}\Bigg],
\end{equation}
where $\nu=2$ is the number of free parameters in the model and $\log{\phi}(\,)$ is the interpolated resolution-corrected luminosity function predicted by our
model. For the $i^{\rm th}$ observation data point, $\log{\phi}_{\rm ob}^{\rm i}$ is the logarithm of the observed galaxy number
density at $M_{\rm ob}^{\rm i}$, $\sigma_{\rm ob}^{\rm i}$ is the uncertainty of the
observational measurement. Only observational data points with observed $M_{\rm
UV}\in[M_{\rm limit},-24\mmag]$ and within the magnitude coverage of the
interpolated luminosity function predicted by the model are taken as effective
points. $M_{\rm limit}$ is set to $-18\mmag$ at $z=3$, $-19\mmag$ at $z=2,4,5$ and $-20\mmag$ at $z\geq6$. The
number of these effective points is $n_{\rm eff}$. The parameter choice
resulting in the smallest $\chi^{2}/(n_{\rm eff}-\nu)$ then represents the best-fit Model A dust parameters at a given redshift. The result of this calibration procedure is presented in the top
row panels of Figure~\ref{fig:model_parameter_tuning}, where we also compare our best-fit values to
observationally derived parameters. We note that at $z=9,10$, the number of luminosity function data points of TNG100 is not sufficient to derive reliably best-fit values for Model A. We therefore neglect dust attenuation for $z=9$ and $z=10$. This assumption is justified by the calibration of Model B, and also by the results of Model C as we will discuss below. These calibrations demonstrate that dust attenuation is indeed negligible at $z=9$ and $z=10$.

{\it Model B}: For Model B, the only free dust parameter is $\tau_{\rm
dust}(z)$.  This optical depth normalisation parameter is a free scaling
factor, which reflects the unknown dust-to-metal ratio in each simulated galaxy. Many previous works have found that the relation between dust
attenuation or dust optical depth and gas metallicity and column density
evolves through cosmic time reflecting a redshift-dependent dust-to-metal
ratio. Also a fixed relation tends to overpredict the effect of dust attenuation for
high redshift galaxies significantly~\citep[e.g.,][]{ somerville2012,
Wilkins2013, reddy2015, whitaker2017, reddy2018, Yung2018}. Therefore, $\tau_{\rm
dust}(z)$ is expected to be redshift dependent.  Similar to \cite{somerville2012} and \cite{Yung2018}, we calibrate the value of
$\tau_{\rm dust}(z)$ to match the rest-frame UV luminosity functions observed
at $z=2-10$.  We explore $81$ values of $\tau_{\rm dust}$ 
linearly spaced from $0$ to $0.8$ at each redshift. We compare the
resulting UV luminosity functions with observations and calculate $\chi^2$ values using Equation~\ref{eq:chisq} with $\nu=1$.
Since the Model B parameter variation only leads to significant differences at the bright end, we only
consider observational data points with observed $M_{\rm UV}\in[M_{\rm limit},-24\mmag]$ for the calibration process. $M_{\rm limit}$ is set to be $-20\mmag$, except for $-21\mmag$ at $z\geq8$. At each redshift, we then calculate the $\chi^{2}/(n_{\rm eff}-\nu)$ for each parameter
choice. The parameter choice resulting in the smallest $\chi^{2}/(n_{\rm eff}-\nu)$ then represents
the best-fit dust parameter for that redshift. The result of this procedure is
presented in the lower left panel of Figure~\ref{fig:model_parameter_tuning}.  At $z=9,10$ the bright end of TNG100 is not sampled well enough to provide reliable best-fit dust parameters for Model B. Extrapolating the $\tau_{\rm dust}(z)$ scaling to these redshifts results in $\tau_{\rm dust}(z) \lesssim 10^{-2}$, which implies a negligible dust correction. We therefore set $\tau_{\rm dust}(z)=0$ for $z=9,10$. Overall we find that the redshift evolution of $\tau_{\rm dust}(z)$ is well described by a power law relation with $z^{-1.92}$ as indicated in the lower left panel of Figure~\ref{fig:model_parameter_tuning}. The redshift dependence at $z\geq 4$ is similar to the dust attenuation scaling recently presented in~\citet{Yung2018} for a similar dust model.

{\it Model C:} For Model C, the only free dust parameter is the dust-to-metal ratio $f_{\rm z}(z)$.  For the calibration of Model C we 
assume the power law fit of $\tau_{\rm dust}(z)$ from Model B as a baseline since it is computationally expensive to perform radiative transfer calculations for all simulations for a large number of different dust-to-metal ratios. We therefore first perform a dust parameter exploration for $z=2$ to $z=8$ following the redshift dependence of $\tau_{\rm dust}(z)$ found for Model B. We explore ${f_{\rm z}(z)
= [0.3,0.5,0.9,1.3,1.7] \times (z/2)^{-1.92}}$ for the
dust-to-metal ratio and compare the resulting luminosity functions with observations. To accelerate this process, we require a less stringent cell splitting criterion for {\sc Skirt}: $1/N_{\rm gas}$ with $10^{-3}$ as upper limit and $2\times10^{-6}$ as lower limit instead of using a fixed value of $2\times10^{-6}$. We find that ${f_{\rm z}(z) = 0.9\times (z/2)^{-1.92}}$ results in the best-fit UV luminosity functions. This best-fit dust parameter choice is presented in the lower right panel of Figure~\ref{fig:model_parameter_tuning}. Based on this exploration result, we execute the {\sc Skirt} pipeline for this redshift dependent dust-to-metal parameter evolution with the fiducial cell splitting criterion, $2\times10^{-6}$, for all redshifts $z=2-10$ to produce the final dust-attenuated luminosities for Model C. During the calibration processes we have also noticed that the unresolved dust attenuation of the {\sc Mappings-\Rmnum{3}} spectra templates results in too much dust attenuation for $z\geq 6$. This results in a suppression of the abundance of faint and intermediate brightness galaxies. We therefore do not consider the unresolved dust component for $z \geq 6$, i.e. we do not employ the  {\sc Mappings-\Rmnum{3}} spectra templates for young stars for these high redshifts. The low dust-to-metal ratios at these high redshifts furthermore implies a vanishing attenuation due to resolved dust. Therefore, the high redshift luminosities are essentially identical to the intrinsic dust-free magnitudes. This is similar to the high redshift behaviour of Model B. In this case the resolved component also vanishes due to the small normalisation factor of the optical depths.  Similarly, also the unresolved component vanishes towards higher redshifts since it scales linearly with the resolved dust attenuation by construction.  

\begin{figure*}
\begin{center}
{\bf {\it \textbf{JWST}} NIRCam (F115W, F150W, F200W) $\mathbf{z=2}$ galaxies without resolved dust}\\ \vspace{-0.2cm}
\end{center}
\includegraphics[width=0.247\textwidth]{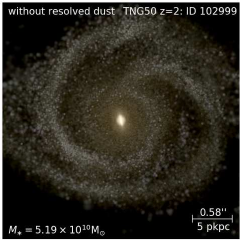}
\includegraphics[width=0.247\textwidth]{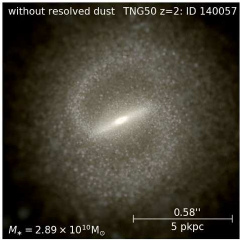}
\includegraphics[width=0.247\textwidth]{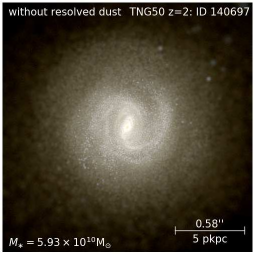}
\includegraphics[width=0.247\textwidth]{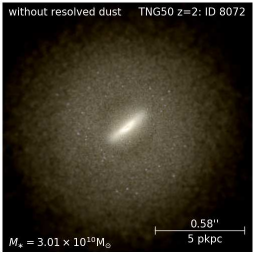}
\includegraphics[width=0.247\textwidth]{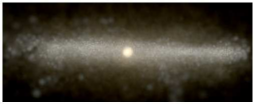}
\includegraphics[width=0.247\textwidth]{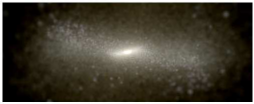}
\includegraphics[width=0.247\textwidth]{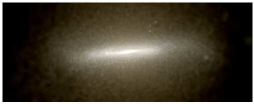}
\includegraphics[width=0.247\textwidth]{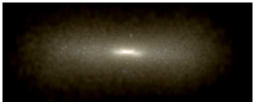}\\
\begin{center}
{\bf {\it \textbf{JWST}} NIRCam (F115W, F150W, F200W) $\mathbf{z=2}$ galaxies with resolved dust (dust-to-metal ratio 0.3 = one third of best-fit value)}\\ \vspace{-0.2cm}
\end{center}
\includegraphics[width=0.247\textwidth]{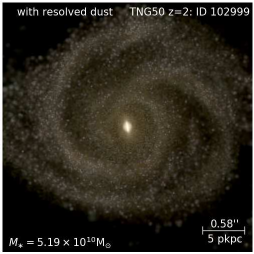}
\includegraphics[width=0.247\textwidth]{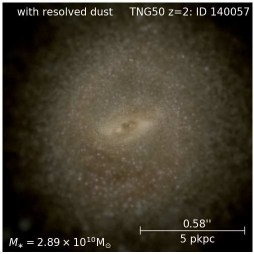}
\includegraphics[width=0.247\textwidth]{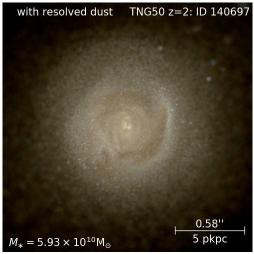}
\includegraphics[width=0.247\textwidth]{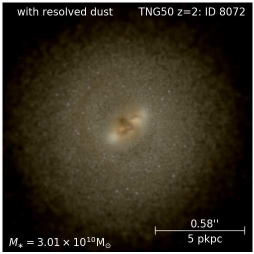}
\includegraphics[width=0.247\textwidth]{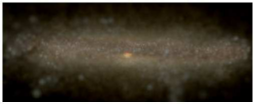}
\includegraphics[width=0.247\textwidth]{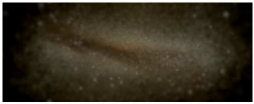}
\includegraphics[width=0.247\textwidth]{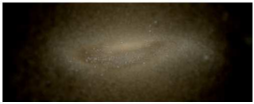}
\includegraphics[width=0.247\textwidth]{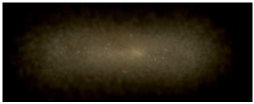}\\
\begin{center}
{\bf {\it \textbf{JWST}} NIRCam (F115W, F150W, F200W) $\mathbf{z=2}$ galaxies with resolved dust (dust-to-metal ratio 0.9 = best-fit value)}\\ \vspace{-0.2cm}
\end{center}
\includegraphics[width=0.247\textwidth]{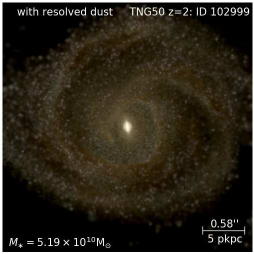}
\includegraphics[width=0.247\textwidth]{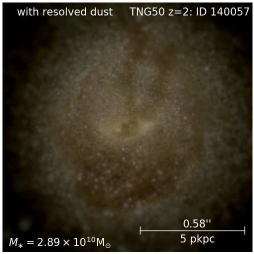}
\includegraphics[width=0.247\textwidth]{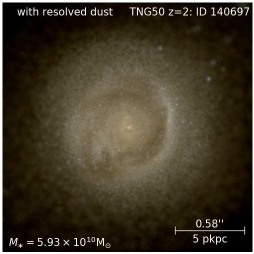}
\includegraphics[width=0.247\textwidth]{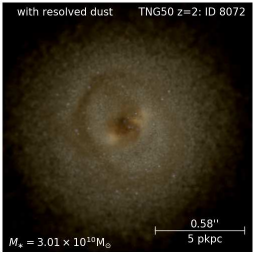}
\includegraphics[width=0.247\textwidth]{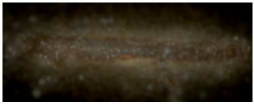}
\includegraphics[width=0.247\textwidth]{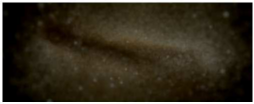}
\includegraphics[width=0.247\textwidth]{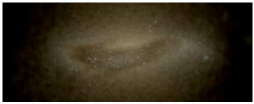}
\includegraphics[width=0.247\textwidth]{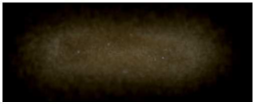}
\caption{{\bf {\it \textbf{JWST}} NIRCam face-on and edge-on images of TNG50 galaxies at $\mathbf{z=2}$}. {\it Top row panels:} Galaxy images without resolved dust. {\it Middle and bottom row panels:} Galaxy images with resolved dust assuming different dust-to-metal ratios as indicated. The images are based on apparent F115W, F150W, F200W filter fluxes. The radiative transfer calculation employed $10^7$ photon packets per wavelength on our reduced wavelength grid with $234$ points in the wavelength range from $0.05\mu m$ to $5\mu m$. This grid is specifically constructed to sample emission lines. The inclusion and sufficient sampling of these emission lines is important for a correct modelling of the appearance of galaxies.  We note that our fiducial wavelength grid to derive galaxy SEDs and magnitudes employs a finer grid with $1168$ points within the same wavelength range. For the dust cases we assume dust-to-metal ratios of $0.9$, which is the best-fit value based on the dust calibration procedure, and $0.3$. In general the appearance of galaxies is severely affected by dust. Most importantly, light from the central bright regions of the various galaxies is strongly absorbed and scattered due to dust.}
\label{fig:image2a}
\end{figure*}

\begin{figure*}
\begin{center}
{\bf {\it \textbf{JWST}} NIRCam (F070W, F090W, F115W) $\mathbf{z=1}$ galaxies without resolved dust}\\ \vspace{-0.2cm}
\end{center}
\includegraphics[width=0.247\textwidth]{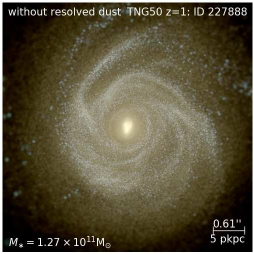}
\includegraphics[width=0.247\textwidth]{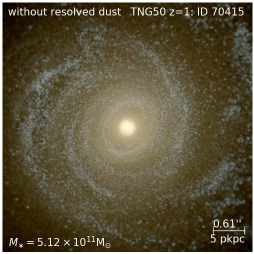}
\includegraphics[width=0.247\textwidth]{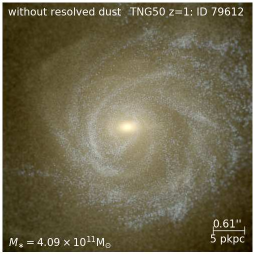}
\includegraphics[width=0.247\textwidth]{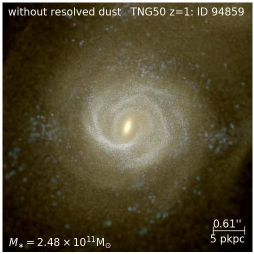}
\includegraphics[width=0.247\textwidth]{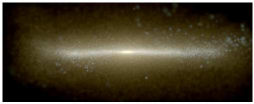}
\includegraphics[width=0.247\textwidth]{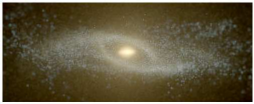}
\includegraphics[width=0.247\textwidth]{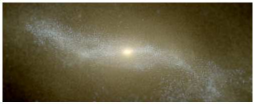}
\includegraphics[width=0.247\textwidth]{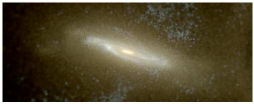}\\
\begin{center}
{\bf {\it \textbf{JWST}} NIRCam (F070W, F090W, F115W) $\mathbf{z=1}$ galaxies with resolved dust (dust-to-metal ratio 0.3)}\\ \vspace{-0.2cm}
\end{center}
\includegraphics[width=0.247\textwidth]{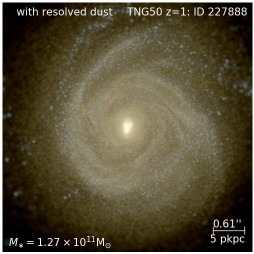}
\includegraphics[width=0.247\textwidth]{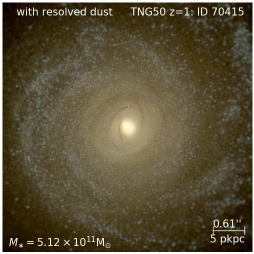}
\includegraphics[width=0.247\textwidth]{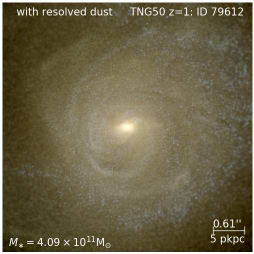}
\includegraphics[width=0.247\textwidth]{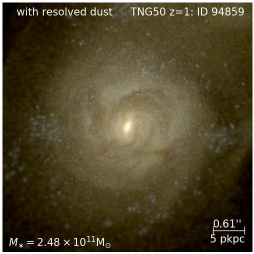}
\includegraphics[width=0.247\textwidth]{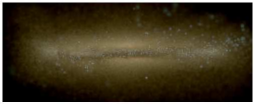}
\includegraphics[width=0.247\textwidth]{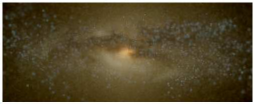}
\includegraphics[width=0.247\textwidth]{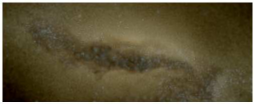}
\includegraphics[width=0.247\textwidth]{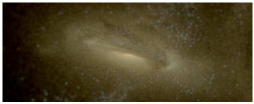}\\
\begin{center}
{\bf {\it \textbf{JWST}} NIRCam (F070W, F090W, F115W) $\mathbf{z=1}$ galaxies with resolved dust (dust-to-metal ratio 0.9)}\\ \vspace{-0.2cm}
\end{center}
\includegraphics[width=0.247\textwidth]{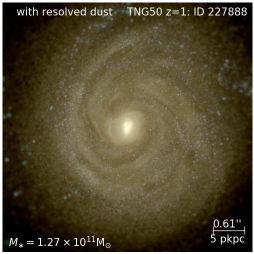}
\includegraphics[width=0.247\textwidth]{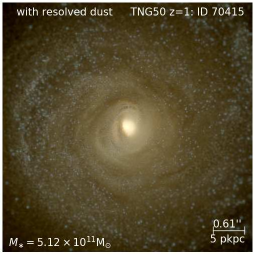}
\includegraphics[width=0.247\textwidth]{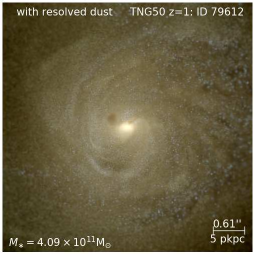}
\includegraphics[width=0.247\textwidth]{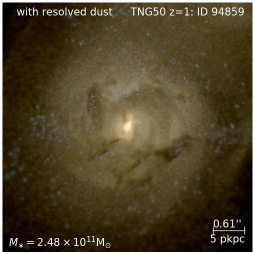}
\includegraphics[width=0.247\textwidth]{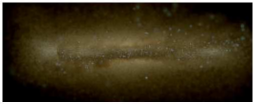}
\includegraphics[width=0.247\textwidth]{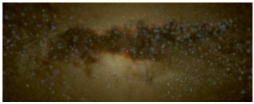}
\includegraphics[width=0.247\textwidth]{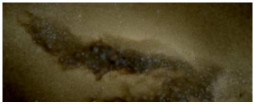}
\includegraphics[width=0.247\textwidth]{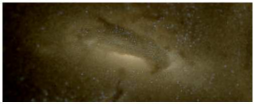}
\caption{{\bf {\it \textbf{JWST}} NIRCam face-on and edge-on images of TNG50 galaxies at $\mathbf{z=1}$}. {\it Top row panels:} Galaxy images without resolved dust. {\it Middle and bottom row panels:} Galaxy images with resolved dust assuming different dust-to-metal ratios as indicated. The images are based on apparent F070W, F090W, F115W filter fluxes. The radiative transfer calculation employed $10^7$ photon packets per wavelength on our reduced wavelength grid with $234$ points in the wavelength range from $0.05\mu m$ to $5\mu m$. This grid is specifically constructed to sample emission lines. The inclusion and sufficient sampling of these emission lines is crucial for a correct modelling of the appearance of galaxies.  We note that our fiducial wavelength grid to derive galaxy SEDs and magnitudes employs a finer grid with $1168$ points within the same wavelength range. For the dust cases we assume dust-to-metal ratios of $0.9$, which is the best-fit value based on the dust calibration procedure, and $0.3$.  In general the appearance of galaxies is severely affected by dust. Most importantly, light from the central bright regions of the various galaxies is strongly absorbed and scattered due to dust.}
\label{fig:image2b}
\end{figure*}
We can compare our findings for the best-fit dust-to-metal ratio with the rather limited amount of observational and theoretical findings. 
Observationally, \citet{Zafar2013} found that the dust-to-metal ratio in a sample of foreground absorbers to Gamma-ray bursts (GRBs) and quasars does not vary significantly over a wide range of redshifts $0.1-6.3$, metallicities, and hydrogen column densities, favouring a universally constant dust-to-metal ratio. \citet{Chen2013} found a weak redshift dependence of dust-to-metal ratios in lensed galaxies. \citet{DeCia2013} studied dust-to-metal ratios of GRB damped Lyman-alpha absorbers. They found a large scatter in the dust-to-metal ratio distribution at $z>1.7$ and also found several systems with low dust-to-metal ratio at high redshift. \citet{RemyRuyer2014} found that dust-to-metal ratios can be lower in low metallicity systems with a large dispersion. \citet{Wiseman2017} studied the evolution of dust-to-metal ratios through cosmic time also via GRB damped Lyman-alpha absorbers. They found that dust-to-metal ratios increase with metallicity~\citep[see also][]{DeCia2016} but did not observe an obvious redshift dependence of dust-to-metal ratio. Theoretically, the analytical model of~\citet{Inoue2003} demonstrated that dust-to-metal ratios are lower in younger galaxies and have a significant redshift dependence. They suggested that the Lyman break galaxies at $z\sim 3$ have a low dust-to-metal ratio, $<0.1$. \citet{Kimm2013} found that the dust-to-metal ratio is $0.06$ $(0.4)$ at $z=7$ assuming an escape fraction of light of $0\%$ $(10\%)$, which represent a lower (upper) limit, respectively. \citet{Yajima2014} explored dust-to-metal ratios in the range from $0.05$ to $0.5$, with $0.5$ as their fiducial model. They found that a dust-to-metal ratio $>0.4$ is required at $z=3$ to match observational data. They also suggest that the dust-to-metal ratio at $z=3$ is close to that of the local Universe. \citet{Bekki2015} found that dust-to-metal ratios can be different between different galaxies at different redshifts and even varies within a single galaxy. \citet{McKinnon2016} used hydrodynamical simulations of Milky Way-like galaxies to study the evolution of
dust and also found no evolution in dust-to-metal ratio at $z>1$. However, they predicted a much lower dust-to-metal ratio at $z>2$ compared to the value at $z=0$. \citet{Popping2017} studied the dust content of galaxies from $z=0$ to $z=9$ and found that dust-to-metal ratios can vary in systems with different metallicity and stellar mass. One of their models shows a redshift dependence for the dust-to-metal ratio. \citet{Aoyama2017} found that the assumption of fixed dust-to-metal ratio is no longer valid once galaxies are older than $0.2\Gyr$.  \citet{Behrens2018} found in simulations that the dust-to-metal ratios for galaxies at $z\sim 8$ are only $0.08$, i.e. far lower than the value in the local Universe. These various theoretical and observational results demonstrate the difficulty of measuring and modelling the dust content of galaxies. We note that it is expected that the dust-to-metal ratios vary with redshift since the metal content of galaxies also changes with redshift. Our dust calibration procedure is therefore necessary since definite predictions and measurements of dust abundances at high redshifts are currently not available. 

We summarise the results of our dust calibration for the different dust models and redshifts in Table~\ref{tab:modelA_fit}. We demonstrate the impact of different amounts of dust in Figure~\ref{fig:image2a} for a few more TNG50 $z=2$ galaxies. All galaxies are presented in face-on and edge-on projections. Galaxy images in the top row do not consider the impact of dust. The other two rows include the effects of dust for two different dust-to-metal ratio: $0.3$ for the middle row, and $0.9$ for the bottom row. The latter represents the best-fit $z=2$ dust-to-metal ratio derived based on the calibration procedure discussed above. In Figure~\ref{fig:image2b} we present also four $z=1$ galaxies of TNG50 for the dust-free case and for two different dust-to-metal ratios. We note that we employed here a different set of NIRCam wide filters: F070W, F090W, F115W instead of the F115, F150W, and F200W filters that we used at $z=2$. For all galaxy images, it is obvious that the inclusion of dust attenuation severely affects the stellar light morphology of galaxies.

\begin{figure}
\includegraphics[width=0.50\textwidth]{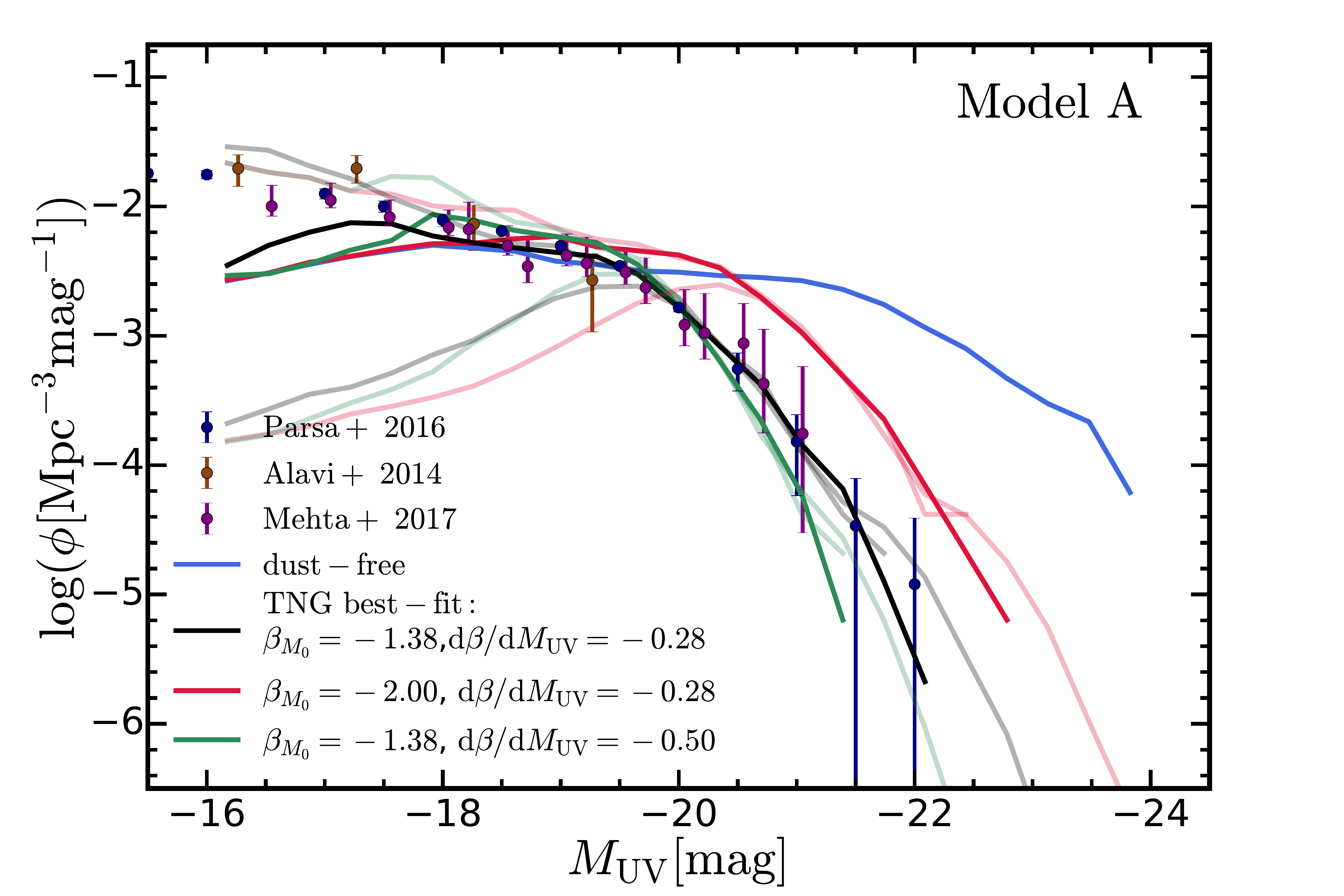}
\includegraphics[width=0.50\textwidth]{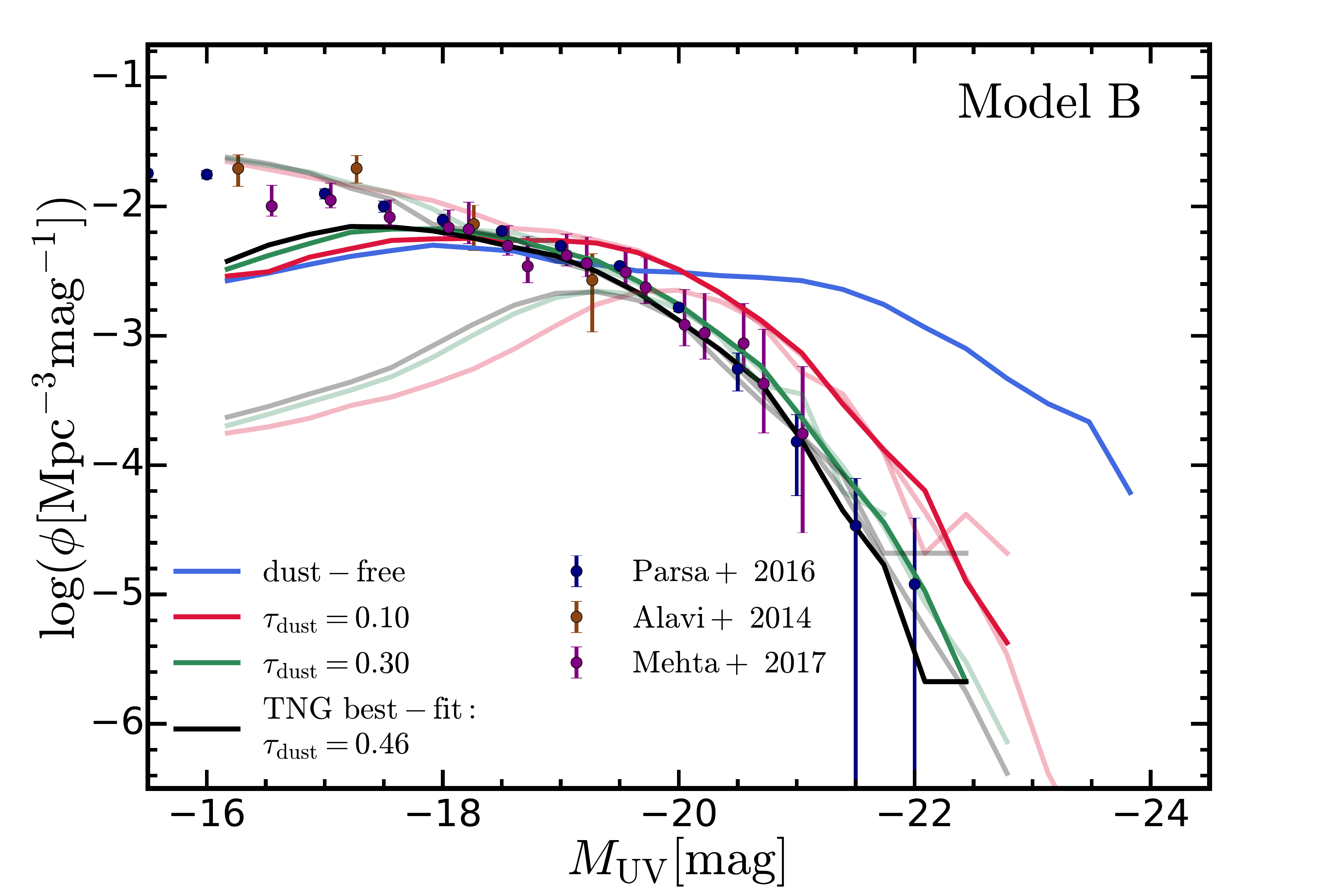}
\includegraphics[width=0.50\textwidth]{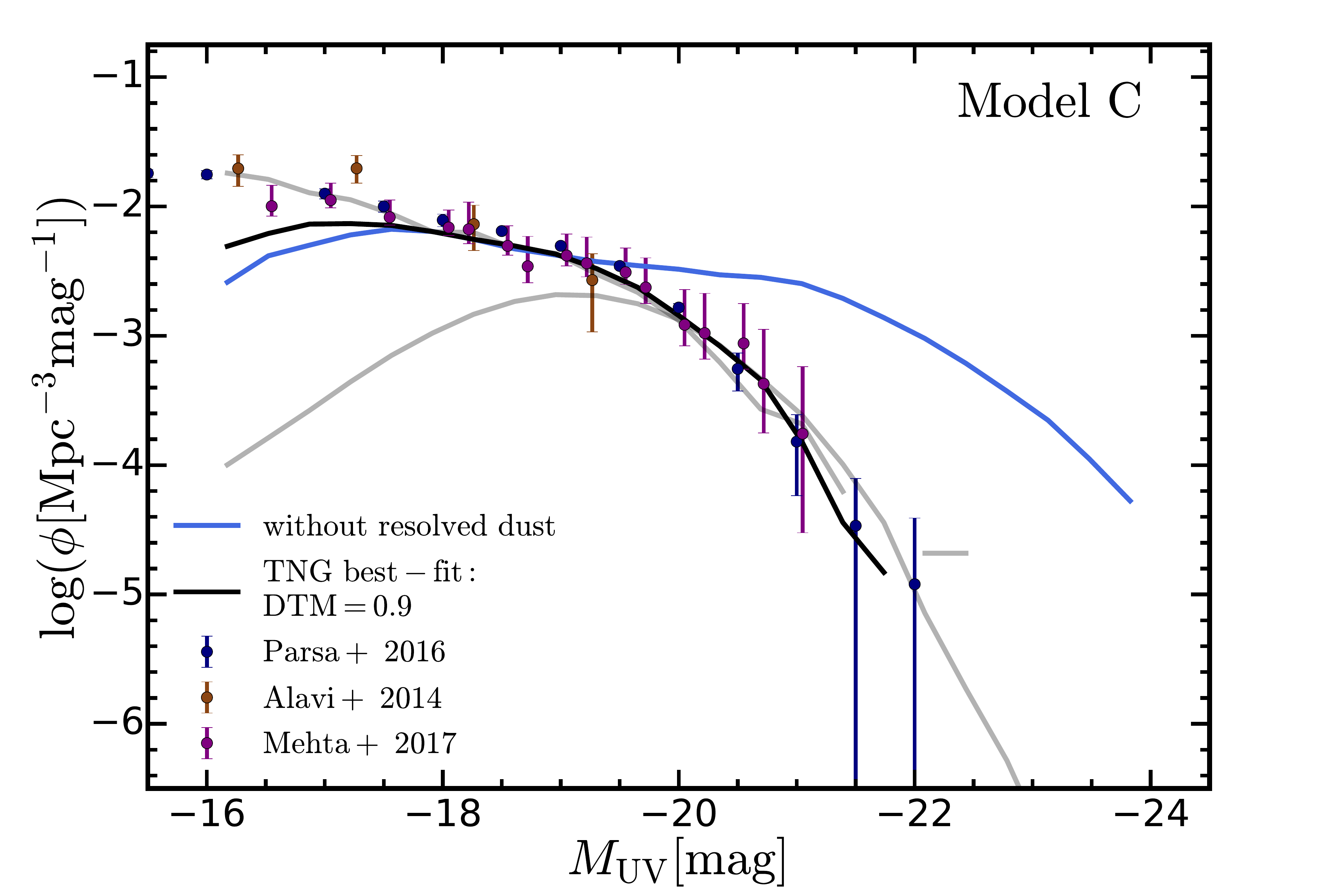}
\caption{{\bf Comparison between rest-frame UV luminosity functions for different parameter choices for the three different dust models at $\mathbf{z=2}$.} The resolution corrected luminosity functions of TNG100 are shown with solid lines. The luminosity functions of TNG50 and resolution corrected TNG300 are shown with transparent lines. Besides the best-fit values, we also show for Models A and B dust parameter variations around these best-fit values. }
\label{fig:lf_compare}
\end{figure}

\begin{figure*}
\includegraphics[width=0.99\textwidth]{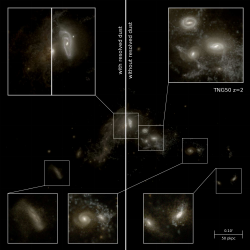}
\caption{{\bf TNG50 protocluster {\it \textbf{JWST}} NIRCam image at ${\mathbf z=2}$ with and without dust attenuation.} The image has a physical extend of $450\pkpc$ and is based on apparent F115W, F150W, F200W filter fluxes. The field of view is sampled by $2000 \times 2000$ pixels and $10^7$ photon packets on the reduced wavelength grid with $234$ points in the range $0.05\mu m$ to $5\mu m$, which is five times coarser than our fiducial wavelength grid employed for the integrated galaxy fluxes. In addition to the large field of view, we place higher resolution detectors, each with $500\times 500$ pixels, on the zoom-in regions as indicated. These additional detectors show the zoom-in images of individual galaxies marked in the larger field of view. The vertical line divides the image in a left half that includes dust attenuation, and a right half does not include dust instead shows stellar light without resolved dust attenuation.}
\label{fig:cluster}
\end{figure*}

\subsection{Combining TNG50, TNG100 and TNG300 predictions} 
So far we have described how to resolution correct the different simulations to
the TNG50 resolution level, and how we calibrate our three dust models based on the resolution-corrected TNG100 simulation results. Next we describe how we can derive a combined galaxy
luminosity function based on the resolution-corrected and dust-corrected magnitudes of all three IllustrisTNG simulations: TNG50, TNG100 and TNG300. This will then lead to predicted galaxy luminosity functions that probe a wide range of galaxy luminosities. 

The combination procedure is based on the following steps. For each redshift, we first study the resolution-corrected luminosity functions of all three simulations. An example of those is presented in Figure~\ref{fig:lf_compare}. The panels in this figure demonstrate that the resolution-corrected luminosity functions of the different simulations overlap within some luminosity range but deviate from each other towards the faint end once the corresponding simulation fails to resolve small galaxies. The magnitude where
the simulations start to deviate can be seen as a sharp drop in the predicted abundance for simulations with a lower resolution. This marks a lower magnitude limit over which a given simulation is considered to be valid. Higher resolution simulations probe the faint end down to lower
magnitudes. Towards the bright end, the galaxy number densities fluctuate due to Poisson sampling noise set by the simulation volume. Larger simulation volumes lead
to less noise towards the bright end. This sets for a given simulation an upper magnitude limit above which a simulation is not considered valid anymore. In order to reliably combine the luminosity functions, we therefore first determine at each redshift and for each simulation these lower and upper magnitude limits within which the luminosity function is reliably sampled. These points are marked as those magnitudes where the different resolution corrected simulations predict galaxy abundances that differ more than $0.2\,{\rm dex}$. Beyond these boundaries, the luminosity function does not contribute to the construction of the combined luminosity function. 

\begin{figure*}
\includegraphics[width=1.02\textwidth]{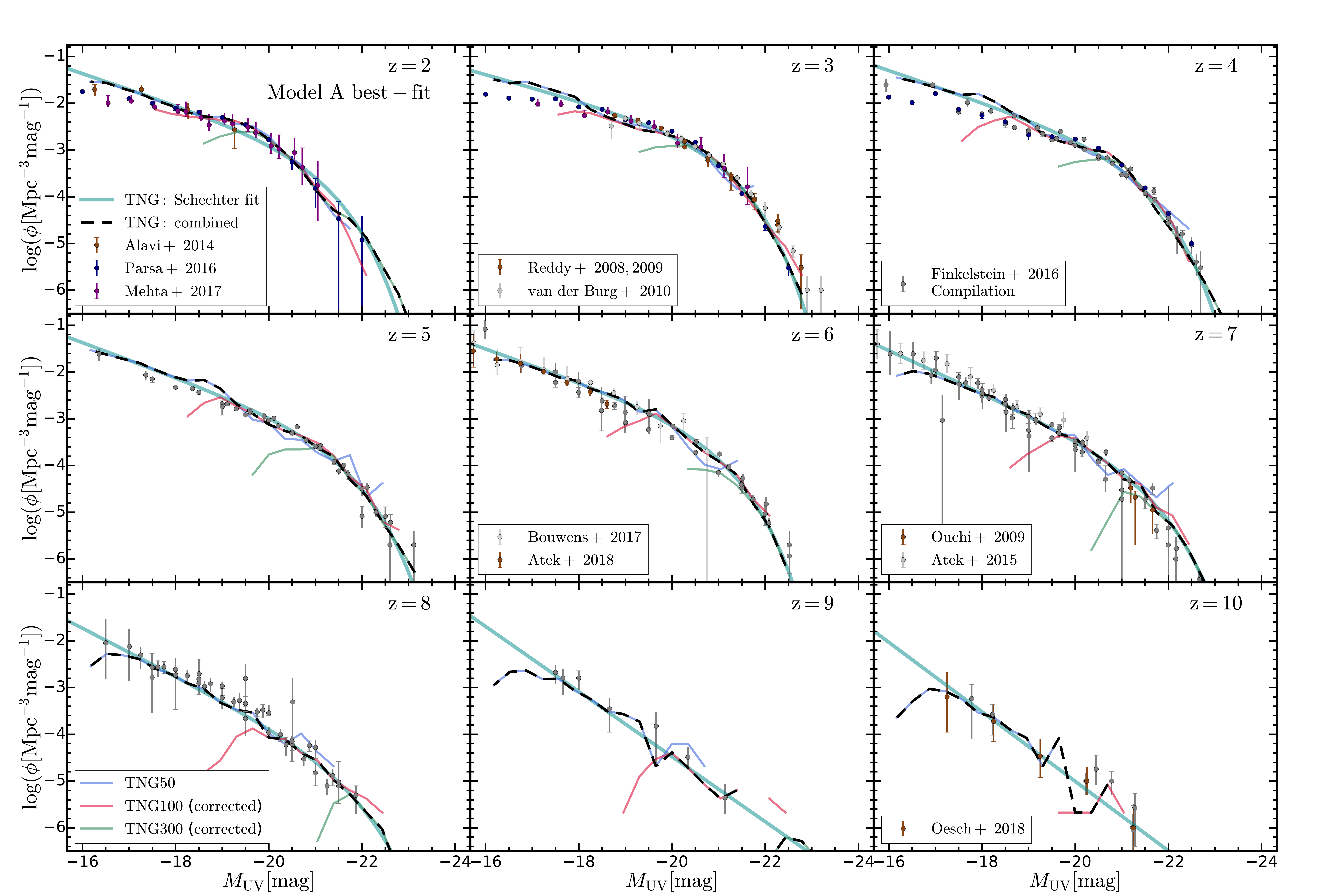}
\caption{{\bf Best-fit Model A UV luminosity functions at $\mathbf{z=2-10}$.} We show resolution corrected luminosity functions of TNG100 in solid red. Comparing those with the luminosity functions of TNG50 (blue) and the resolution corrected luminosity function of TNG300 (green) demonstrates that we achieve a good agreement between the different simulations over their common luminosity range. Combining the three simulations yields the combined luminosity functions (black dashed lines). Schechter fits for the combined luminosity functions are presented with cyan lines. In general the combined luminosity functions agree with  the observational data very well. At $z\lesssim4$, the combined luminosity function predicts slightly higher number densities and steeper faint-end slopes compared to observational data. Towards higher redshifts we find that IllustrisTNG predicts slightly lower abundances of galaxies. This is most noticeable at $z=8$, where we predict lower abundances over a quite wide range of rest-frame UV luminosities from $-18\mmag$ to $-21\mmag$.}
\label{fig:UVLF_A}
\end{figure*}

For each galaxy magnitude bin within these limits we then calculate an average galaxy abundance, which represents the combined galaxy luminosity function at that specific magnitude. The average is performed with weights based on the number of galaxies in each magnitude bin.
This gives more weight to simulations that sample a given bin with more galaxies. Specifically,
we calculate the average for the combined luminosity function with:
\begin{equation}
\phi_{\rm combined}  =  \frac{\phi_{\rm 50} \, N_{\rm 50}^2 + 
      \phi_{\rm 100} \, N_{\rm 100}^2  + 
      \phi_{\rm 300} \, N_{\rm 300}^2}{N_{\rm 50}^2  +  N_{\rm 100}^2 + N_{\rm 300}^2},
\label{eq:combine}
\end{equation}
where $N_{\rm X}$ and $\phi_{\rm X}$ refer to the number of galaxies of TNGX in each bin and the resolution-corrected luminosity function of TNGX, respectively. This procedure then leads to one combined galaxy luminosity function for each redshift and for all bands. In the following we will study those combined luminosity functions at different redshifts, for different bands, and for different dust models. We note that this combination procedure can also be applied to other quantities like the galaxy stellar mass functions.

\section{Results}
\label{sec:Section4}

After discussing in detail our methods, we will present our results in this section. We start with a first image of the high redshift universe as predicted by the TNG50 simulation on larger scales. To this end, we show in Figure~\ref{fig:cluster} a {\it JWST} mock image of the most massive protocluster within the TNG50 volume at a redshift of $z=2$. This image covers a field of view of $450\times 450\,{\rm pkpc}^2$. In addition we also zoom in onto some galaxies by placing high resolution detectors on some regions of the field of view. We can see a wide range of galaxy morphologies in this image. This protocoluster turns into a $z=0$ cluster with a mass of $M_{\rm 200,crit} \sim 2 \times 10^{14}\msun$. The vertical line divides the image in a left half that includes dust attenuation based on dust Model C, and a right half does not include resolved dust attenuation.

\begin{figure*}
\includegraphics[width=1.02\textwidth]{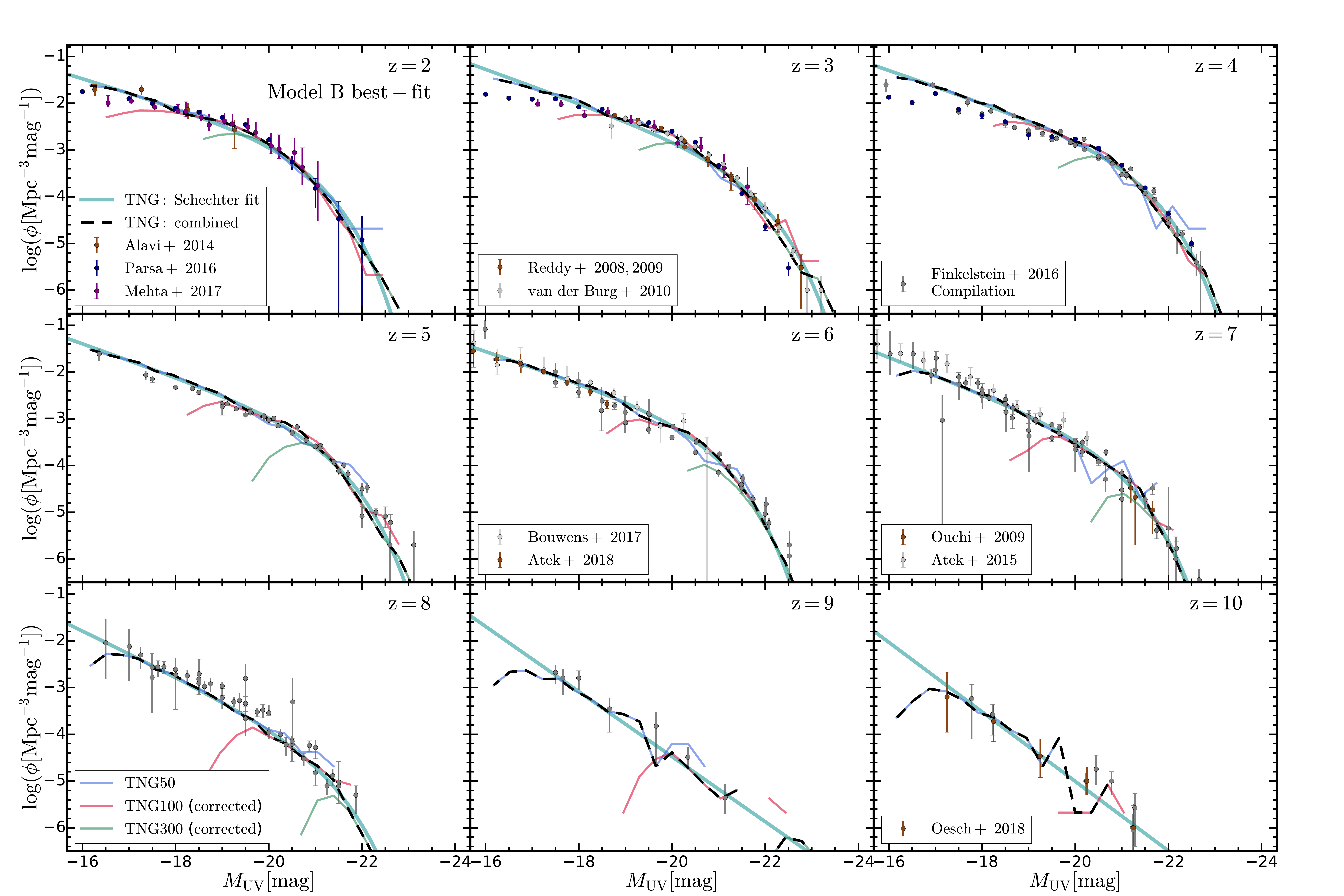}
\caption{{\bf Best-fit Model B UV luminosity functions at $\mathbf{z=2-10}$.} The combined luminosity functions agree better with observational data than the combined luminosity functions of Model A. However, at $z\lesssim 4$, the combined luminosity function still predicts slightly higher number densities and steeper slopes towards the faint end compared with observations. This is an inherent feature of the galaxy population predicted by TNG50. Similarly, towards higher redshifts we still find that IllustrisTNG predicts slightly lower abundances of galaxies for dust Model B. }
\label{fig:UVLF_B}
\end{figure*}

\begin{figure*}
\includegraphics[width=1.02\textwidth]{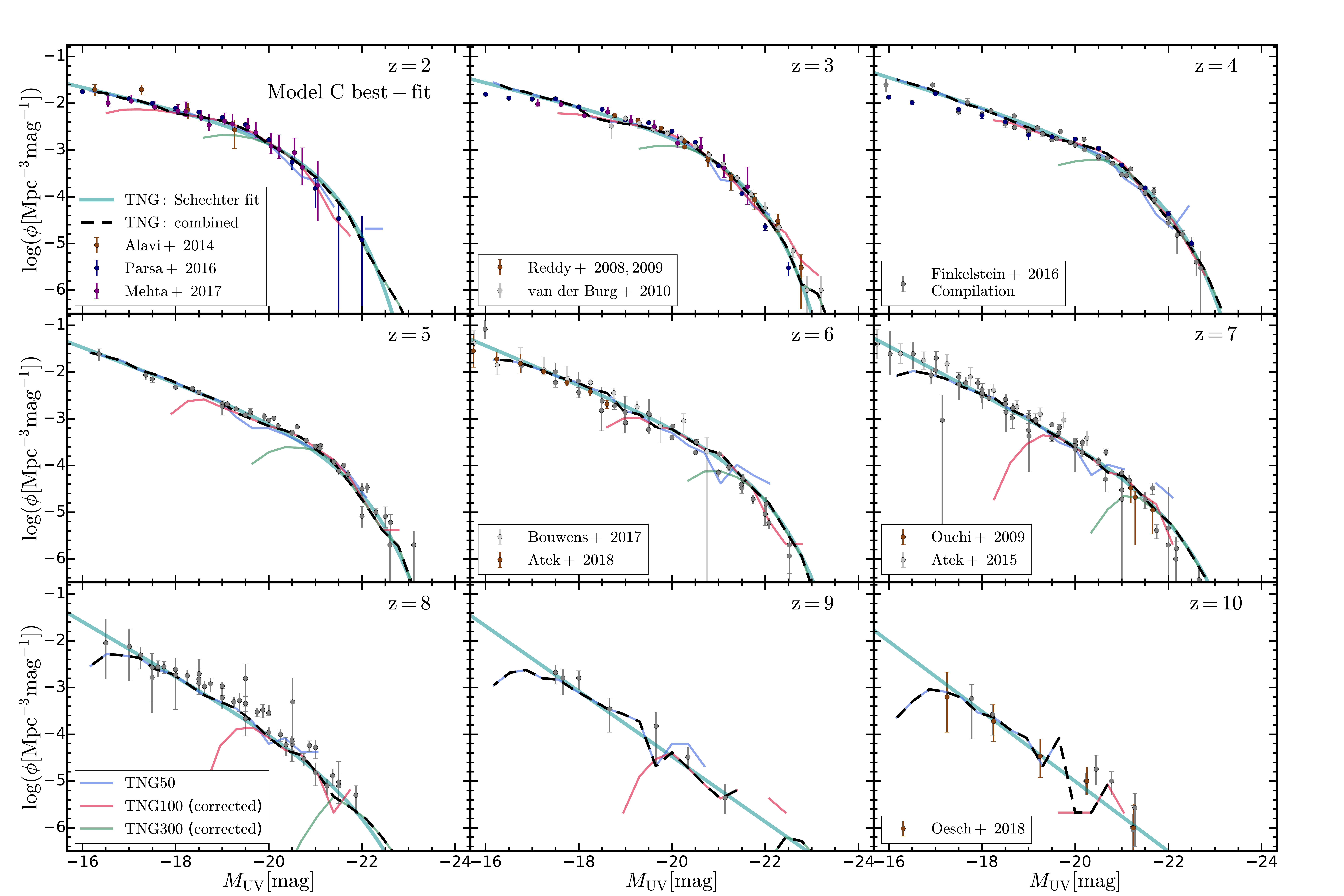}
\caption{{\bf Best-fit Model C UV luminosity functions at $\mathbf{z=2-10}$.} The combined luminosity functions agree with observational data as well as Model B although Model C leads to a slightly better agreement with observational data towards the faint end.
}
\label{fig:UVLF_C}
\end{figure*}

\begin{figure*}
\includegraphics[width=0.48\textwidth]{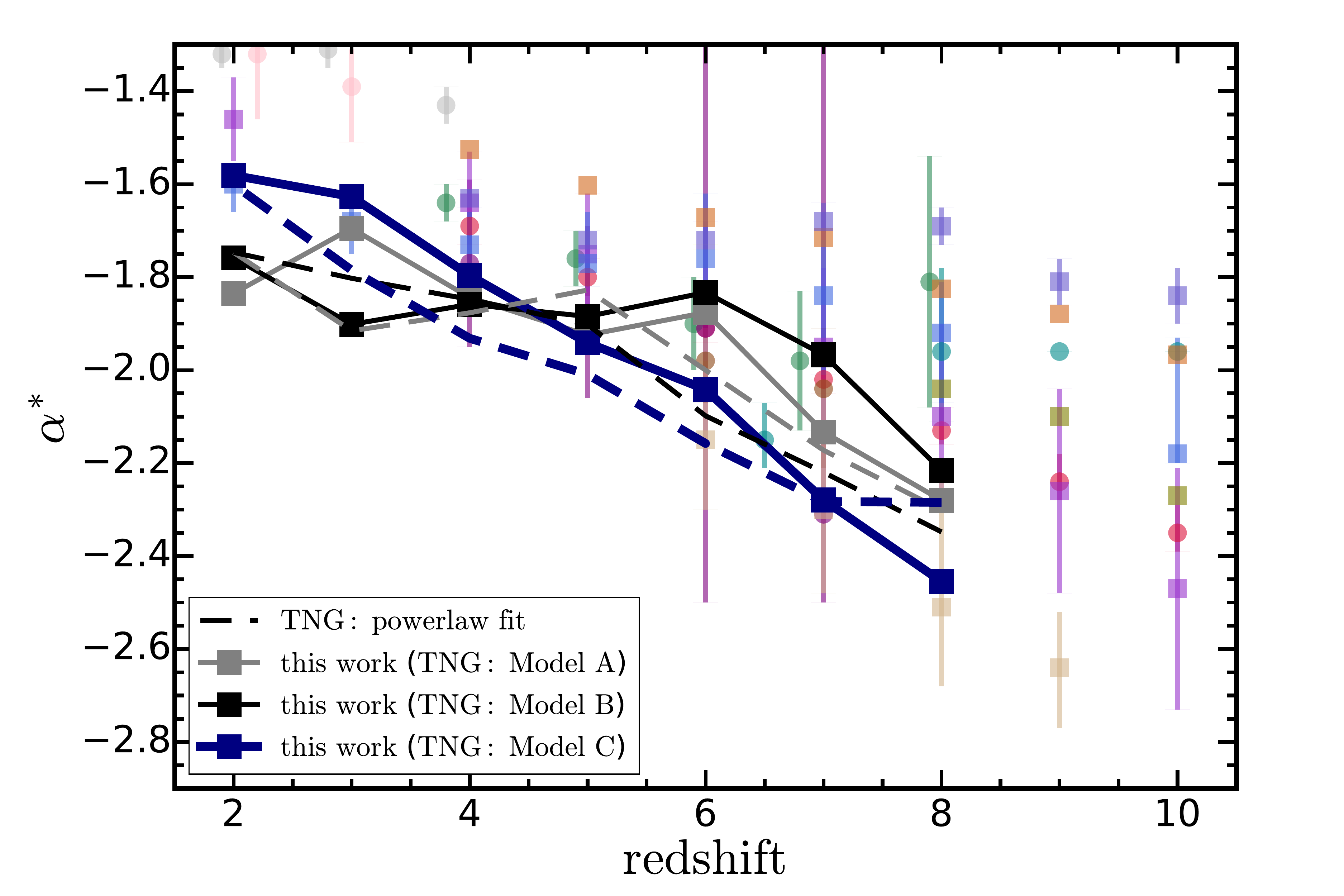}
\includegraphics[width=0.48\textwidth]{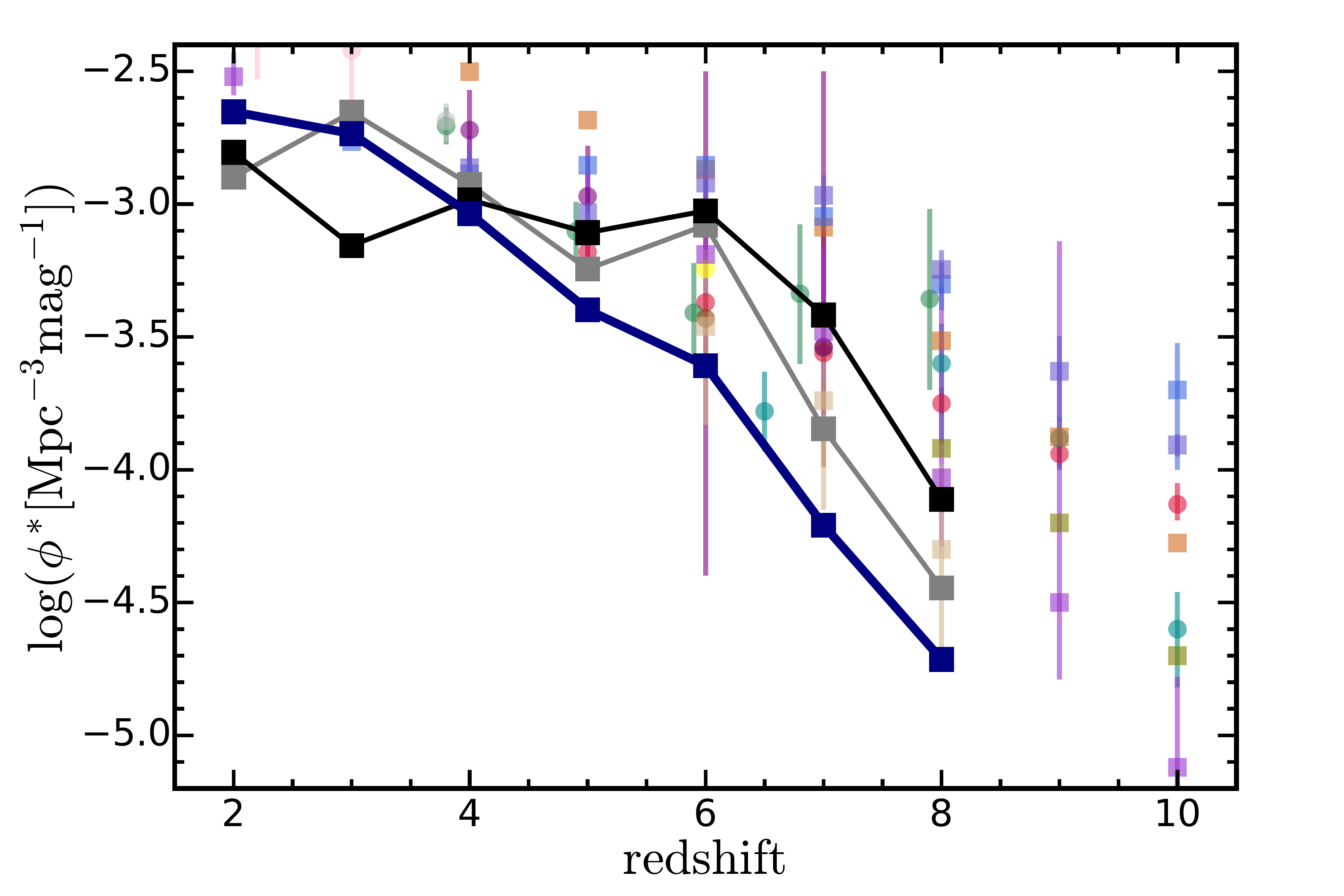}\\
\hspace{0.07cm}\includegraphics[width=0.48\textwidth]{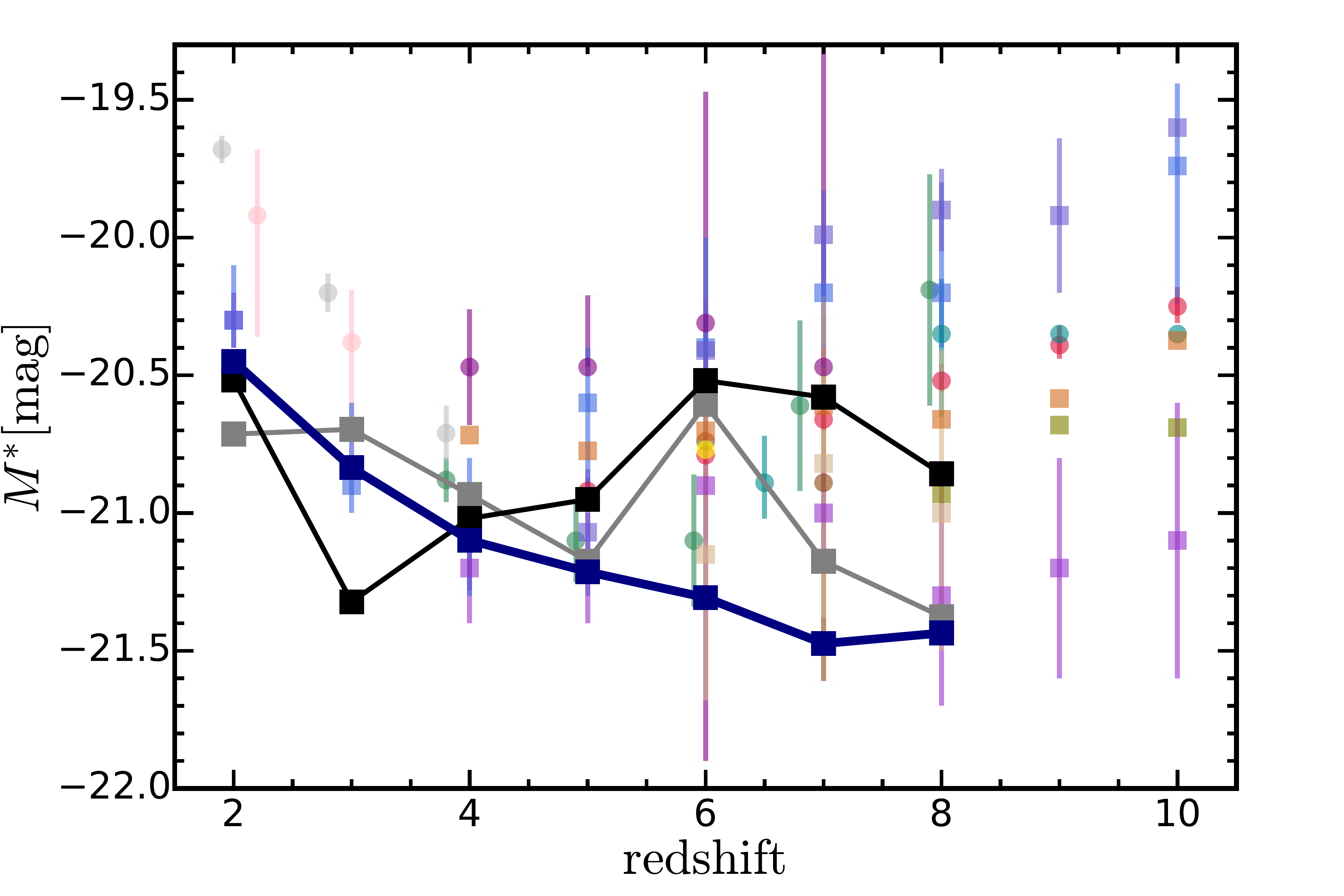}
\includegraphics[width=0.48\textwidth]{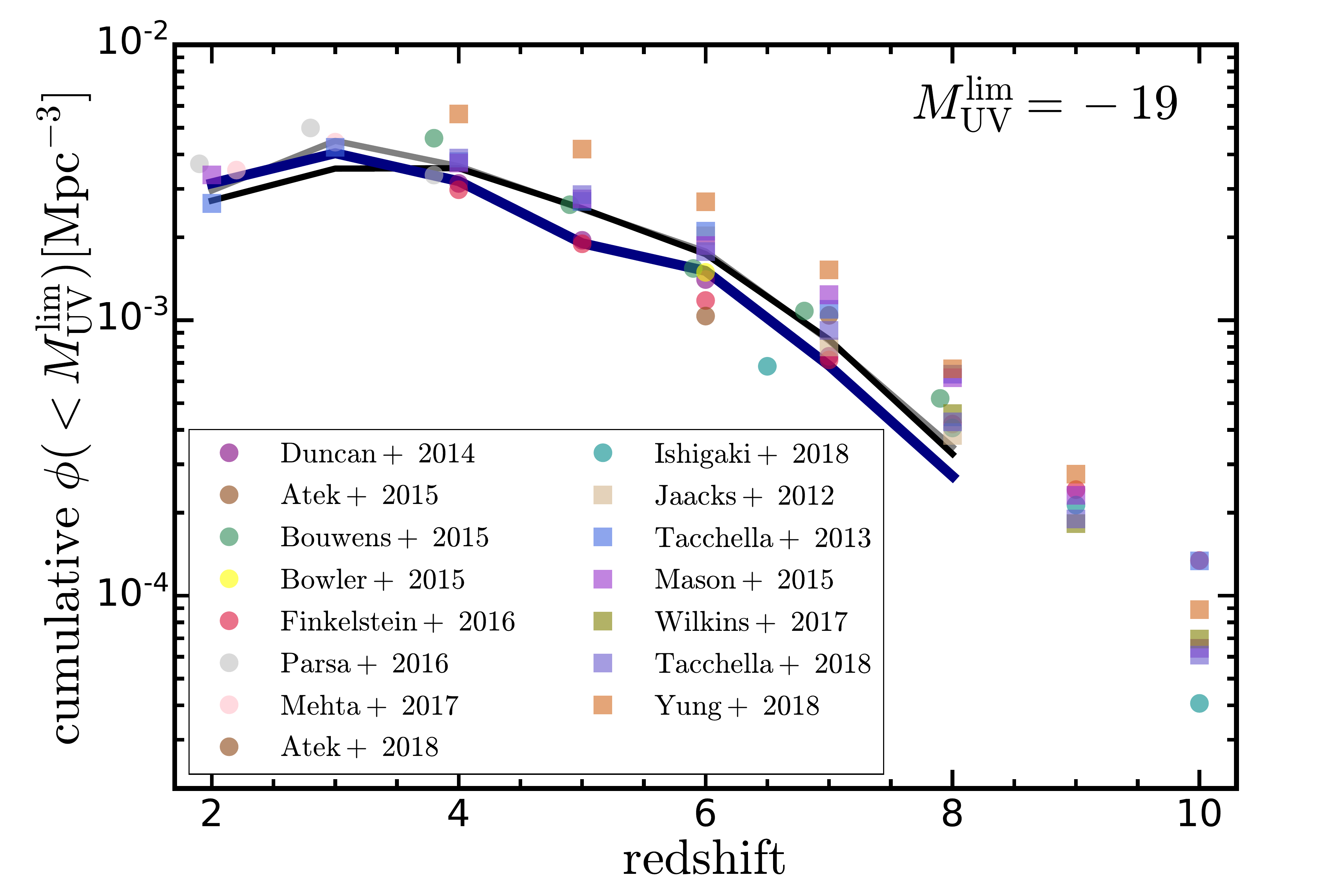}
\caption{{\bf Best-fit Schechter function parameters and cumulative number densities of galaxies above a certain UV luminosity threshold.} {\it Upper left, upper right, lower left panels:} Redshift evolution of the three derived Schechter fit parameters. We compare those with previous results from observations~\citep{Duncan2014, Atek2015b,Bowler2015,  Bouwens2015,  Finkelstein2016,Parsa2016,Mehta2017, Atek2018,Ishigaki2018} (circles), and analytical models and simulations predictions~\citep{jaacks2012, tacchella2013, mason2015, Wilkins2017, tacchella2018, Yung2018} (squares). The Schechter fit parameters of Model A, B, C are shown in gray, black, dark blue squares respectively. The dashed lines in the upper left panel show the faint-end slopes based on single power law fits for $M_{\rm UV}>-19.5\mmag$, which yield consistent results with the Schechter slopes. {\it Lower right panel:} Cumulative number densities analytically calculated based on best-fit Schechter fits of our work and fits presented in previous work. Our results tend to agree slightly better with existing observational data compared to most previous theoretical studies.}
\label{fig:UV_schfit}
\end{figure*}

\subsection{Rest-frame UV luminosity functions}

More quantitatively, we next discuss our result for the rest-frame UV luminosity functions. 
The UV luminosity functions at $z=2$ with different dust models and different dust parameters are presented in Figure~\ref{fig:lf_compare} that we briefly discussed already in the previous section. We present the best-fit resolution-corrected luminosity function of TNG100 in black lines along with the corresponding luminosity functions of TNG50 and the resolution-corrected luminosity function of TNG300 with transparent lines. Figure~\ref{fig:lf_compare} demonstrates
that the resolution correction works reliably over the magnitude ranges where the individual simulations sample the galaxy population sufficiently in terms of resolution and simulation volume.
In addition to the luminosity functions based on the best-fit dust parameters, we also present the luminosity functions for slightly different dust  parameters (coloured lines) for Model A and Model B to highlight their impact on the shape of the UV luminosity functions. The variation of the Model A intercept value $\beta_{M_{0}}$ dust parameter mainly affects the offset of the luminosity function while the slope ${\rm d}\beta/{\rm d}M_{\rm UV}$ mostly affects the shape of the bright end. For Model B, as expected, increasing the dust parameter $\tau_{\rm dust}(z=2)$ shifts the luminosity function away from the dust-free luminosity function at the bright end while preserving the faint end shape. This behaviour is generic also at other redshifts, i.e. as expected dust attenuation mostly affects the bright end of the galaxy population but not the faint end.
Towards the bright end the impact of dust is quite severe leading to large corrections on top of the dust-free luminosity function, which is also presented in the different panels. We also note that given the fact that the faint end is not significantly affected by dust attenuation, any predictions for the faint end, like the faint-end slope, are not affected by our dust modelling in any significant way. Furthermore, the faint end is mostly sampled by TNG50, which also does not require any resolution corrections. This implies that the faint-end slope is a quite robust prediction of our analysis given that it does not depend on dust- or resolution-corrections. Of course, TNG50 itself might still not be fully converged, but we expect that resolution impacts are comparatively small for TNG50.
We stress here however that only the faint-end prediction of Model B and C are not sensitive to dust parameter choices since those directly model the dust based on the metal content of galaxies. However, this is not the case for Model A, which is only an empirical relation providing a mapping between the dust-free and dust-attenuated rest-frame UV magnitudes. Therefore parameter variations of Model A can in principle also shift the galaxy number densities at the faint-end of the luminosity function.

\begin{table*}
\centering
\begin{tabular}{ p{0.08\textwidth}|p{0.08\textwidth}|p{0.08\textwidth}|p{0.08\textwidth}|p{0.08\textwidth}|p{0.08\textwidth}|p{0.08\textwidth}|p{0.08\textwidth}|p{0.08\textwidth}|p{0.06\textwidth} }
\hline 
{\bf redshift} & Model A &&& Model B &&& Model C\\
$\mathbf{z}$ & $\log{\phi^{\ast}}$ & $M^{\ast}$ & $\alpha^{\ast}$ & $\log{\phi^{\ast}}$ & $M^{\ast}$ & $\alpha^{\ast}$ & $\log{\phi^{\ast}}$ & $M^{\ast}$ & $\alpha^{\ast}$\\
& $[{\rm Mpc}^{-3}{\rm mag}^{-1}]$ & $[{\rm mag}]$ &  & $[{\rm Mpc}^{-3}{\rm mag}^{-1}]$ & $[{\rm mag}]$ &  & $[{\rm Mpc}^{-3}{\rm mag}^{-1}]$ & $[{\rm mag}]$ & \\
\hline
\hline
{\bf 2} & -2.898 & -20.71 & -1.835 & -2.805 & -20.52 & -1.759 & -2.652 &  -20.45 & -1.581\\
{\bf 3} & -2.654 & -20.70 & -1.695 & -3.156 & -21.32 & -1.901 & -2.735 & -20.84 & -1.627\\
{\bf 4} & -2.926 & -20.93 & -1.844 & -2.983 & -21.02 & -1.859 & -3.035 & -21.10 & -1.796\\
{\bf 5} & -3.244 & -21.17 & -1.924 & -3.107 & -20.95 & -1.884 & -3.398 & -21.21 & -1.941\\
{\bf 6} & -3.079 & -20.61 & -1.876 & -3.025 & -20.52 & -1.833 & -3.608 & -21.31 & -2.042\\
{\bf 7} & -3.846 & -21.18 & -2.133 & -3.418 & -20.58 & -1.967 & -4.209 & -21.47 & -2.279\\
{\bf 8} & -4.445 & -21.38 & -2.280 & -4.111 & -20.86 & -2.216 & -4.714 & -21.44 & -2.455\\
\hline
\end{tabular}
\caption{{\bf Best-fit UV luminosity function Schechter parameters for the three dust models (A, B and C).} We list for each dust model the three Schechter parameters: the normalisation $\Phi^\ast$, the transition magnitude $M^\ast$, and the faint-end slope parameter $\alpha^\ast$.}
\label{tab:UV_schfit}
\end{table*}

Next we explore the best-fit dust-attenuated galaxy luminosity functions at all redshifts of interest beyond $z=2$.
In Figures~\ref{fig:UVLF_A},\,\ref{fig:UVLF_B} and \ref{fig:UVLF_C}, we present the rest-frame UV luminosity functions from $z=2$ to $z=10$ for the best-fit dust parameters of Models A, B, and C, respectively. Here we include all
simulation volumes, TNG50, TNG100, and TNG300, with the latter two being resolution-corrected. We stress again that the dust parameter calibration for all dust models is solely based on the resolution-corrected TNG100 simulation to avoid any mis-calibration due to imperfections in combining the three simulation volumes. In addition to the individual resolution-corrected luminosity functions, we also present the combined luminosity function of the best-fit dust model for each redshift. The combined luminosity functions are shown with black dashed line and have been derived as discussed above.  These combined luminosity functions can be well described by Schechter functions~\citep[][]{Schechter1976}
\begin{equation}
\phi(M)=\dfrac{0.4\ln{(10)}\,\phi^{\ast}}{10^{0.4(M-M^{\ast})(\alpha^{\ast}+1)}}e^{-10^{-0.4(M-M^{\ast})}},
\end{equation}
shown in cyan transparent lines in each panel. We perform a $\chi^{2}$ fit to derive the best-fit Schechter parameters. To accelerate this process we employ the observationally derived Schechter parameters of~\cite{Finkelstein2016} as initial guesses for the fitting procedure. We note that even with TNG50 resolution, it is challenging to sample the faint end of the luminosity function ($M_{\rm UV}\gtrsim-17\mmag$) at higher redshifts. Specifically, we observe a drop of the faint end luminosity function for redshifts above $z=6$. This drop can affect the derived Schechter fits, especially in terms of the faint-end slope parameter $\alpha^{\ast}$. We therefore exclude these magnitude ranges of the luminosity function from the fitting procedure to avoid contamination of the derived best-fit parameters. We only fit the luminosity function in the range $M_{\rm UV}<-17\mmag$ for $z = 6$ and $M_{\rm UV}<-17.5\mmag$ for $z>6$. Furthermore, we ignore bins with less than $10$ galaxies at the bright end of TNG300, roughly corresponding to $\log{(\phi)}=-6$, for the fitting procedure. Our derived best-fit Schechter parameters for the rest-frame UV luminosity functions at different redshifts are listed in Table~\ref{tab:UV_schfit}.

We compare our model predictions for the rest-frame UV luminosity function also with observational data. Specifically, we have added at each redshift the compilation of observational datasets described in Section~\ref{sec:calibration}. Most of these observational points are taken from the \citet{Finkelstein2016} compilation with the original sources listed in Section~\ref{sec:calibration}. This compilation contains observations mainly from the CANDELS, HUDF, UltraVISTA and UKIDSS UDS surveys. In addition we also include a few more observations that are not included in this large compilation. Specifically, \citet{Atek2018} used gravitational lensing data of the {\it HST} Frontier Fields program to constrain the extreme faint end UV luminosity function at $z=6$. \citet{Oesch2018} presented the UV luminosity function at $z=10$ observed in all {\it HST} Legacy fields. \citet{Mehta2017} provided constraints on the UV luminosity functions at $z=2,3$ from the {\it HST} Ultra-Violet Ultra Deep Field program. We note that some of the observations at $z=2$~\citep[e.g.,][]{reddy2008,reddy2009,Oesch2010,Hathi2010} are not fully consistent with the most recent observations~\citep{Parsa2016,Alavi2014,Mehta2017}. We therefore do not consider these constraints in our analysis.
Generally our model predictions agree well with the different observational datasets, which is largely a consequence of the dust calibration procedure. At higher redshifts, $z\sim 9-10$, the predictive power of the IllustrisTNG simulation suite is limited due to the limited simulation resolution of TNG100 and TNG300. We are therefore not able to derive reliable best-fit Schechter parameters for these two high redshifts.

Inspecting our predictions in more detail, we find however also some discrepancies between observations and our theoretical predictions.  
For example, towards lower redshifts ($z<4$), the predicted faint end galaxy number densities are slightly higher than observationally found. This also goes along with a slightly steeper faint-end slope. As discussed above, we note that the faint-end behaviour of the luminosity function is expected to be more reliably described by Models B and C, where the dust-free and dust-attenuated results should agree nearly perfectly. This is not necessarily the case for Model A due to its empirical nature. The higher galaxy number densities of faint galaxies for these two dust models are therefore a direct consequence of the galaxy formation model employed in IllustrisTNG and are independent of resolution and dust corrections. Towards the bright end we find that our luminosity function predictions agree well with observational data. We stress here that the dust corrections for the bright end are crucial and obviously our predictions here are sensitive to the dust parameter calibration. We also note that at each redshift the dust attenuation is described  only by one normalisation constant for Models B and C, which is the same for all galaxies at the given redshift. The luminosity functions for the three dust models also reveal that for $z \sim 8$ the abundance of galaxies around $M_{\rm UV}\sim -19$ is slightly underpredicted compared to observations. To some degree this is also the case for $z\sim 7$. Despite these minor discrepancies, we conclude that the rest-frame UV luminosity functions of IllustrisTNG agree well with observational data over a wide range of luminosities and redshifts.

\begin{figure*}
\includegraphics[width=1.02\textwidth]{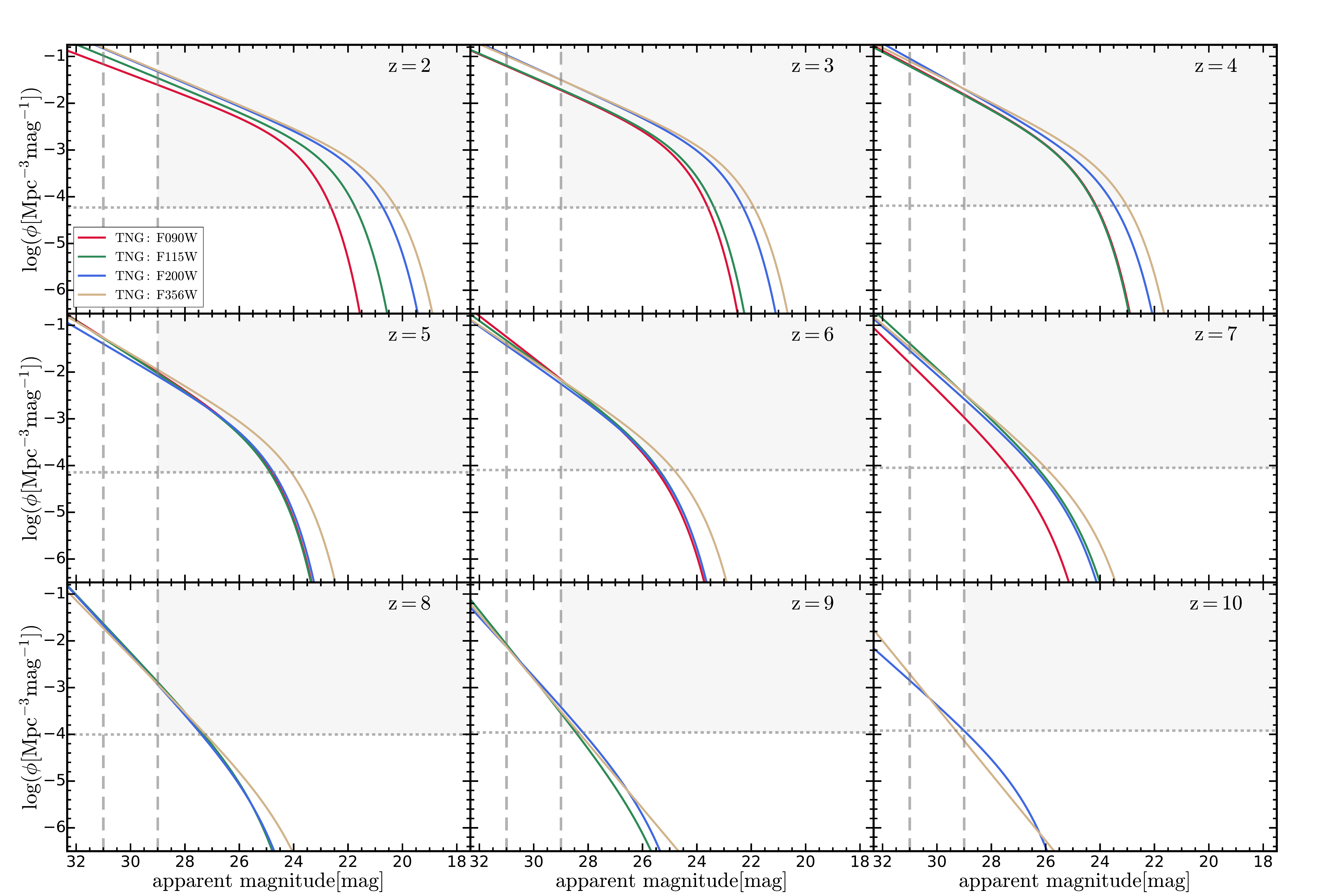}
\caption{{\bf Schechter fits from the TNG apparent luminosity functions of four selected NIRCam wide filters (F090W, F115W, F200W, F356W) at $\mathrm{\mathbf{z=2-10}}$.} Vertical dashed line show two approximate {\it JWST} detection limits: $\sim29\mmag$ for an exposure time of $10^{4}{\rm s}$ and a target signal-to-noise ratio of $10$; $\sim31\mmag$ for an exposure time of $10^{5}{\rm s}$ and a target signal-to-noise ratio of $5$. Horizontal dashed lines represent the number density required for the detection of one object per magnitude per unit redshift in the {\it JWST} NIRCam field of view.}
\label{fig:jwstlf}
\end{figure*}

In Figure~\ref{fig:UV_schfit}, we compare the derived best-fit UV luminosity function Schechter parameters with other values taken from the literature.  Observational constraints~\citep{Duncan2014,Bowler2015,Bouwens2015,Parsa2016,Finkelstein2016,Mehta2017,Atek2015b,Atek2018,Ishigaki2018} are presented with circles, theoretical predictions~\citep{jaacks2012,tacchella2013,mason2015,Wilkins2017,tacchella2018,Yung2018} are shown as squares. In general, we find slightly steeper faint-end slopes, $\alpha^{\ast}$, and lower $\phi^{\ast}$ values while the predictions for $M^{\ast}$ are broadly consistent with previous theoretical results and observational constraints.  We have also performed a pure power law fit to the faint-end of the luminosity functions to find the faint-end slope. These power law fits have been performed for $M_{\rm UV}>-19.5\mmag$. The result for these slopes are presented as dashed lines in the upper left panel of Figure~\ref{fig:UV_schfit}. Those are broadly consistent with the Schechter fit faint-end slopes.
Based on the best-fit Schechter functions we can also estimate the cumulative number density of galaxies above a certain magnitude limit. 
In the bottom right panel of Figure~\ref{fig:UV_schfit}, we show this cumulative number density of galaxies above a certain UV luminosity limit $M_{\rm UV}^{\rm lim}=-19\mmag$. This cumulative number density is calculated by integrating the best-fit Schechter functions as:
\begin{align}
\label{eq:phi_cum}
\phi_{\rm cum}\left (<M_{\rm UV}^{\rm lim}\right )&=\int^{\infty}_{L^{\rm lim}}\!\!\!\!\phi^{\ast}\left(\dfrac{L}{L^{\ast}}\right)^{\alpha^{\ast}}\exp\left(-\dfrac{L}{L^{\ast}}\right)\dfrac{{\rm d}L}{L^{\ast}}\\
&=\phi^{\ast}\,\Gamma_{\rm inc}(\alpha^{\ast}+1,10^{-0.4(M_{\rm UV}^{\rm lim}-M^{\ast})}),\nonumber
\end{align}
where $\Gamma_{\rm inc}(a,z)=\int^{\infty}_{z}t^{a-1}e^{-t}{\rm d}t$ is the incomplete upper gamma function. We process the best-fit Schechter parameters from the literature similarly to derive cumulative number densities for comparison. Over the range of redshifts where we have valid Schechter fits, we find consistent cumulative number densities compared to previous studies. Our results tend to be in slightly better agreement with observational data than the predictions of some analytical models. We note that this cumulative number density above $M_{\rm UV}^{\rm lim}=-19\mmag$ mostly probes the bright end of the rest-frame UV luminosity function. As discussed above, our theoretical predictions are in good agreement with observational data at the bright end. It is therefore not surprising that we find such a good agreement for this cumulative number density given our calibration procedure.

\begin{figure*}
\includegraphics[width=0.48\textwidth]{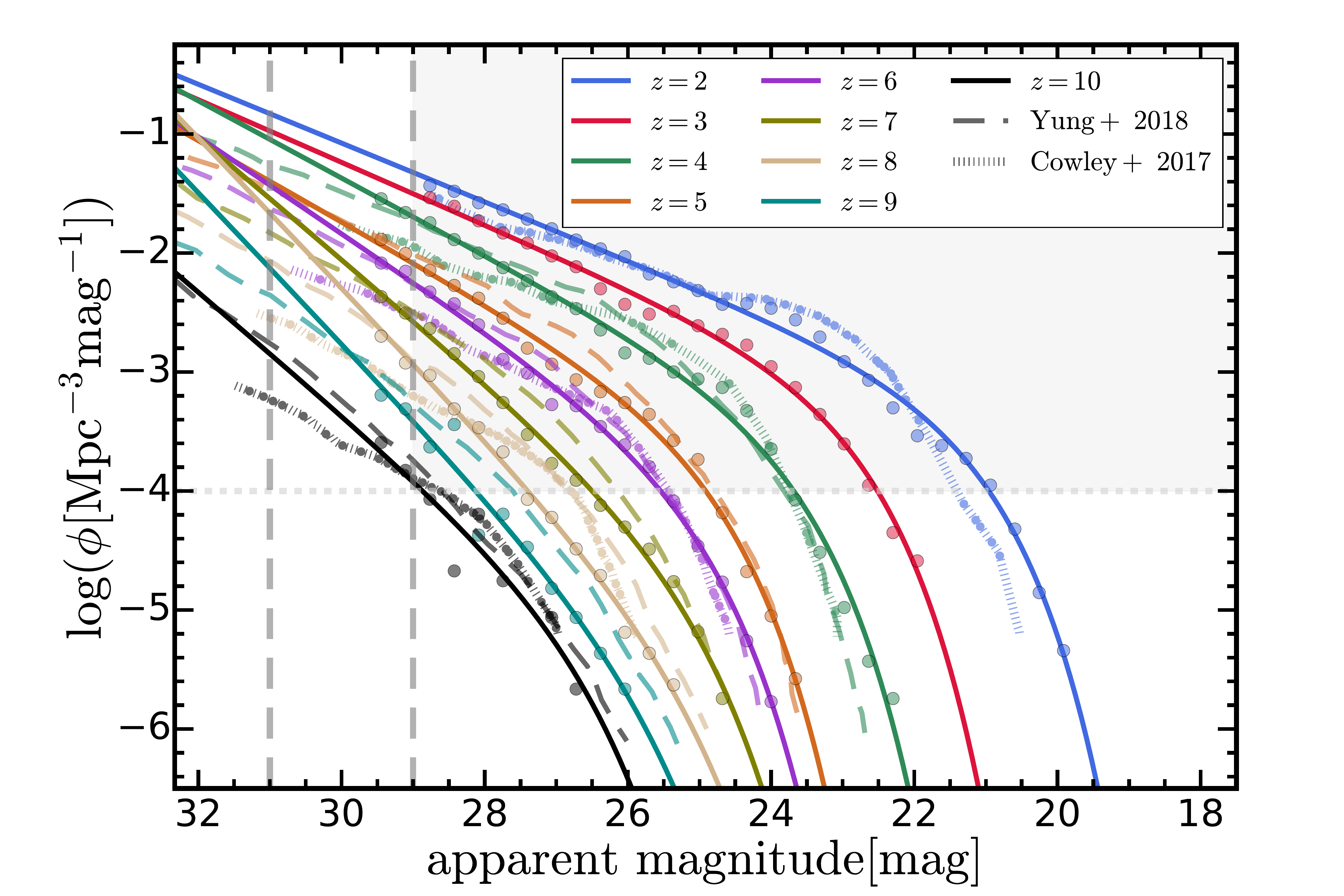}
\includegraphics[width=0.48\textwidth]{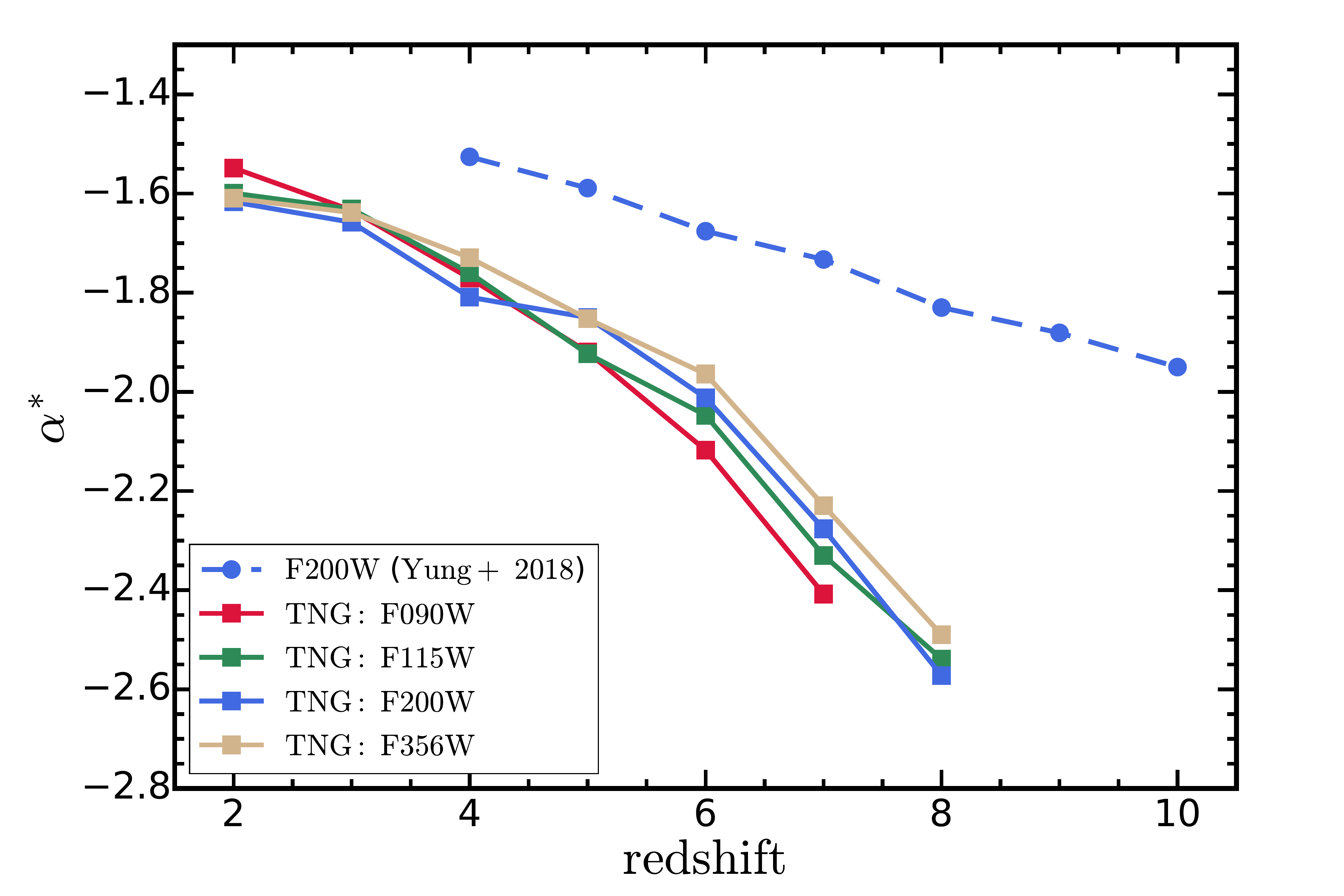}\\
\hspace{0.07cm}\includegraphics[width=0.48\textwidth]{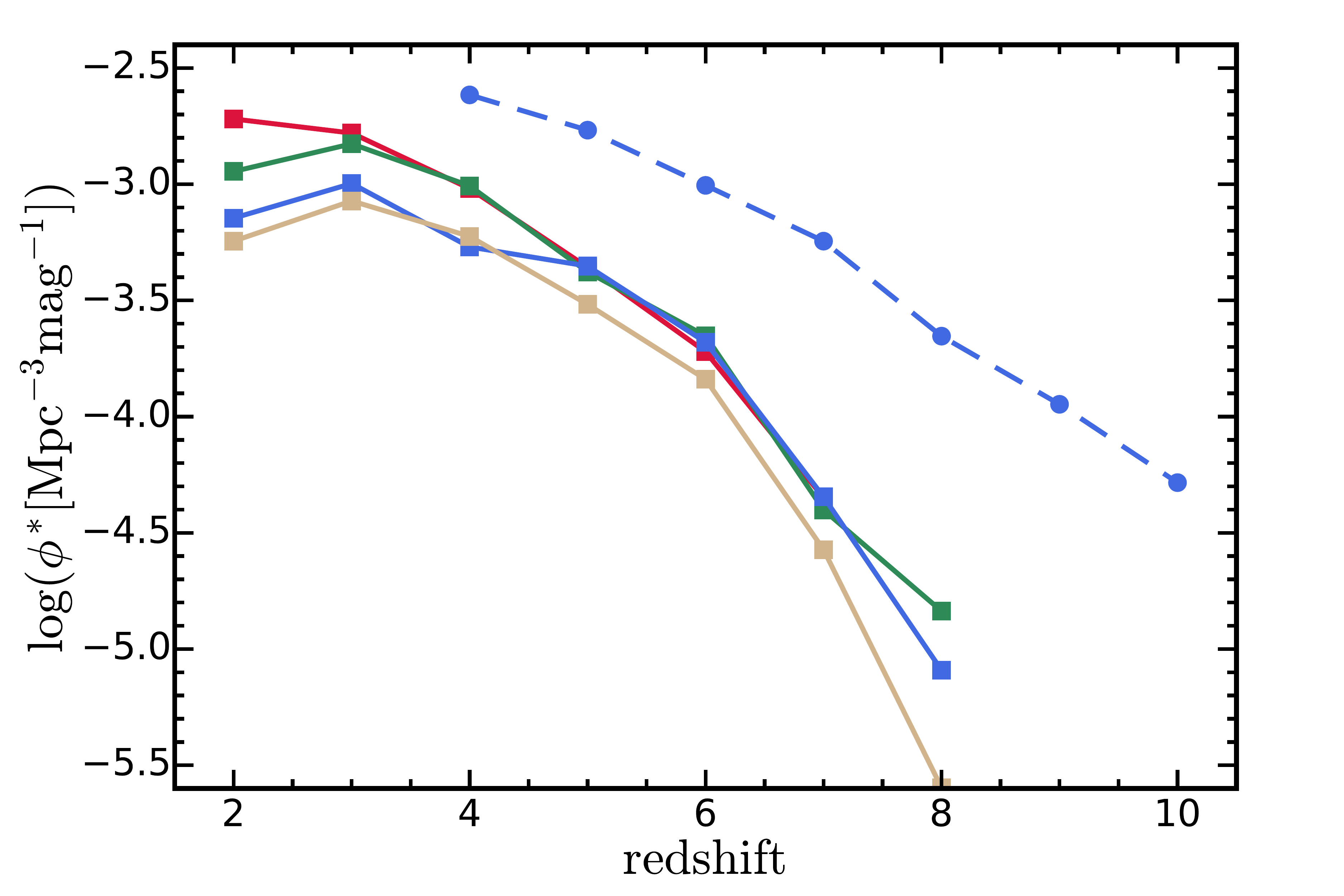}
\includegraphics[width=0.48\textwidth]{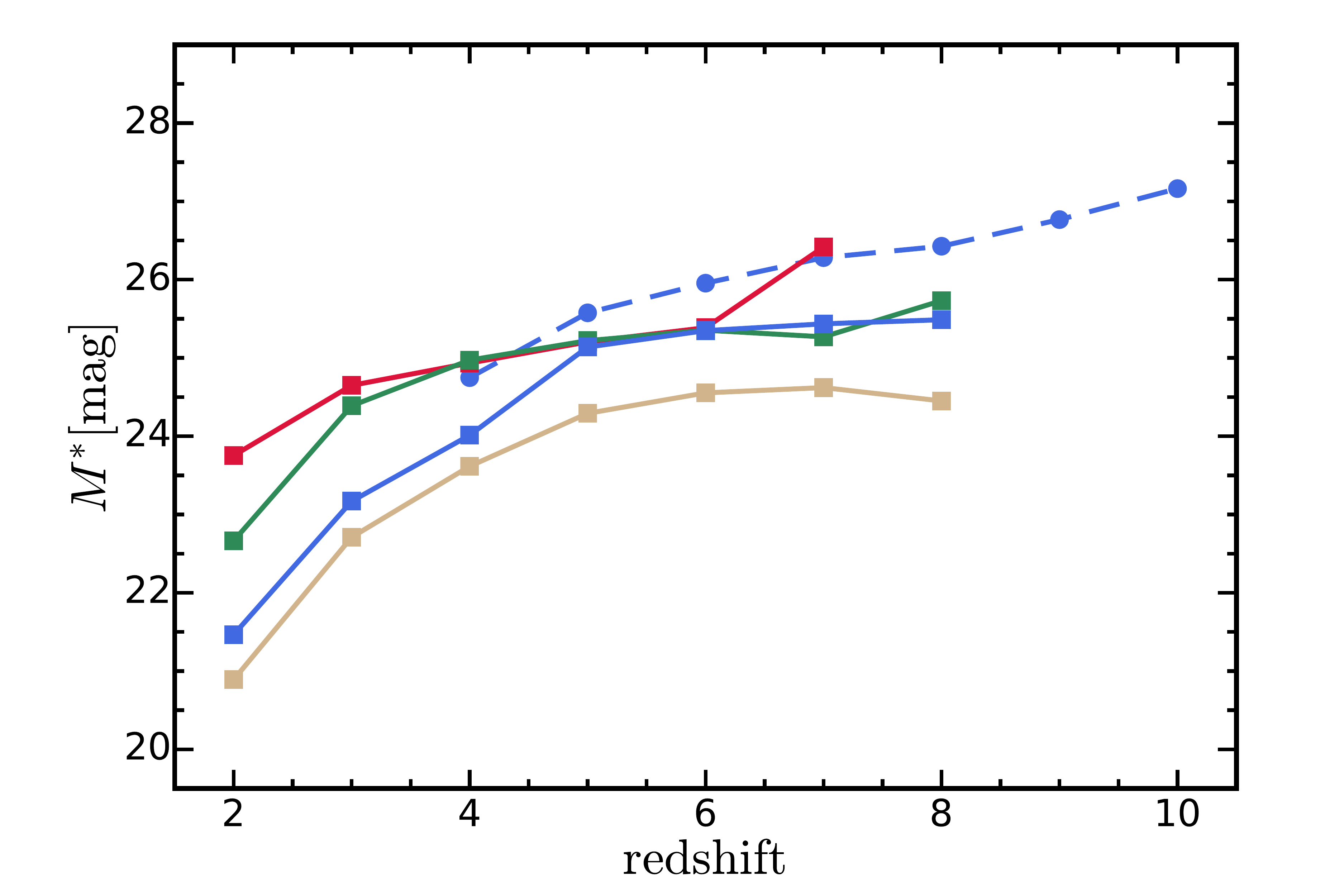}
\caption{{\bf F200W band luminosity functions and best-fit Schechter function parameters for the predicted NIRCam wide filter apparent luminosity functions.} {\it Upper left:} {\it JWST} NIRCam F200W band apparent luminosity functions at different redshifts. Symbols show the actual simulation data that determines the fits. We compare IllustrisTNG Schechter fits of apparent {\it JWST} luminosity function predictions with Schechter fits  of~\citet{Yung2018} and predictions from~\citet{Cowley2018}. Vertical and horizontal dashed lines have the same meaning as in Figure~\ref{fig:jwstlf} but the detection limit in number density is a rough value $\sim-4$ for $z=2-10$. {\it Upper right, lower left, lower right panels:} Best-fit Schechter function parameters at each redshift for selected wide filters. The best-fit Schechter parameters of \citet{Yung2018} for the F200W band are presented for comparison. 
}
\label{fig:f200w_compare}
\end{figure*}

\subsection{Apparent {\it JWST} band luminosity functions}

After studying the rest-frame UV band luminosity functions, we will now explore the {\it JWST} apparent band magnitude luminosity functions. Hereafter, we will only use predictions of our dust Model C in all subsequent analysis. The {\it JWST} band luminosity functions are resolution corrected and combined in the same way as the UV luminosity functions presented above. The only difference is that we employ here $47$ linearly spaced bins from apparent magnitude $16\mmag$ to $32\mmag$ for the analysis. Schechter fits for the combined luminosity functions of four selected {\it JWST} NIRCam wide filter bands at each redshift are presented in Figure~\ref{fig:jwstlf}.
For the Schechter fitting procedure we set the apparent magnitude range to $<29\mmag$ at $z=2,3$ and $<29.5\mmag$ at $z\geq4$. Also similar to the procedure used to derive the UV luminosity function fits, we ignore bins with less than $10$ galaxies, roughly corresponding to $\log{\phi}=-6$ for TNG300. 
Towards lower redshifts, the luminosity functions separate clearly. Towards higher redshifts, the luminosity functions of different bands begin however to overlap due to the closer rest-frame wavelength range that is probed. Each panel of Figure~\ref{fig:jwstlf} also contains rough and typical detection limits in apparent magnitude and number density: $\sim 29\mmag$ is the rough detection limit for a $10^{4}{\rm s}$ exposure time with signal-to-noise ratio $10$, and $\sim 31\mmag$ is the rough detection limit for a $10^{5}{\rm s}$ exposure time with signal-to-noise ratio $5$. We note that the exact detection limit depends on the actual band filter, and $\sim 29\mmag$, $\sim31\mmag$ are here only  approximate values. The precise detection limits for each of the bands are listed in Table~\ref{tab:jwst_filters}. The luminosity functions of F070W, which is not shown here, and F090W quickly drop out of the detectable regime towards higher redshifts due to IGM absorption. The horizontal detection line corresponds to one object in the NIRCam field of view per unit redshift and per magnitude calculated as described below. 

\begin{table*}
\centering
\begin{tabular}{ p{0.10\textwidth}|p{0.10\textwidth}|p{0.12\textwidth}|p{0.125\textwidth}|p{0.125\textwidth}|p{0.125\textwidth}|p{0.125\textwidth}|p{0.125\textwidth}|p{0.00\textwidth} }
\hline 
{\bf NIRCam} & wavelength & bandwidth & $m^{\rm lim}_{{\rm SNR=10}, T_{\rm exp}=10^4{\rm s}}$ & $m^{\rm lim}_{{\rm SNR=10}, T_{\rm exp}=10^5{\rm s}}$ & $m^{\rm lim}_{{\rm SNR=5}, T_{\rm exp}=10^4{\rm s}}$ & $m^{\rm lim}_{{\rm SNR=5}, T_{\rm exp}=10^5{\rm s}}$  \\
{\bf wide filter} & [$\mu{\rm m}$] & [$\mu{\rm m}$] & [${\rm mag}$] & [${\rm mag}$] & [${\rm mag}$] & [${\rm mag}$]\\
\hline
\hline
{\bf F070W} & 0.704 & 0.132 & 28.46 & 29.69 & 29.24 & 30.46\\
{\bf F090W} & 0.902 & 0.194 & 28.79 & 30.02 & 29.57 & 30.78\\
{\bf F115W} & 1.154 & 0.225 & 29.02 & 30.24 & 29.80 & 31.00\\
{\bf F150W} & 1.501 & 0.318 & 29.27 & 30.49 & 30.05 & 31.25\\
{\bf F200W} & 1.989 & 0.457 & 29.34 & 30.55 & 30.12 & 31.31\\
{\bf F277W} & 2.762 & 0.683 & 29.06 & 30.25 & 29.83 & 31.01\\
{\bf F356W} & 3.568 & 0.781 & 29.06 & 30.25 & 29.83 & 31.01\\
{\bf F444W} & 4.408 & 1.029 & 28.61 & 29.78 & 29.37 & 30.55\\
\hline
\end{tabular}
\caption{{\bf {\it \textbf{JWST}} NIRCam wide filter characteristics and detection limits.} The table contains the pivot wavelengths, bandwidths and detection limits of the {\it JWST} NIRCam wide filters adopted in this work. The detection limits are calculated assuming $10^{4}{\rm s}$ and $10^{5}{\rm s}$ exposure time with target signal-to-noise ratios of $5$ and $10$ based on the {\it JWST} Exposure Time Calculator. }
\label{tab:jwst_filters}
\end{table*}

The exact {\it JWST} detection limits of Table~\ref{tab:jwst_filters} are derived based on calculations using the {\it JWST} Exposure Time Calculator (ETC)\footnote{\url{https://jwst.etc.stsci.edu/}} with the following configuration details. Sources are treated as point sources, and the exposure time is set to either $10^{4}\rm s$ or $10^{5}\rm s$ as indicated in the table. Furthermore, the readout pattern is set to DEEP8, which yields a high signal-to-noise ratio and can efficiently reach a  maximum survey depth. For the $10^{4}{\rm s}$ exposure time we employ the full sub-array with $13$ groups per integration, $1$ integration per exposure, $4$ exposures per specification. For the $10^{5}{\rm s}$ exposure time we employ $20$ groups per integration, $1$ integration per exposure, $24$ exposures per specification. 
The aperture radius is set to $0.04"$ for the short-wavelength filters (F070W,F090W,F115W,F150W,F200W) and $0.08"$ for the long-wavelength filter (F277W,F356W,F444W). The aperture used for imaging aperture photometry is circular, which specifies the radius of the circular aperture used for extraction. The background subtraction is performed with a sky annulus with inner and outer radii of $0.6"$ and $0.99"$ for short-wavelength bands, $1.2"$ and $1.98"$ for long-wavelength bands. The ETC background model includes celestial sources (zodiacal light, interstellar medium, and cosmic infrared background) along with telescope thermal and scattered light. This background model varies with the target coordinates (RA, Dec) and time of year. Here, we choose the fiducial background at RA = 17:26:44, Dec = -73:19:56 on June 19, 2019\footnote{\url{ttps://jwst-docs.stsci.edu/display/JTI/NIRCam+Imaging+Sensitivity}}. We note that some authors employ background configurations with a pre-calculated percentile of the range of backgrounds predicted at the specified sky position over the window of visibility. For example, \citet{Yung2018} chose the percentile to be $10\%$ referred to as ``low'' level configuration. For each band, we then set up all the ETC parameters as described and then vary the apparent magnitude of the source until a target signal-to-noise ratio is reached. This then sets the corresponding apparent magnitude detection limit for this band for various exposure times and target signal-to-noise ratios. The final result of this procedure are the detection limits listed in Table~\ref{tab:jwst_filters}.

In the upper left panel of Figure~\ref{fig:f200w_compare}, we compare the {\it JWST} F200W band luminosity functions at all redshifts with previous theoretical predictions. The results from \citet{Yung2018} are based on a semi-analytical galaxy formation model along with a dust model similar to our Model B to provide {\it JWST} predictions for {\it JWST} NIRCam wide filters in the redshift range from $z=4$ to $z=10$. 
\citet{Cowley2018} combined the spectrophotometric radiative transfer code {\sc Grasil}~\citep{silva1998} with the semi-analytic {\sc Galform} code to model the impact of dust emission and absorption. They provide predictions for deep galaxy surveys with the {\it JWST} NIRCam and MRI instruments.  Our results are roughly consistent with those of~\citet{Yung2018} but we predict lower number densities at the intermediate luminosity range ($-28\mmag$ to $-26\mmag$) at $z\geq5$. Our results are consistent with those of~\citet{Cowley2018} towards the bright and intermediate luminosity range but differ at the faint end, where we predict much higher number densities. We note however that the predictions of \citet{Cowley2018} have a strong dependence on their feedback model. For example, their evolving feedback model predicts much higher galaxy number densities at the faint end. In the upper left panel of Figure~\ref{fig:f200w_compare}, we also plot the approximate detection limits: $\sim31\mmag$ for $10^{5}{\rm s}$ and signal-to-noise ratio $5$, $\sim29\mmag$ for $10^{4}{\rm s}$ and signal-to-noise ratio $10$. We note that the numerical resolution of TNG50 is not sufficient to probe the luminosity function down to $\sim31$ or beyond.

\begin{figure*}
\includegraphics[width=0.48\textwidth]{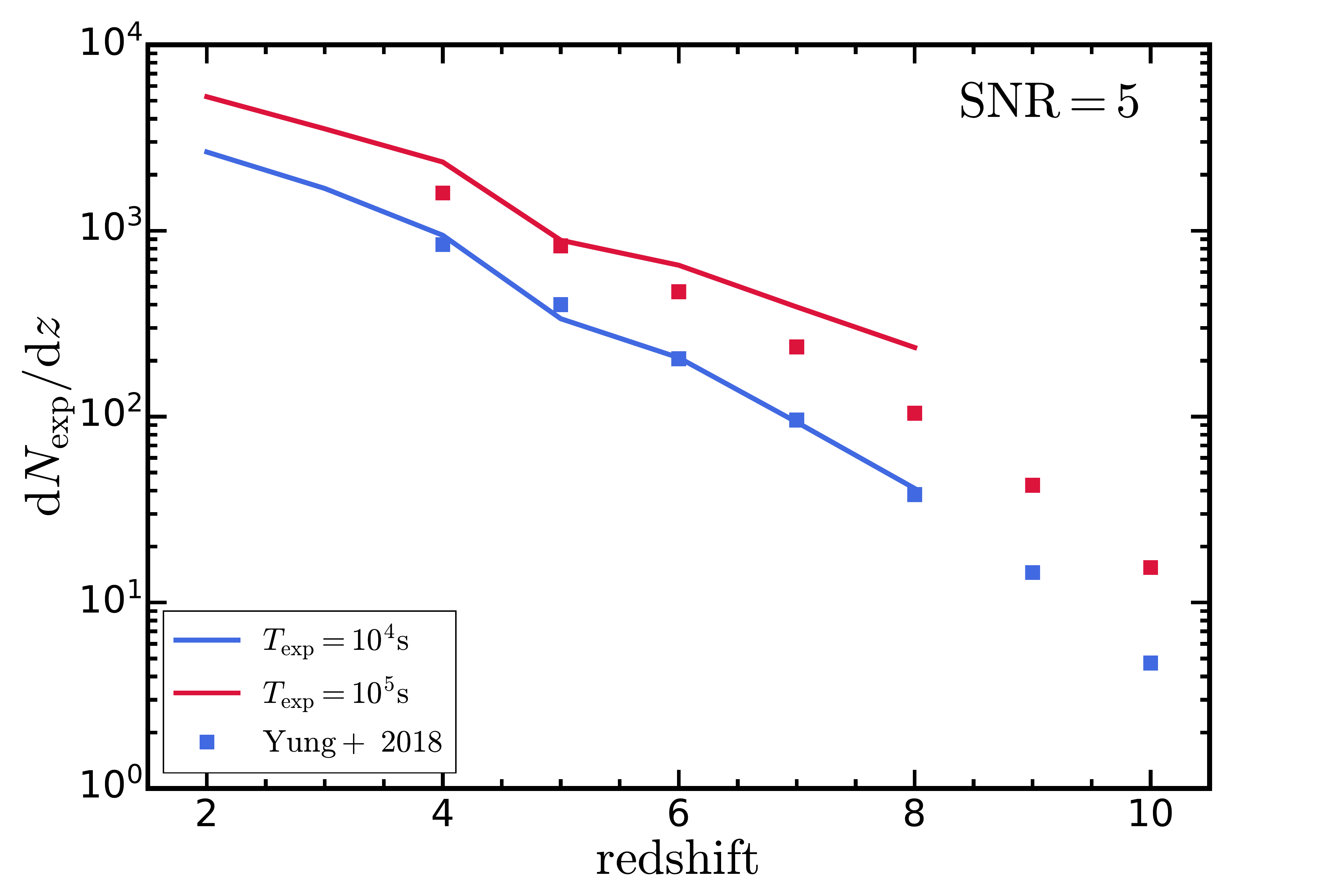}
\includegraphics[width=0.48\textwidth]{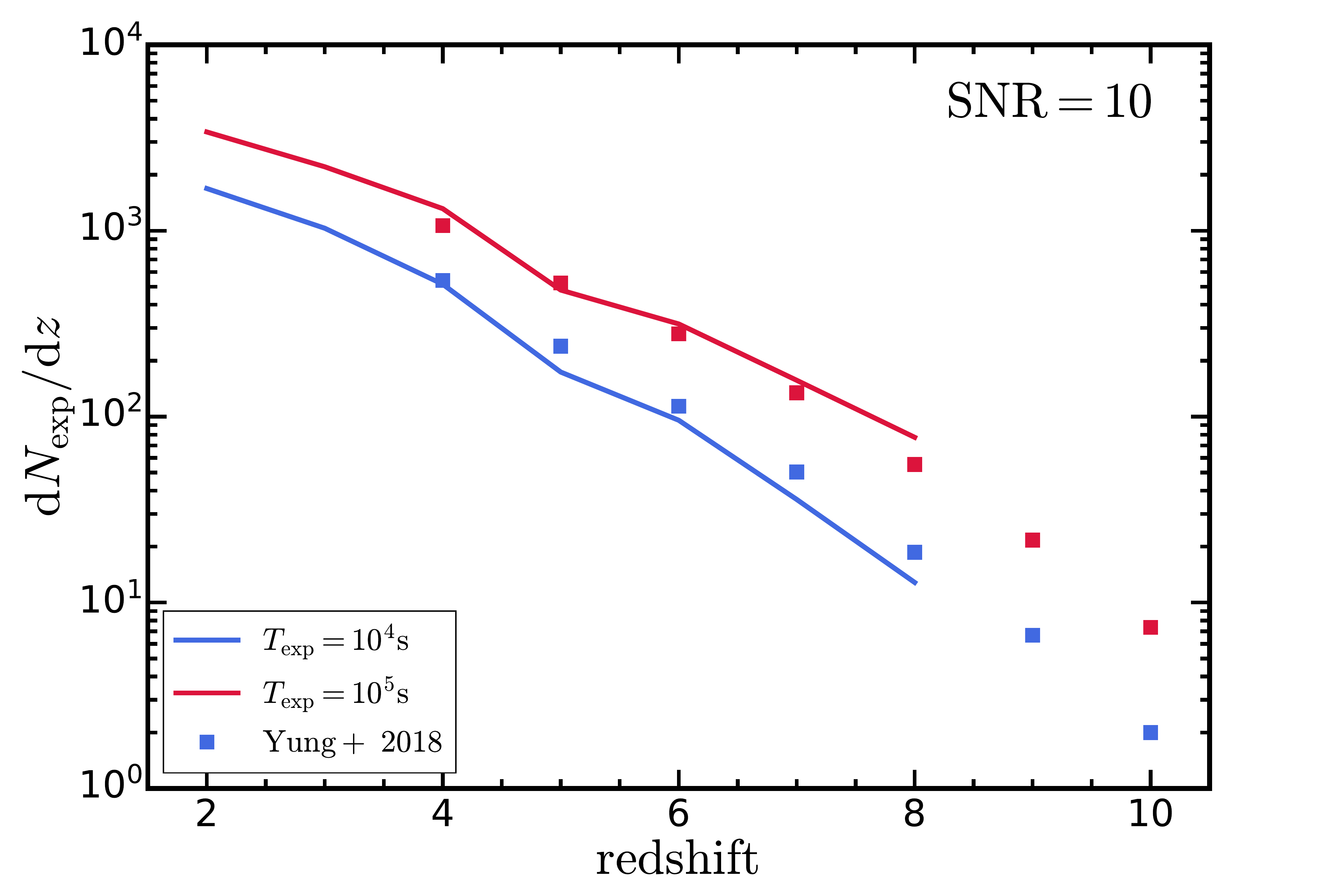}
\caption{{\bf Expected number of galaxies per unit redshift observed in the NIRCam F200W band in the NIRCam field of view ($\mathbf{2.2\times2.2 \,{\rm \bf arcmin}^{2}}$).} We present the results as a function of redshift with exposure times of $10^{4}{\rm s}$ and $10^{5}{\rm s}$. The left and the right panel show the expected numbers assuming two different target signal-to-noise ratios: $5$ ({\it left panel}) and $10$ ({\it right panel}). Our predictions are roughly consistent with those from \citet{Yung2018}. However, for a signal-to-noise ratio of $5$ and $T_{\rm exp}=10^{5}{\rm s}$ we predict a larger number of galaxies for the different redshifts. We note however that this setup has the faintest detection limit, $\sim 31\mmag$. Since our luminosity function cannot reach that low detection limit, our predictions here largely rely on an extrapolation. The steeper faint end slope we find then leads to the slight over-prediction compared with \citet{Yung2018}.} 
\label{fig:f200w_nexp}
\end{figure*}

The Schechter fits of the apparent {\it JWST} luminosity functions are reliable up to $z=8$. At $z=9$ and $z=10$, we are not able to derive reliable Schechter fit parameters due to the limited statistical sample similar to the limitation when deriving UV luminosity function Schechter fits. The redshift evolution of the Schechter fit parameters are presented in Figure~\ref{fig:f200w_compare}. 
We note that once the luminosity function of a given band drops out of the detectable regime, the shaded regions in Figure~\ref{fig:jwstlf}, we do not show the parameters anymore. 
We predict a steeper faint-end slope than the theoretical predictions of \citet{Yung2018},  and we also find lower $\phi^{\ast}$ normalisation values.  We note here however that the UV faint-end slope predicted by \cite{Yung2018} is also shallower than most observational data and other theoretical predictions.

\subsection{{\it JWST} predictions}

Based on the best-fit Schechter functions, we can integrate the apparent {\it JWST} luminosity functions down to the detection limit to derive the cumulative number density above that detection limit following Equation~\ref{eq:phi_cum}. We can then calculate the expected number of galaxies per unit redshift in the NIRCam field of view  as:
\begin{equation}
\dfrac{{\rm d}N_{\rm exp}}{{\rm d}z}=\phi_{\rm cum}(<m^{\rm lim}) \,\, \dfrac{{\rm d}V_{\rm com}}{{\rm d}\Omega \, {\rm d}z}(z) \,\, \Delta \Omega,
\end{equation}
where $\phi_{\rm cum}(<m^{\rm lim})$ is the cumulative number density calculated with Equation~\ref{eq:phi_cum}, $\Delta \Omega$ is the solid angle corresponding to the field of view of the {\it JWST} NIRCam instrument ($2.2\times 2.2\,{\rm arcmin}^{2}$) and ${\rm d}V_{\rm com}/{\rm d}\Omega{\rm d}z$ is the differential comoving volume element calculated as:
\begin{equation}
\dfrac{{\rm d}V_{\rm com}}{{\rm d}\Omega\,{\rm d}z}(z)=\dfrac{c\,(1+z)^{2}\,\,d_{\rm A}(z)^{2}}{H_{0}\,E(z)},
\end{equation}
where $d_{\rm A}(z)$ is the angular diameter distance and $H(z) = H_{0}E(z)$ is the Hubble parameter at redshift $z$. We note that we can also use this equation to inversely derive the number density detection limit corresponding to only one object per magnitude per unit redshift in the field of view of the {\it JWST} NIRCam instrument. This limit is plotted in Figure~\ref{fig:jwstlf} and the upper left panel of Figure~\ref{fig:f200w_compare} as horizontal dashed lines. The predicted number counts of galaxies for the NIRCam F200W band in the {\it JWST} NIRCam field of view are presented in Figure~\ref{fig:f200w_nexp}. Here we also compare those values with results derived from the best-fit Schechter functions of \citet{Yung2018}. For the left (right) panel, we assume a target signal-to-noise ratio of $5$ ($10$). For a short exposure time, $10^{4}{\rm s}$, we predict consistent numbers with \citet{Yung2018}, despite some deficit at $z>7$, where our luminosity function is lower. For a longer exposure time, $10^{5}{\rm s}$, we predict slightly higher numbers since we find a steeper faint-end slope in the derived Schechter fits. However, here we also start approaching a regime where the detection limit reaches or is already beyond our numerical resolution limit. Due to this limitation in numerical resolution, we do not provide predictions for longer exposure times, e.g., $10^{6}{\rm s}$, which would result in even lower apparent magnitudes as detection limits. We note that the expected number of detected galaxies in the field of view rises monotonically towards lower redshifts.

\begin{figure}
\includegraphics[width=0.51\textwidth]{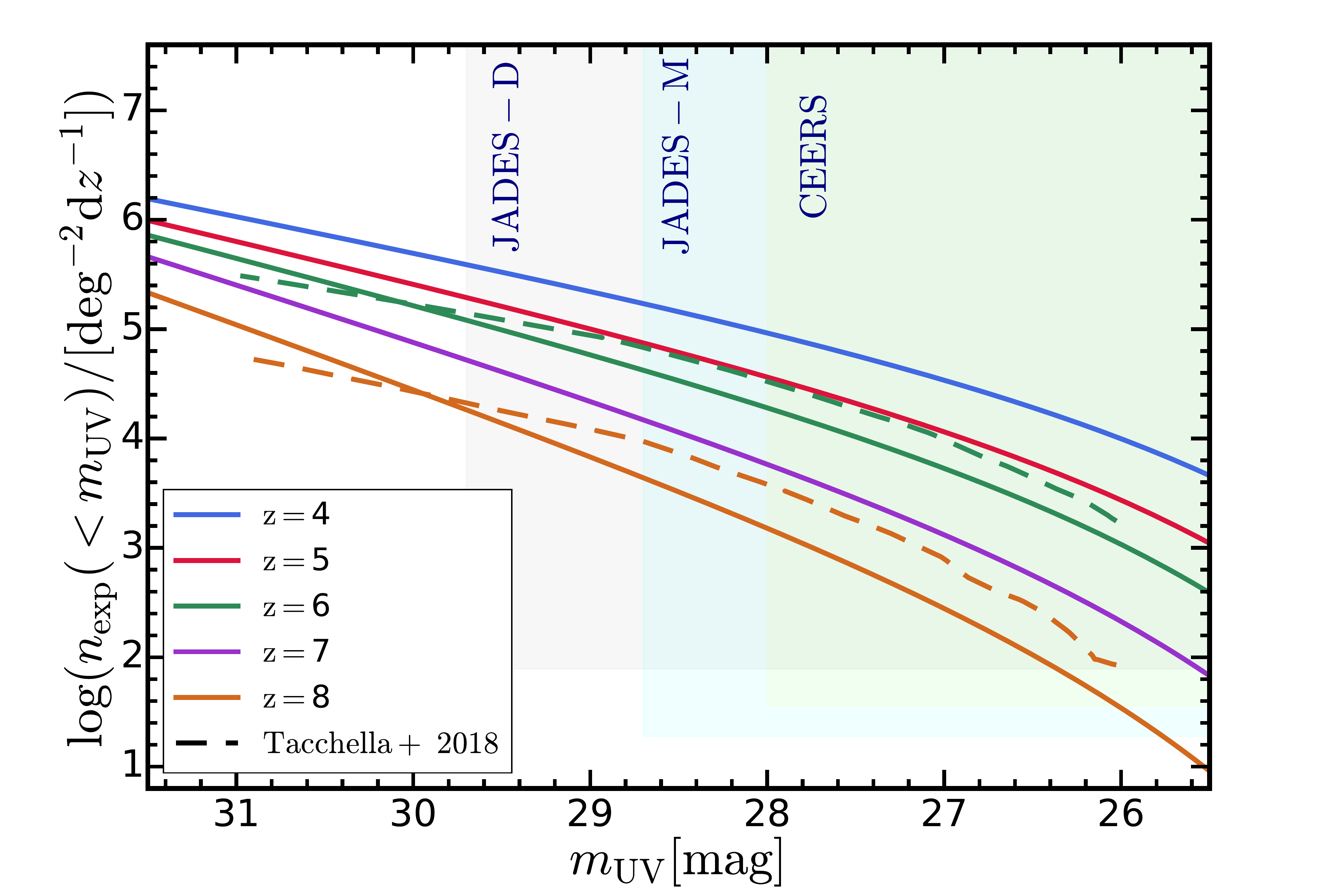}
\caption{{\bf Predictions for future {\it \textbf{JWST}} NIRCam high redshift dropout
surveys.} We present the predicted number of galaxies per square degree per unit redshift at $z=4-8$ for galaxies brighter than a given apparent magnitude limit $m_{\rm UV}$, which is the apparent magnitude evaluated at the wavelength where the rest-frame UV light is redshifted to. The shaded regions indicate the coverage of planned {\it JWST} surveys. The dashed lines show the prediction from \citet{tacchella2018} for comparison.}
\label{fig:apparent_uv}
\end{figure}

We can also use the {\it JWST} galaxy luminosities to make concrete predictions for
the number counts of {\it JWST} NIRCam high redshift dropout
surveys. Following earlier theoretical work by \cite{tacchella2018}, we focus here on two extragalactic surveys
that are currently planned for {\it JWST}. The first survey is
a large ($∼720$ hours) observational program, the {\it JWST}
Advanced Deep Extragalactic Survey (JADES). The second is the Cosmic
Evolution Early Release Science Survey (CEERS). We use our apparent galaxy UV magnitudes, $m_{\rm UV}$, to make predictions for galaxy number counts for these future {\it JWST} surveys. 
Using the estimates in~\cite{tacchella2018}, the deep component of JADES (JADES-D) will have an averaged survey depth of $\sim29.7\mmag$ for all bands assuming a $10\sigma$ point source limit with a sky coverage of $46\,{\rm arcmin}^{2}$. The medium component JADES-M will have an averaged depth of $28.7\mmag$  with a sky coverage of $190\,{\rm arcmin}^{2}$. The CEERS program will have an averaged depth of $28.0\mmag$ with a sky coverage of $100\,{\rm arcmin}^{2}$. 
In Figure~\ref{fig:apparent_uv}, we compare our predicted galaxy number counts with the coverage of these {\it JWST} surveys. The horizontal lower boundaries of the survey coverage regions are calculated assuming that there is only one object per unit redshift in the entire sky coverage of the surveys. We find that JADES-D can detect $\sim 3$ galaxies per unit redshift at the bright end up to $m_{\rm UV}\sim 27\mmag$ at $z=8$, while having a detection limit $m_{\rm UV}\sim29.7\mmag$ at the faint end. The total number of detectable galaxies can reach $235$ per unit redshift at $z=8$. JADES-M has a larger sky coverage and can detect a $\sim4$ times larger number of galaxies at the bright end. The total number of detectable objects is $238$ comparable to that of JADES-D at $z=8$. 
Our results are roughly consistent with the predictions of~\cite{tacchella2018} but we find a slightly different slope of the expected number of detected galaxies as a function of apparent magnitude.
We conclude that the IllustrisTNG simulation suite can provide important and insightful predictions for these kind of upcoming {\it JWST} surveys.

\begin{figure}
\includegraphics[width=0.51\textwidth]{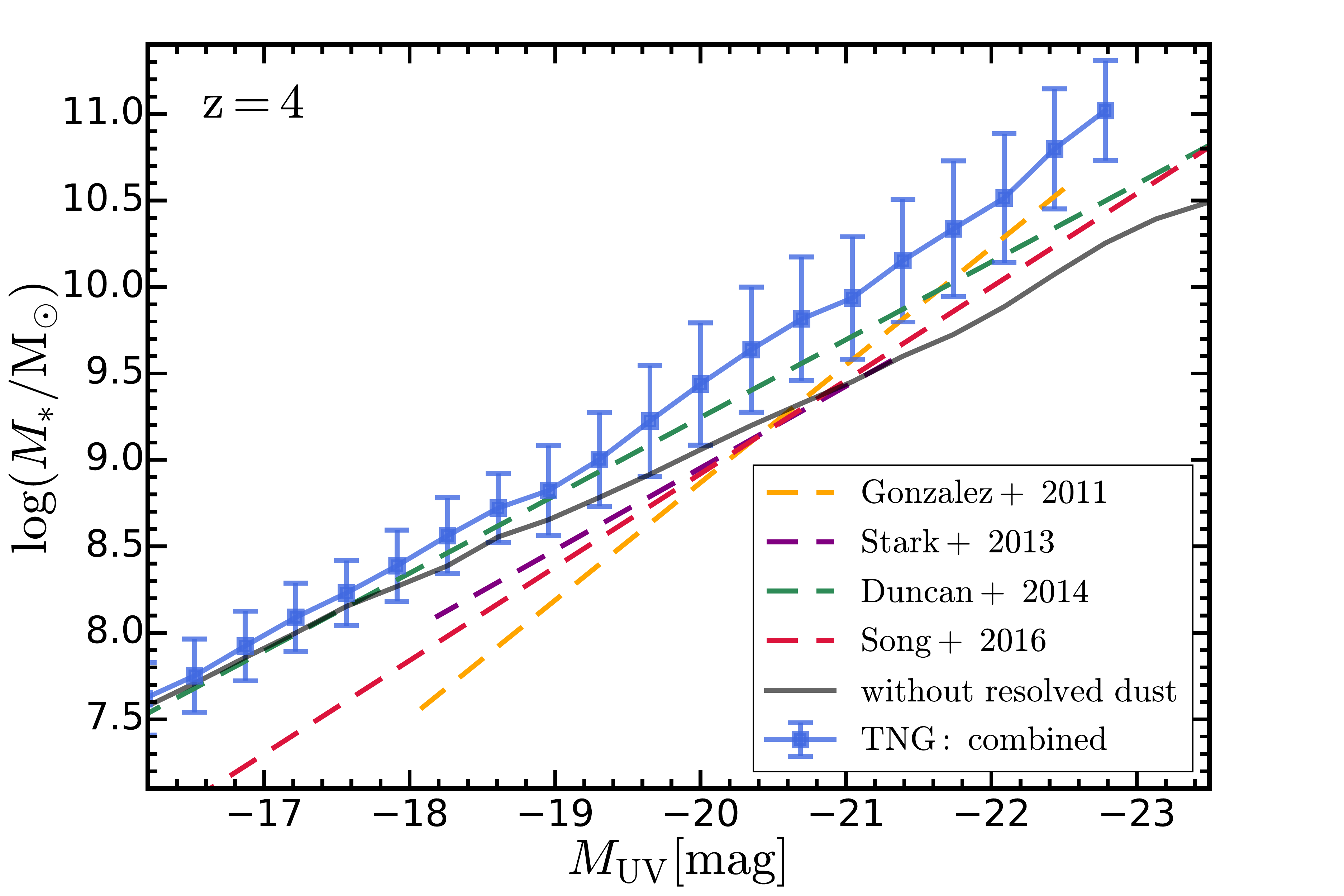}
\includegraphics[width=0.51\textwidth]{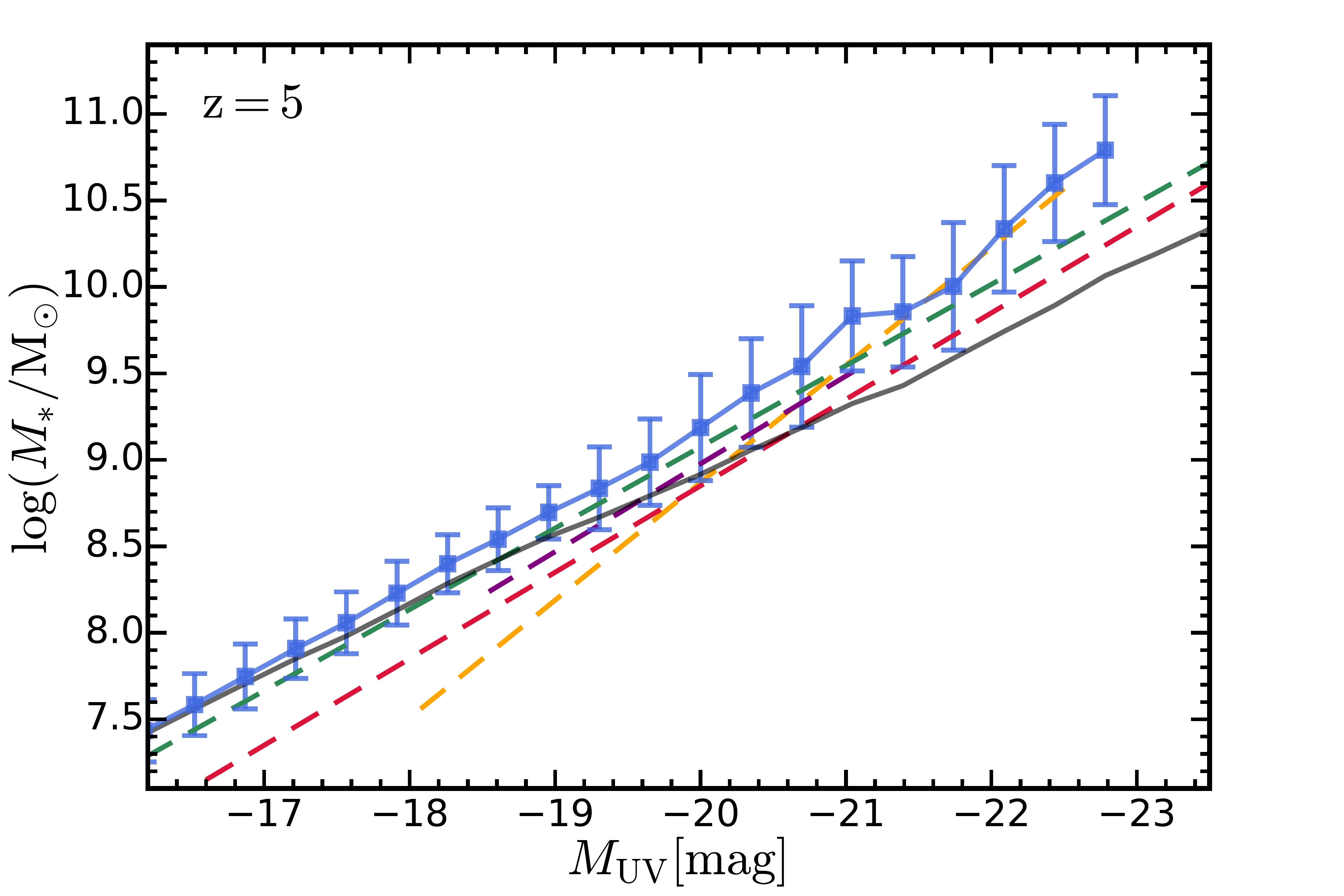}
\includegraphics[width=0.51\textwidth]{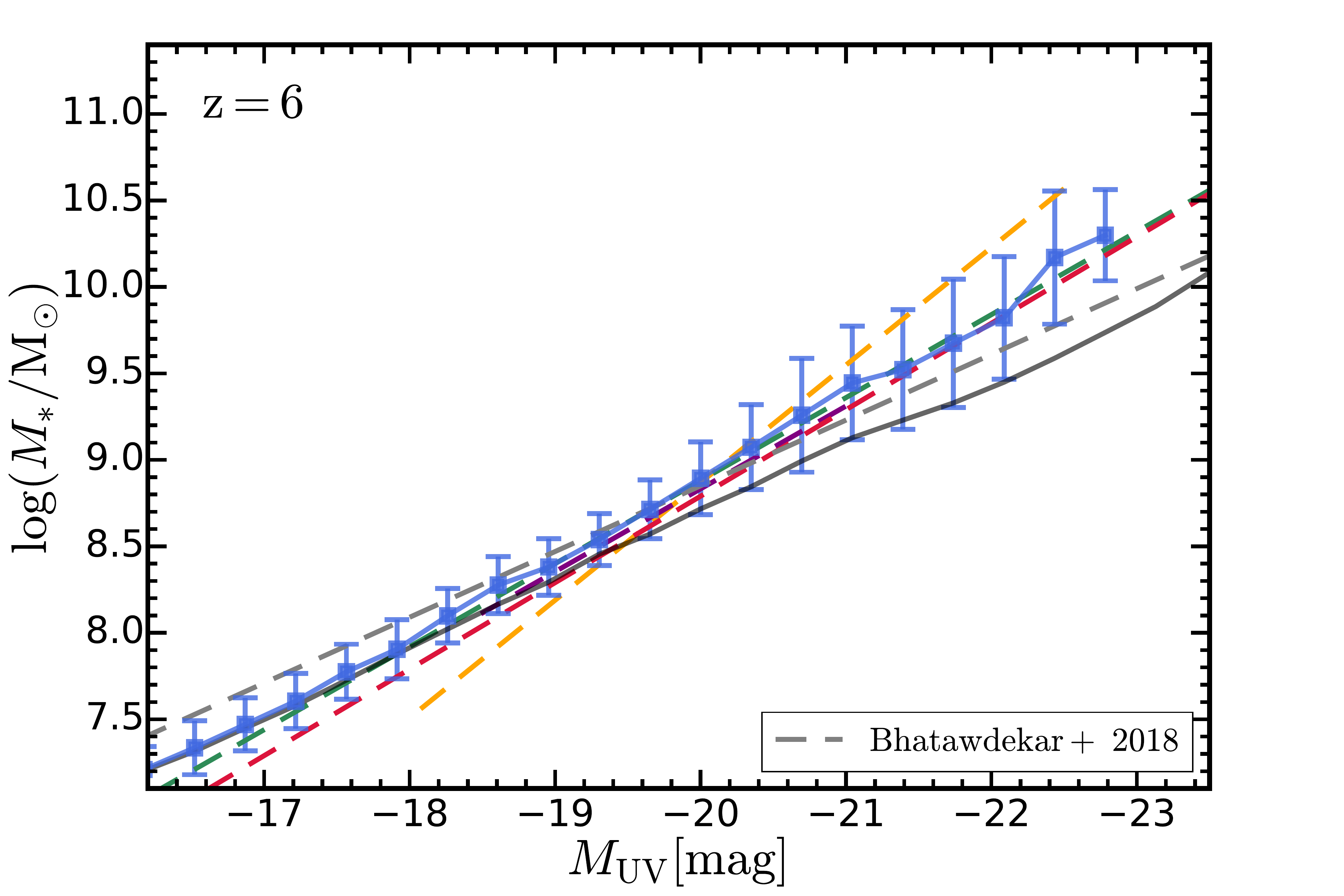}
\caption{{\bf UV luminosity versus stellar mass relations at different redshifts.} The galaxy stellar mass is defined here within a fixed physical aperture of $30\pkpc$. The blue line shows the median $\log{(M_{\ast}/{\rm M}_{\odot})}$ within each dust-attenuated UV magnitude bin. The error bars represent the $\log{(M_{\ast}/{\rm M}_{\odot})}$ standard deviation  of the samples in each bin. The black line indicates the relation when no resolved dust attenuation is taken into account. Observational constraints are taken from \citet{Gonzalez2011,Stark2013,Duncan2014,Song2016,Bhatawdekar2018}. For \citet{Gonzalez2011,Stark2013,Song2016}, the initial stellar mass function is converted from a Salpeter to a Chabrier initial stellar mass function by decreasing $\log{(M_{\ast}/{\rm M}_{\odot})}$ by $0.24$.}
\label{fig:Muv_Mstar}
\end{figure}

\subsection{UV luminosity versus stellar mass relation}
So far we have only discussed galaxy luminosity functions, but the galaxy stellar mass function is actually the more fundamental property to quantify the galaxy population, and also a direct prediction of any galaxy formation model. However, deriving galaxy stellar mass functions from observed luminosity functions is a challenging task since the conversion from stellar light to stellar mass depends on multiple assumptions for photometry, redshift determination, stellar population synthesis modelling,
star formation histories, metallicities, dust
characteristics, and initial stellar mass functions. This introduces significant systematic uncertainties in the whole conversion process~\citep[e.g.,][]{Conroy2010}.
Especially, measuring the stellar masses of galaxies at
high redshifts is challenging, because the rest-frame wavelengths probed by optical and near-infrared surveys cover only to the UV continuum requiring mid-IR observations to extend past a rest-frame wavelength of $4000$\AA. In addition, selection effects, contamination and incompleteness of the galaxy sample make it much more difficult to constrain the galaxy stellar mass function at higher redshifts. 
An alternative approach is to convert well-constrained galaxy UV luminosity functions to a stellar mass function assuming a certain mass-to-light ratio relation. The advantage of UV luminosity functions is that they can typically be corrected for selection effects and provide good statistics at both the bright and faint end. 
Many observations at high redshifts revealed that the galaxy UV luminosity correlates well with the stellar mass of the galaxy although the relation contains also quite significant scatter~\citep[e.g.,][]{Stark2009,Stark2013,Duncan2014,Song2016}. Once the mass-to-light ratio relation has been derived, the relation can be used to map observed UV luminosity functions to galaxy stellar mass functions without going through the demanding and uncertain SED modelling task. Specifically, one can construct galaxy stellar mass functions conveniently by convolving the $M_{\ast}-M_{\rm UV}$ distribution with the observed rest-frame UV luminosity function. For example, \citet{Gonzalez2011} derived the stellar masses for $401$ star forming galaxies by fitting the observed optical to UV SEDs. With these samples, they constructed a stellar mass versus UV luminosity relation, which they then convolved with well-constrained UV luminosity functions to derive galaxy stellar mass functions for redshifts $z=4-7$.

\begin{figure}
\includegraphics[width=0.51\textwidth]{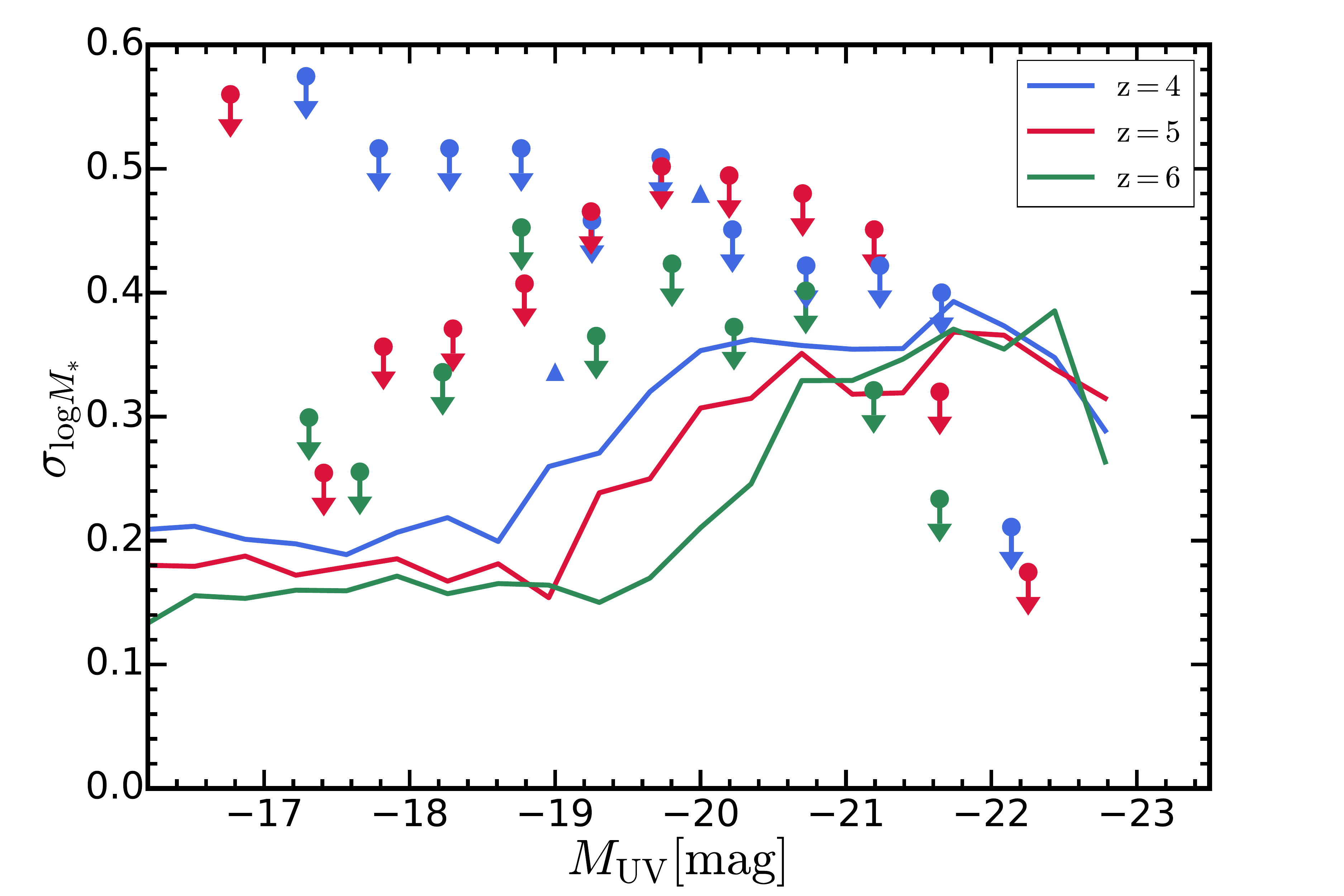}
\caption{{\bf Scatter in the UV luminosity versus stellar mass relation at $\mathbf{z=4,5,6}$.} The observational upper limits are taken from \citet{Song2016} where the scatter in each UV magnitude bin is measured and includes both intrinsic scatter and photometric and model uncertainties. The solid triangles are taken from \citet{Gonzalez2011} where the photometric uncertainties at $M_{\rm UV}=-20\mmag$ and at $M_{\rm UV}=-19\mmag$ are quantified. The intrinsic scatter is then calculated as $\sigma_{\rm int}=(\sigma_{\rm tot}^{2}-\sigma_{\rm pho}^{2})^{1/2}$. The scatter is larger towards lower redshifts. }
\label{fig:Muv_Mstar_scatter}
\end{figure}

Based on our simulation predictions, we can also construct
a ${M_{\ast}-M_{\rm UV}}$ relation by combining the IllustrisTNG simulations, TNG50, TNG100 and TNG300. We can then compare this predicted relation to the observed relation as a further test of our galaxy formation and dust model. In Figure~\ref{fig:Muv_Mstar}, we present the predicted relation between galaxy stellar masses and their UV magnitudes at $z=4,5,6$. To derive this relation, we employ the same binning in UV magnitudes as for the rest-frame UV luminosity function analysis presented above. For each bin, we then calculate the median $\log{(M_{\ast}/{\rm M}_{\odot})}$ and the standard deviation of $\sigma_{\log{M_{\ast}}}$ for the TNG50, TNG100 and TNG300 simulations respectively. Here, the stellar masses have been resolution-corrected following the same procedure that has also been used for the galaxy luminosities discussed above. Then similar to the combination step for the luminosity functions, we combine the median $\log{(M_{\ast}/{\rm M}_{\odot})}$ or $\sigma_{\log{M_{\ast}}}$ using Equation~\ref{eq:combine} in the luminosity range covered by two or three simulations, replacing $\phi$ with the median of $\log{(M_{\ast}/{\rm M}_{\odot})}$ or $\sigma_{\log{M_{\ast}}}$. At the faint end, where TNG100 (TNG300) does not provide sufficient resolution and therefore deviates from TNG50 (TNG100) predictions for the relation, we only use TNG50 (TNG50 and TNG100) to construct the combined ${M_{\ast}-M_{\rm UV}}$ relation. In the combination step we do not consider bins with less than $10$ galaxies for TNG300. The combined median $\log{(M_{\ast}/{\rm M}_{\odot})}$ versus $M_{\rm UV}$ relation is shown as the solid line in Figure~\ref{fig:Muv_Mstar}. The error bars represent the standard deviation in $\log{(M_{\ast}/{\rm M}_{\odot})}$ of the samples in each bin. We note that stellar mass is here defined as the total mass of stellar particles within a radial aperture of $30\pkpc$ to be consistent with the UV luminosity calculation discussed above. Figure~\ref{fig:Muv_Mstar} reveals a strong correlation between stellar mass and UV luminosity. At the faint end, our results are consistent with \citet{Duncan2014}, but not with \citet{Gonzalez2011}, \citet{Stark2013} and \citet{Song2016} at $z=4,5$. However, the number of galaxies at the faint end $M_{\rm UV}>-18$ in these observations are very limited. The observational relations presented here have also some significant scatter, which we do not show for clarity of the figure. 
The discrepancies at the bright end are likely related to the fact that observations are based on UV selected galaxies and have a potential bias missing dusty star-forming galaxies and quiescent galaxies. This can lead to a shift in the relation towards UV brighter galaxies at a given stellar mass. At the bright end, the relation is no longer linear and deviates towards lower UV luminosity due to the influence of dust. This can be appreciated in the figure through a comparison with the dust-free lines. The relation is tighter towards higher redshifts due to younger stellar populations with dominant UV emission. Towards lower redshifts, the scatter increases since the stellar populations become older, which weakens the correlation between stellar mass and UV luminosity causing a visible larger scatter.

\begin{figure}
\includegraphics[width=0.51\textwidth]{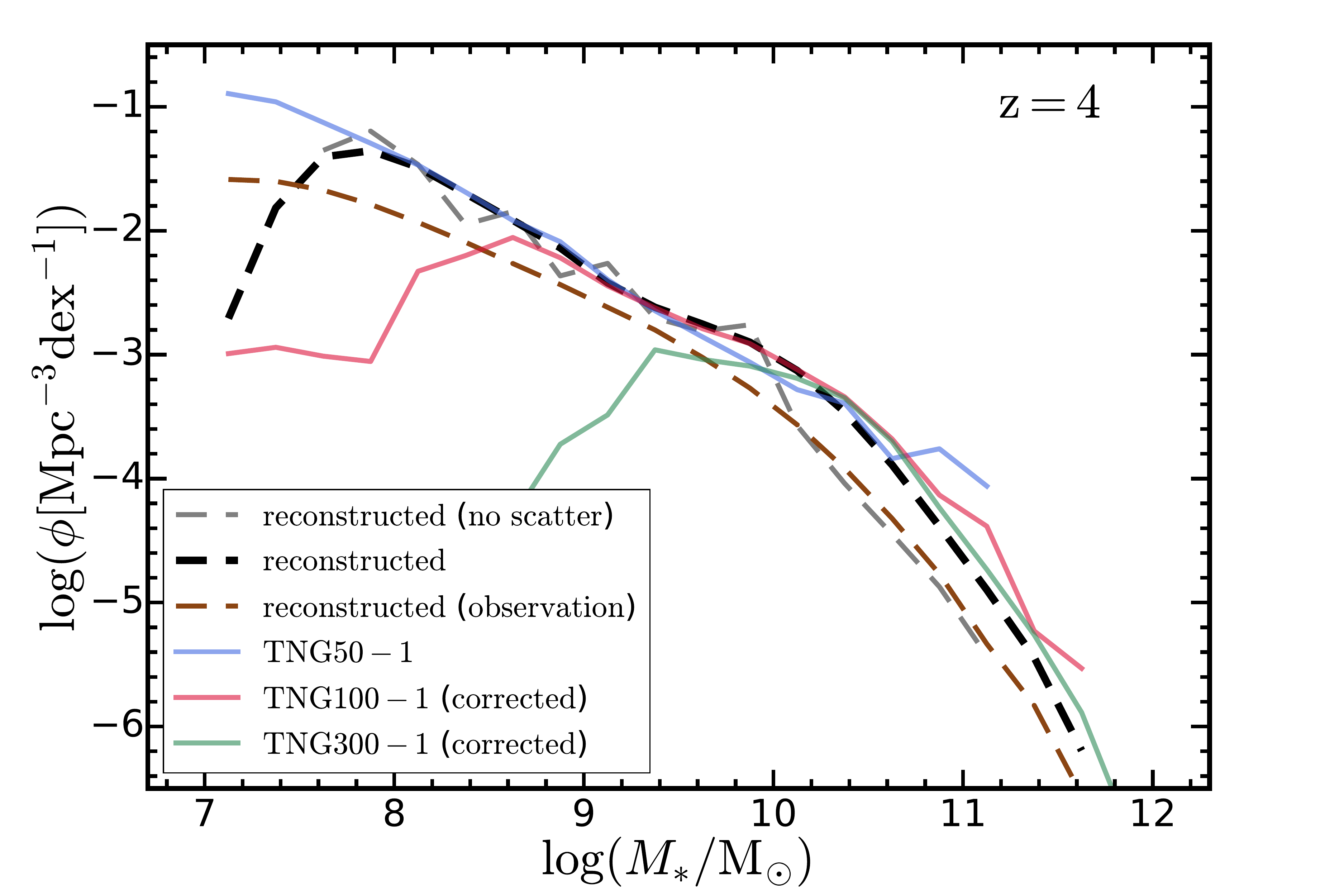}
\caption{{\bf Reconstructed galaxy stellar mass function through a convolution of the rest-frame UV luminosity function with the UV luminosity versus stellar mass relation at $\mathbf{z=4}$.} Solid curves show the stellar mass function directly from resolution corrected simulation outputs. The black dashed line represents the reconstructed stellar mass function using the $M_{\ast}-M_{\rm UV}$ relation including its scatter. We also show the reconstructed stellar mass function using same relation without this scatter as a gray dashed line. The brown dashed line shows the reconstructed stellar mass function using an observationally determined $M_{\ast}-M_{\rm UV}$ relation and scatter taken from \citet{Song2016}.}
\label{fig:SMF_reconstructed}
\end{figure}

In Figure~\ref{fig:Muv_Mstar_scatter} we quantify this scatter in the mass-to-light ratio relation in more detail by presenting the standard deviation in the stellar mass to UV magnitude relation at $z=4,5,6$. This scatter increases from the faint end towards the bright end. Observations typically find that the scatter is around $0.3-0.5\,{\rm dex}$ at $z=4-7$, which is generally larger than our values. For example, \cite{Song2016} found a stellar mass standard deviation of about $0.4\,{\rm dex}$. Specifically, they found standard deviations of $0.52, 0.42, 0.36, 0.40$, and $0.30\,{\rm dex}$ at $z = 4, 5, 6, 7$, and $8$, respectively. Those were measured as the mean standard deviations in
stellar mass bins with more than five galaxies. They found no noticeable correlation of the scatter with redshift or UV luminosity.  The scatter at the bright end
(measured at $-21.5\mmag < M_{\rm UV} < -20.5\mmag$), where the effect of
observational uncertainty should be minimal, is $0.43, 0.47,
0.36, 0.52$, and $0.40\,{\rm dex}$ at $z = 4, 5, 6, 7$, and $8$, respectively, similar to the quoted value above and the scatter at the faint end (at $-19.0\mmag < M_{\rm UV} < -18.0\mmag$) of $0.51, 0.39, 0.39$, and $0.41\,{\rm dex}$ at $z = 4, 5, 6$, and $7$, respectively. We note that these scatter measurements represent upper limits since there are significant photometric and model uncertainties on top of the intrinsic scatter. \citet{Gonzalez2011} found that the scatter in this relation at $z=4$ is about $0.5\,{\rm dex}$. They also estimated that the photometric uncertainty is about $0.14\,{\rm dex}$ at $M_{\rm UV}=-20\mmag$ and $0.37\,{\rm dex}$ at $M_{\rm UV}=-19\mmag$. Considering these photometric uncertainties, the intrinsic scatter is close to what we find. \citet{Stark2013} estimated the intrinsic scatter to be in the range $0.2-0.5\,{\rm dex}$ at $z=4$ due to photometric and model uncertainties. 

\begin{figure}
    \includegraphics[width=0.51\textwidth]{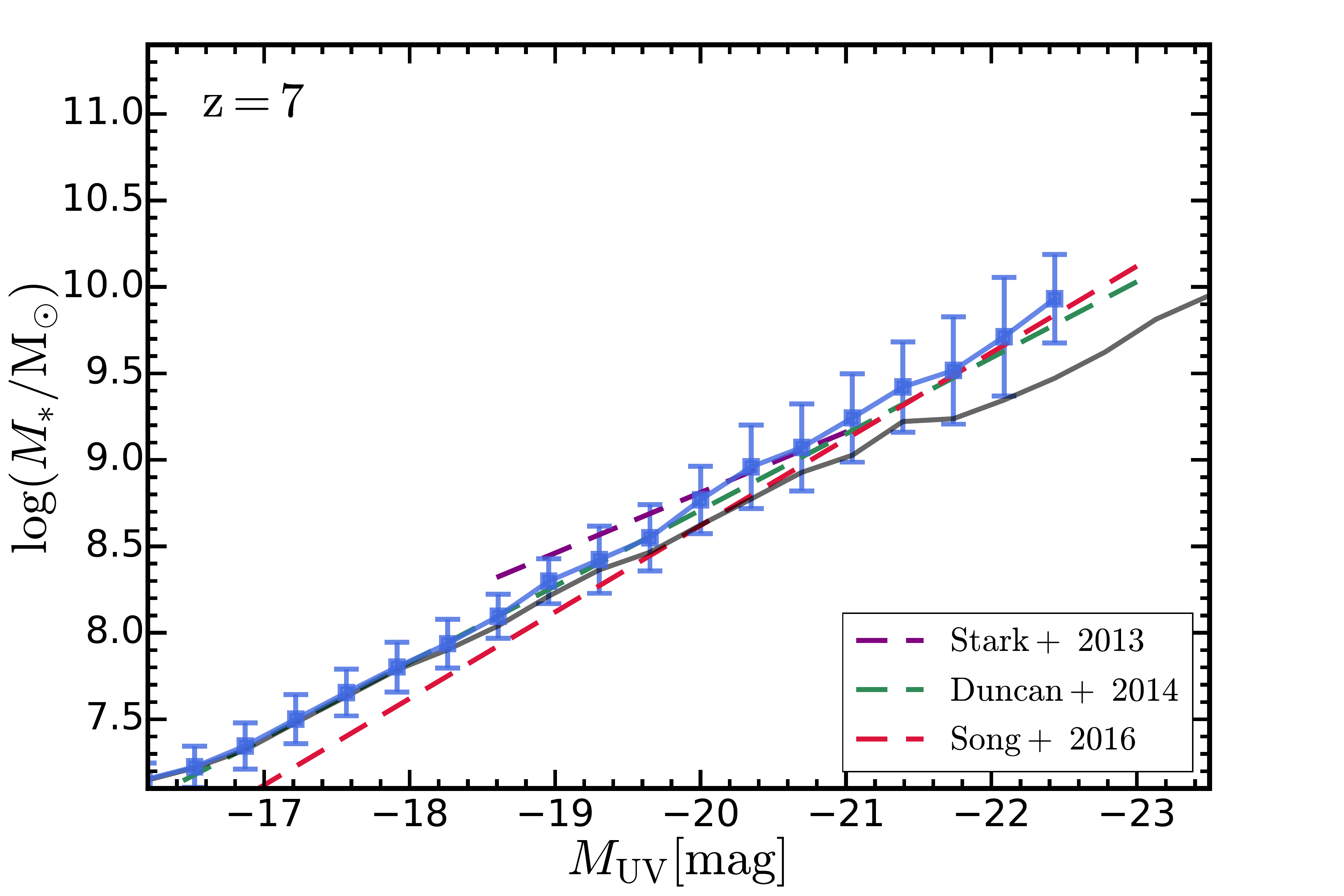}
    \includegraphics[width=0.51\textwidth]{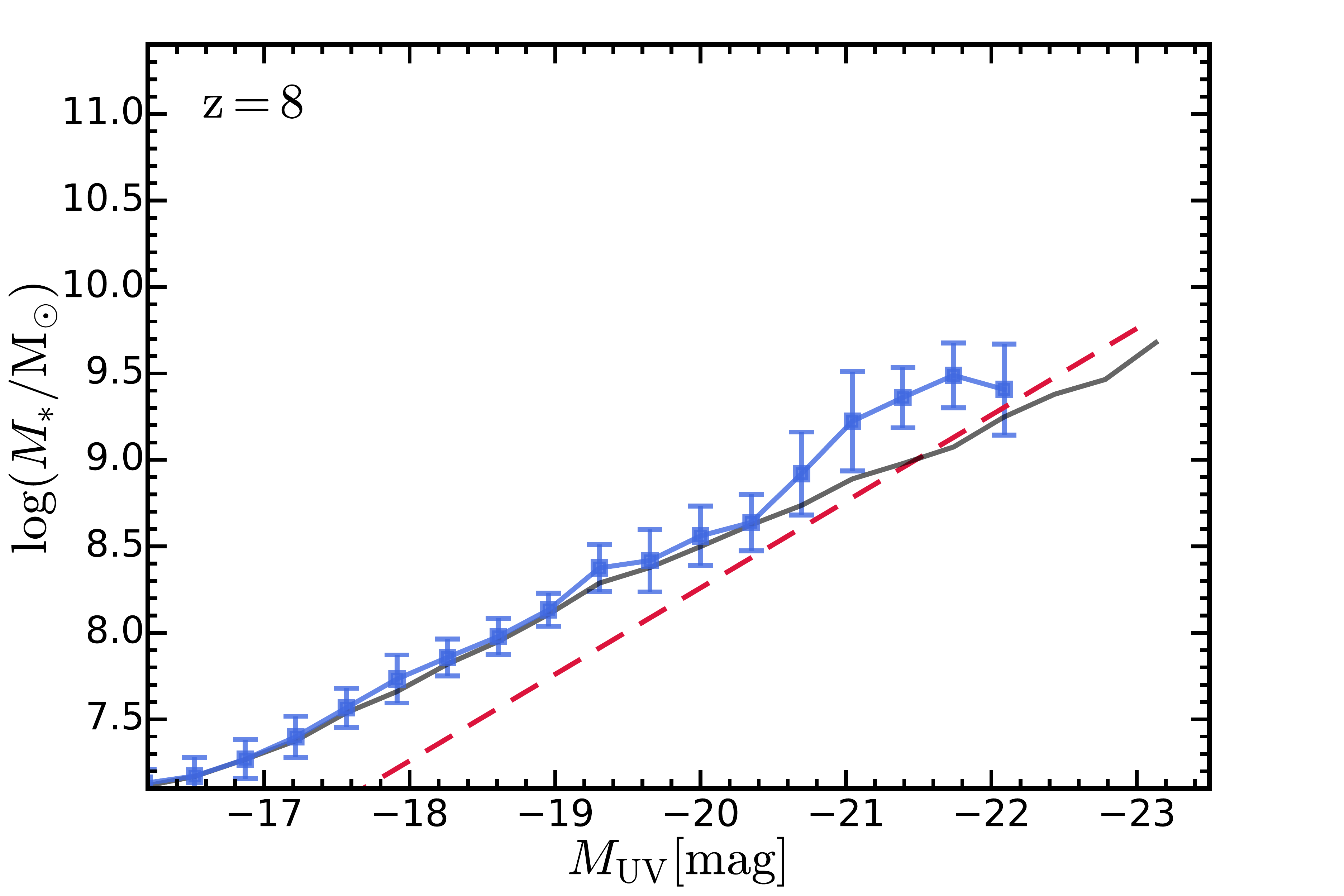}
    \caption{{\bf UV luminosity versus stellar mass relations at high redshifts.} The galaxy stellar mass is defined here within a fixed physical aperture of $30\pkpc$. The blue line shows the median $\log{(M_{\ast}/{\rm M}_{\odot})}$ within each dust-attenuated UV magnitude bin. The error bars represent the $\log{(M_{\ast}/{\rm M}_{\odot})}$ standard deviation  of the samples in each bin. The black line indicates the relation when no dust attenuation is taken into account. Observational constraints are taken from \citet{Stark2013,Duncan2014,Song2016}.}
    \label{fig:Muv_Mstar_highz}
\end{figure}

Based on the mass-to-light ratio relation, we can also reconstruct the galaxy stellar mass function by convolving the derived UV luminosity function with the derived stellar mass versus UV magnitude relation. This follows the common approach to convert observed rest-frame UV luminosity functions into stellar mass functions. To this end, we take the binned $\log{(M_{\ast}/{\rm M}_{\odot})}$ versus $M_{\rm UV}$ relation with its scatter and convolve it with the combined luminosity functions derived above. Specifically, in each UV magnitude bin, we generate stellar mass samples with a normal distribution in $\log{(M_{\ast}/{\rm M}_{\odot})}$ centred around the median value of the $\log{(M_{\ast}/{\rm M}_{\odot})}$ versus $M_{\rm UV}$ relation with its scatter. The total number of the samples is proportional to the galaxy number density at the given UV magnitude. By combining the samples derived from each UV magnitude bin, we generate a reconstructed galaxy stellar mass function, with $21$ bin edges linearly spaced from $7$ to $12$ for $\log{(M_{\ast}/{\rm M}_{\odot})}$. In Figure~\ref{fig:SMF_reconstructed}, we present the reconstructed galaxy stellar mass function and compare it with the directly derived resolution-corrected stellar function. We find good agreement between those two galaxy stellar mass functions over the entire dynamic range. We also plot the reconstructed function without including the scatter in the $\log{(M_{\ast}/{\rm M}_{\odot})}$ versus $M_{\rm UV}$ relation. In this case we find that neglecting the scatter significantly underestimates the galaxy number densities at the massive end. Using the observationally determined $\log{(M_{\ast}/{\rm M}_{\odot})}$ versus $M_{\rm UV}$ relation along with its scatter~\citep{Song2016} to reconstruct the stellar mass function, we find that the stellar mass is underestimated in the entire dynamic range consistent with the discrepancy we find in $\log{(M_{\ast}/{\rm M}_{\odot})}$ versus $M_{\rm UV}$ relation discussed above.

Finally, we can also use the simulation results to make predictions for the mass-to-light ratio relation towards higher redshifts, where observational data becomes more scarce and uncertain. Figure~\ref{fig:Muv_Mstar_highz} presents the $z=7$ and $z=8$ mass-to-light ratio relations and compares those with some observational data. At $z=7$ we find good agreement between the simulation predictions and observational data. At $z=8$ observations have not yet fully explored the mass-to-light ratio relation, and only limited data is available. This data is, as can be seen in Figure~\ref{fig:Muv_Mstar_highz}, not consistent with our predictions at this high redshift.

\subsection{H$\alpha$ luminosity versus star formation rate relation}

The UV luminosity of a galaxy is an indicator for the stellar mass of the galaxy through the mass-to-light ratio relation as demonstrated above. Furthermore, the UV luminosity can also be employed as a star formation rate indicator. Spectral indices like the H$\alpha$ emission line luminosity, on the other hand, provide further indicators for star formation rates of galaxies~\citep[][]{Kennicutt1998}. 
Young, massive stars produce large amounts of ionising photons that ionise the surrounding gas. Hydrogen recombination cascades then produce line emission, including the well-known Balmer series lines of ${\rm H}\alpha$ ($656.3\nm$) and ${\rm H}\beta$ ($486.1\nm$). These lines are strong and  represent the most traditional star formation rate indicators.
Above we have explored the correlation between UV luminosities and stellar masses. Here we now explore how H$\alpha$ luminosities correlate with star formation rates of galaxies~\citep[e.g.,][]{Murphy2011}.  Probing these UV and H$\alpha$ luminosity relations and comparing them to the observed relations provides important tests of our galaxy formation model and the employed luminosity and dust modelling.  Specifically, they probe relations of the continuum, UV, and emission lines,  H$\alpha$.

\begin{figure}
\includegraphics[width=0.51\textwidth]{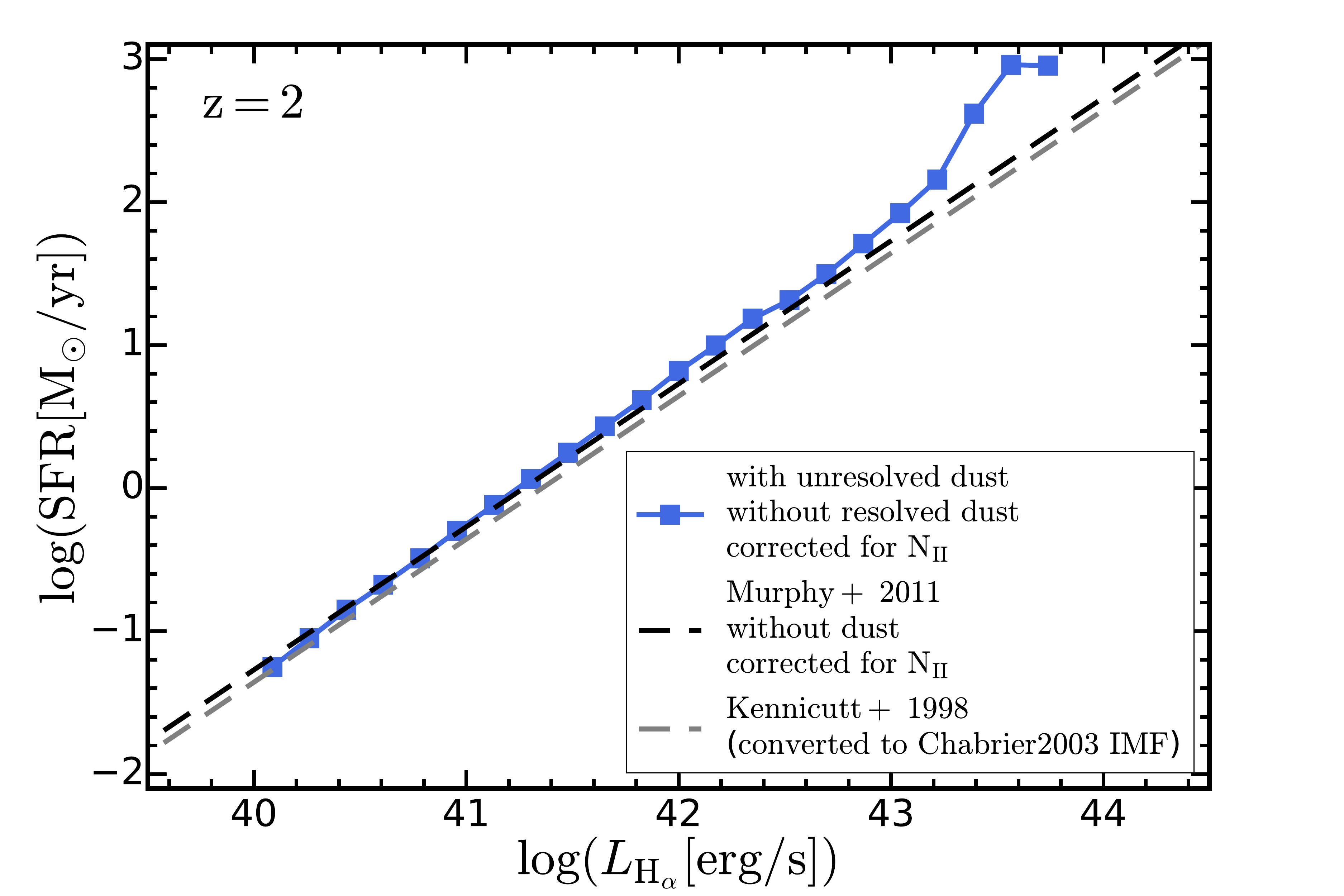}
\caption{{\bf H$\alpha$ luminosity versus star formation rate relation at $\mathbf{z=2}$.} We compare the relation to the observational results from \citet{Murphy2011}, which is based on the observation of local star-forming regions in nearby galaxies. The simulation results are shown at $z=2$. In addition we also show the \citet{Kennicutt1998} observational relation converted to a Chabrier initial stellar mass function.}
\label{fig:sfr_halpha}
\end{figure}

To derive the H$\alpha$ emission line luminosity of a given galaxy, we follow an approach commonly adopted in observations~\citep[e.g.][]{Lee2012,Sobral2013,Sobral2015,Matthee2017}.
We first construct broad band and narrow band tophat filters around the H$\alpha$ emission line. The broad band filter is centred at $656.3\nm$ with a band width of $\Delta\lambda_{\rm BB}=80\nm$. The narrow band filter is also centred at $656.3\nm$ but with a band width of $\Delta\lambda_{\rm NB}=20\nm$. Both of these filters cover the entire rest-frame H$\alpha$ emission line. We then convolve the galaxy SED with these two filters and get the filter averaged fluxes $f_{\rm BB}$ (broad band) and $f_{\rm NB}$ (narrow band). We can then derive the line luminosity and equivalent width (EW) as:
\begin{equation}
L_{{\rm H}\alpha}=\Delta\lambda_{\rm NB}\,\,\dfrac{f_{\rm NB}-f_{\rm BB}}{1-\dfrac{\Delta\lambda_{\rm NB}}{\Delta\lambda_{\rm BB}}}, \quad
{\rm EW}=\Delta\lambda_{\rm NB}\,\,\dfrac{f_{\rm NB}-f_{\rm BB}}{f_{\rm BB}-f_{\rm NB}\dfrac{\Delta\lambda_{\rm NB}}{\Delta\lambda_{\rm BB}}}.
\end{equation}
We note that the two ${\rm N}_{\,\rm \Rmnum{2}}$ emission lines, located at $654.8\nm$ and $658.3\nm$, contaminate the H$\alpha$ luminosity and equivalent width. To correct for this, we assume that the ratio between the ${\rm N}_{\,\rm \Rmnum{2}}$ luminosity and the H$\alpha$ luminosity depends linearly on the total measured equivalent width with contributions from the lines of ${\rm N}_{\,\rm \Rmnum{2}}$ and H$\alpha$~\citep{Sobral2015}:
\begin{equation}
f({\rm N}_{\,\rm \Rmnum{2}}/{\rm H}\alpha)=-0.296\times \log{({\rm EW}_{{\rm N}_{\,\rm \Rmnum{2}}+{\rm H}\alpha})}+0.8,
\end{equation}
where the equivalent width is measured in units of \AA. Negative $f$ are manually set to $0$. We then calculate the ${\rm  H}\alpha$ line luminosity as~\citet{Sobral2015}:
\begin{equation}
L_{{\rm H}\alpha}=L_{{\rm N}_{\,\rm \Rmnum{2}}+{\rm H}\alpha}\times \dfrac{1}{1+f({\rm N}_{\,\rm \Rmnum{2}}/{\rm H}\alpha)}.
\end{equation}
To probe the H$\alpha$ luminosity versus star formation rate relation, we compute the instantaneous star formation rate of each galaxy within a physical aperture of $30\pkpc$ consistent with the aperture applied above for galaxy luminosities and stellar masses. We present the result of this calculation in Figure~\ref{fig:sfr_halpha} for $z=2$ along with the observationally inferred linear relation taken from ~\citet{Murphy2011}: 
\begin{equation}
\left (\frac{\text{SFR}_{\rm H\alpha}}{{\rm M}_\odot\,{\rm yr}^{-1}} \right ) = 5.37 \times 10^{-42}\, \left (\frac{L_{\rm H\alpha}}{{\rm erg}\, {\rm s}^{-1}} \right )
\end{equation}
This relation is a recalibration of the relation from \cite{Kennicutt1998}
and  is dust-corrected. We therefore do not consider the impact of resolved dust, when deriving H$\alpha$ luminosities. We note that the emission lines, and so the H$\alpha$ line, are modelled through {\sc Mappings-\Rmnum{3}} assuming a Kroupa initial stellar mass function. This is consistent with the \cite{Murphy2011} relation that assumes the same initial stellar mass function. Figure~\ref{fig:sfr_halpha} demonstrates that our results reproduce the observed relation. We also include in this figure the \cite{Kennicutt1998} relation converted to a Chabrier initial stellar mass function, which has a marginally different normalisation compared to the \cite{Murphy2011} relation.

\begin{figure}
\includegraphics[width=0.51\textwidth]{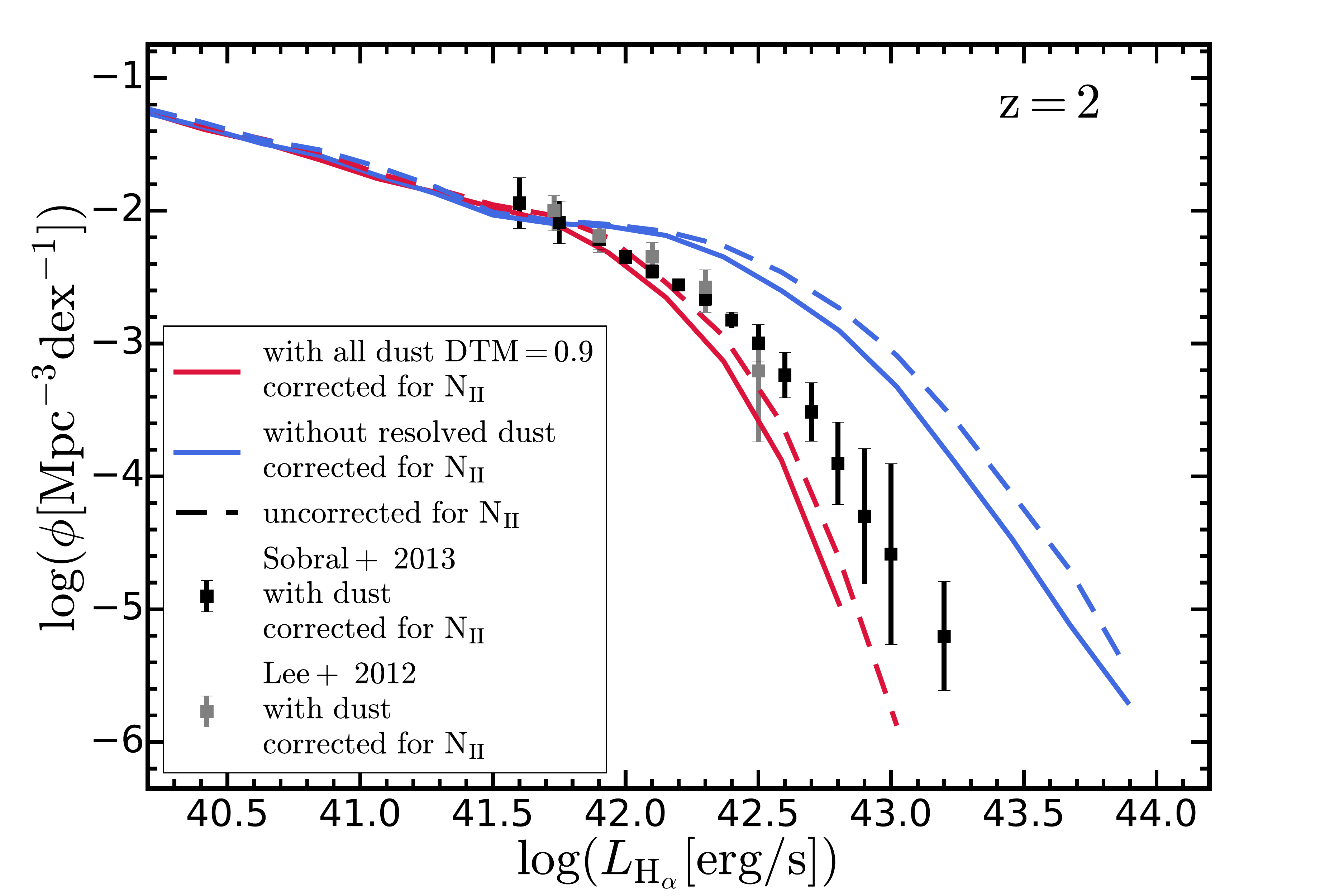}
\caption{{\bf The H$\alpha$ luminosity function at $\mathbf{z=2}$.} We present the luminosity functions of the H$\alpha$ emission evaluated at $z=2$ and compare to observational data from \citet{Lee2012,Sobral2013}. The solid blue line shows the luminosity function without resolved dust attenuation. The solid red line shows the luminosity function with resolved dust attenuation assuming a dust-to-metal ratios of $0.9$. Not including the ${\rm N}_{\,\rm \Rmnum{2}}$ line correction results in the dashed lines. }
\label{fig:Halpha_lf}
\end{figure}

\begin{figure}
\includegraphics[width=0.51\textwidth]{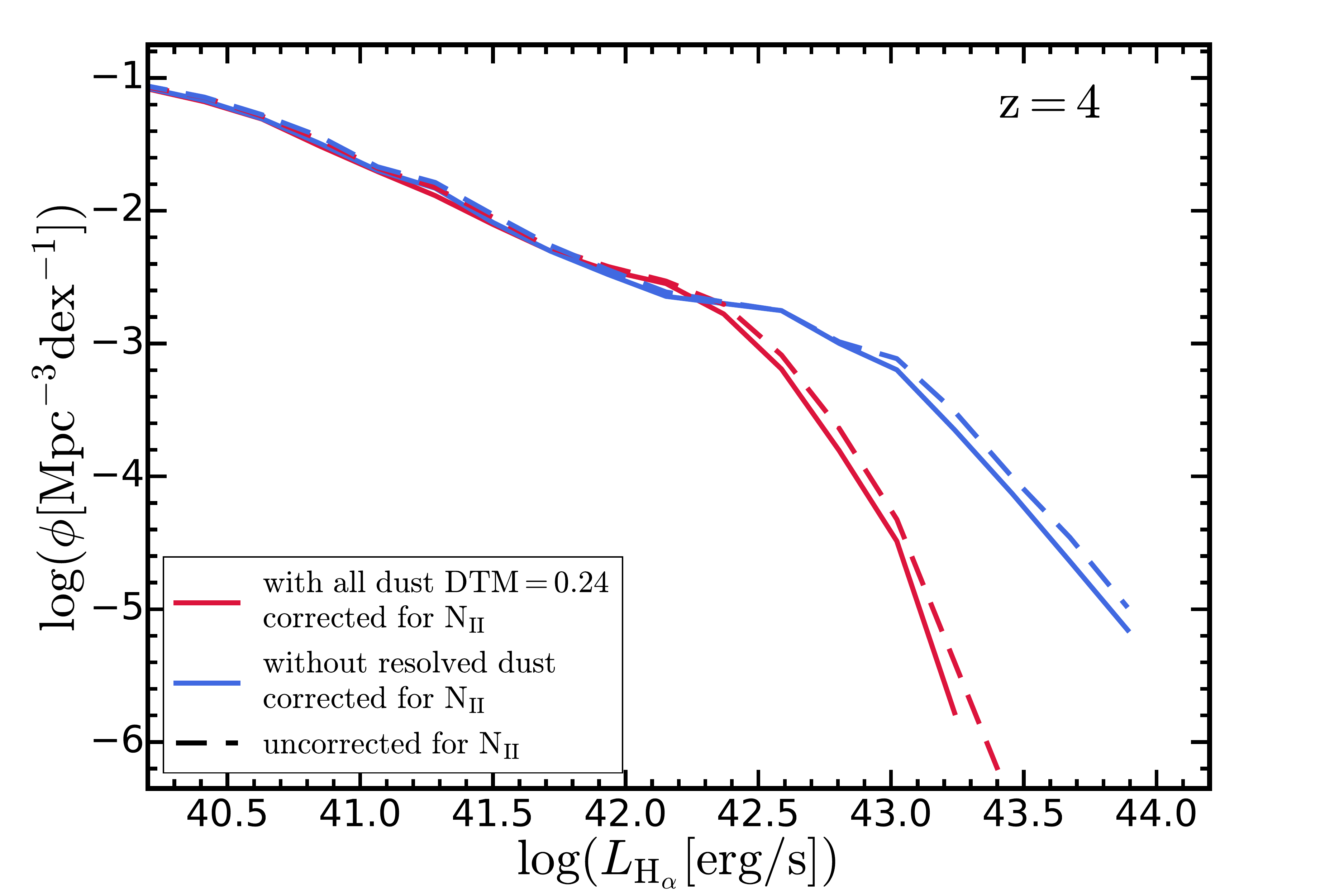}
\caption{{\bf The H$\alpha$ luminosity function at $\mathbf{z=4}$.}  The solid blue line shows the luminosity function without resolved dust attenuation. The solid red lines shows the luminosity function with resolved dust attenuation assuming dust-to-metal ratios of $0.24$. Not including the ${\rm N}_{\,\rm \Rmnum{2}}$ line correction results in the dashed lines.}
\label{fig:Halpha_lf_z4}
\end{figure}

To explore the impact of dust on the H$\alpha$ emission, we next present the H$\alpha$ luminosity function at $z=2$ along with a comparison to observational results in Figure~\ref{fig:Halpha_lf}. 
 The solid blue line shows the luminosity function without resolved dust attenuation. The solid red line shows the luminosity function with resolved dust attenuation assuming dust-to-metal ratios of $0.9$, i.e. the best-fit value based on the UV calibration. Not including the ${\rm N}_{\,\rm \Rmnum{2}}$ line correction results in the dashed lines. Neglecting the resolved dust component clearly leads to an inconsistent H$\alpha$ luminosity function. Including the resolved dust on the other hand, moves the luminosity function towards significantly lower luminosities. We note that a dust-to-metal ratio of $0.9$ leads to slightly too much attenuation of the H$\alpha$ fluxes, which can be seen in Figure~\ref{fig:Halpha_lf} towards the bright end, where the simulation predictions start deviating from the observational data. Nevertheless, our results are still marginally consistent within the observational uncertainties. We also stress that a dust-to-metal ratio of $0.9$ represents the best-fit to the UV luminosity function at $z=2$ as discussed above, and lower dust-to-metal ratios would lead to an inconsistent UV luminosity function. We note that the unresolved attenuation through the {\sc Mappings-\Rmnum{3}} model is calibrated only for the local universe and uncertainties arise when applying this towards higher redshifts. This can possibly lead to a too high attenuation as is shown in Figure~\ref{fig:Halpha_lf}.
Measuring the H$\alpha$ luminosity function towards higher redshifts is observationally challenging. We therefore present in Figure~\ref{fig:Halpha_lf_z4} the predicted H$\alpha$ luminosity function at $z=4$. We show here the luminosity function again with and without resolved dust assuming the best-fit dust-to-metal ratio as derived from the dust calibration step. 

We conclude that our modelling of emission lines leads to consistent H$\alpha$ relations. This provides another important test of our luminosity and dust attenuation modelling.

\section{Summary and Conclusions}
\label{sec:Section5}

{\it JWST} is going to transform our understanding of the high redshift Universe. However, exploiting the unprecedented capabilities of this novel telescope depends on reliable theoretical predictions. Such predictions for the properties of high redshift galaxies are therefore of uttermost importance, and detailed cosmological simulations of the early Universe are required to provide those forecasts. Ideally, these cosmological simulations should be based on galaxy formation models that produce a realistic low redshift galaxy population and cover a wide dynamic range. The low redshift verification of galaxy formation models is necessary for the fidelity and consistency of the simulated high redshift Universe.

In this paper we have studied the high redshift galaxy population of the IllustrisTNG simulation suite consisting of TNG50, TNG100, and TNG300. Most importantly the IllustrisTNG simulations employ a galaxy formation model that produces a realistic low redshift Universe that is consistent with a wide range of observational data. The main focus of our work has been a detailed study of high redshift galaxy luminosity functions in various bands. To this end, we have combined the results of all three IllustrisTNG simulations, TNG50, TNG100, and TNG300, to maximise the dynamic range of these luminosity function predictions. We have devised a resolution correction and luminosity function combination procedure to predict combined luminosity functions over a large dynamic range. The resolution correction scheme  extrapolates the TNG100 and TNG300 results to TNG50 resolution resulting in three simulations with the same effective numerical mass resolution. The luminosity function predictions of these simulations are then combined to achieve a large dynamic range in galaxy luminosities (e.g., $M_{\rm UV}=-17$ to $-22$ at $z=8$). 

Dust attenuation severely affects luminosities of dust rich galaxies. 
We have therefore developed a variety of dust attenuation models for the IllustrisTNG galaxies to derive these luminosity functions in different bands and at different redshifts. Neglecting these corrections results in rest-frame UV luminosity functions that deviate significantly from observational data for redshifts $z \lesssim 5$ towards the bright end.  We have therefore presented in this paper a detailed discussion of dust modelling for high redshift galaxies. Specifically, we have introduced three distinct dust attenuation models of increasing complexity and applied them in post-processing to the simulation output: an empirical dust model (Model A), a resolved dust optical depth model (Model B), and  a resolved dust radiative transfer model (Model C). IllustrisTNG, similar to most other existing large volume galaxy formation simulations, does not follow the evolution of dust in galaxies. Each of the applied dust models therefore depends on some redshift-dependent free dust parameters that have to be determined through a calibration procedure:
\begin{itemize}[leftmargin=*]
\item Dust Model A (empirical model): The two free parameters of Model A are the intercept $\beta_{M_0}$ and the slope ${\rm d}\beta/{\rm d}M_{\rm UV}$ values of the employed ${M_{\rm UV}^{\rm dust} - \langle \beta \rangle}$ relation. 
\item Dust Model B (resolved optical depth model): Model B contains an overall optical depth normalisation parameter for resolved dust. 
\item Dust Model C (resolved dust radiative transfer model):  Similarly to Model B, the dust attenuation of Model C depends on one parameter, the dust-to-metal ratio that sets at each redshift the conversion factor from the metal distribution in the simulated galaxies to the actual dust distribution required for the Monte Carlo radiative transfer calculation. 
\end{itemize}
For Models B and C we consider both the impact of resolved and unresolved dust. The resolved dust component is modelled through resolved gas properties around each stellar particles. For the unresolved dust component we model the impact of birth clouds around young stars. For Model B this is handled through different optical depth normalisations and different extinction curves. The additional optical depth due to unresolved dust is applied to attenuate light from stellar particles younger than $10\,{\rm Myr}$. For Model C, we instead use different stellar spectral energy distributions depending on whether stellar particles are surrounded by birth clouds or not. Specifically, the spectral energy distributions are altered for stellar particles younger than $10\,{\rm Myr}$.

We have calibrated the free dust parameters of the three dust attenuation models based on a best-fit procedure against observed rest-frame UV luminosity functions. This dust model calibration procedure is based on the resolution-corrected TNG100 simulation. Specifically, we have calibrated the free dust parameters of all three dust models to match the observed rest-frame UV galaxy luminosity functions over the redshift range $z=2$ to $z=10$. The calibration process revealed for each dust model the redshift evolution of the free dust parameters constraining the dust models. We note that this calibration procedure results in redshift dependent dust parameters that are then applied to all galaxies at the corresponding redshift. For example, for Model C, we convert the metal abundances in each galaxy to dust abundances using a fixed dust-to-metal ratio at a given redshift. Based on this dust calibration we find:
\begin{itemize}[leftmargin=*]
\item For dust Model A the inferred redshift evolution of the best-fit values for the intercept $\beta_{M_0}$ agree reasonably well with the observationally derived redshift evolution. However, the redshift evolution of the slope ${\rm d}\beta/{\rm d}M_{\rm UV}$ does not agree with observational estimates. Here our calibration prefers steeper slopes than what is observationally found. 
\item For dust Model B the calibration yields the redshift evolution of the optical depth normalisation. The redshift dependence of this normalisation can be described by a rather steep power law $z^{-1.92}$. This redshift evolution differs from other typically employed redshift dependencies of similar dust attenuation models especially towards lower redshifts ($z\lesssim 4$), where we predict a much stronger redshift dependence. 
\item For dust Model C the best-fit redshift evolution of the dust-to-metal ratios follows the same power law as the optical depth normalisation of Model B. For Model C we have also found that the additional unresolved birth cloud dust attenuation results in too much dust attenuation towards higher redshifts, $z \geq 6$. We therefore do not consider unresolved dust for $z \geq 6$ and only include the attenuation due to resolved dust. We also note that the resolved dust attenuation is negligible at higher redshifts due to the strong redshift dependence of the dust-to-metal ratio. For example, at $z=8$  we predict a dust-to-metal ratio of about $\sim 5\%$. This results in high redshift luminosity functions that are essentially identical to the intrinsic dust-free luminosity functions.  
We note that our dust Model B also
predicts that the contribution of unresolved dust is reduced at higher redshifts. For Model B this is a direct consequence of the employed proportionality between
the resolved and unresolved optical depth in the model. Attenuation due to both components quickly vanishes towards higher redshifts due to the rather steep power law dependence of the resolved dust attenuation normalisation. Both dust models, Model B and Model C, therefore predict a vanishing unresolved dust attenuation contribution towards higher redshifts.
\end{itemize}
We stress that the redshift dependence of these dust models is crucial. Assuming, for example, a non-evolving optical depth normalisation or dust-to-metal ratio, results in inconsistent predictions for the rest-frame UV luminosity functions. Any high redshift galaxy luminosity predictions assuming constant dust-to-metal ratios are therefore questionable. 

The dust calibration is the basis for our exploration of galaxy luminosity functions for a variety of bands. We first discussed the rest-frame UV luminosity functions, which agree for all our models well with currently existing UV luminosity function data in the redshift range $z=2$ to $z=10$. This result is largely a consequence of the dust calibration process, which is based on rest-frame UV luminosity functions. Upon closer inspection we find however also some important discrepancies. 
\begin{itemize}[leftmargin=*]
\item For dust Models A and B, towards lower redshifts ($z\lesssim 4$), the faint end of the rest-frame UV luminosity function deviates slightly from observational data, where IllustrisTNG, more specifically, TNG50 overpredicts the abundance of faint galaxies. Model C leads to a better agreement with the data, but the overprediction still occurs at $z=3,4$. For instance, at $z=3,4$ for Model C IllustrisTNG predicts about a $\sim 0.2\,{\rm dex}$ too high galaxy number density at $M_{\rm UV} = -16\mmag$. These results are not affected by the employed resolution correction since this portion of the rest-frame UV luminosity function is probed solely by TNG50, where we do not apply any resolution correction. 
\item Towards higher redshifts we find that IllustrisTNG predicts slightly lower abundances of galaxies. This is most noticeable at $z=8$, where we find lower abundances over a quite wide range of rest-frame UV luminosities from $-18\mmag$ to $-21\mmag$. Dust attenuation is negligible at these redshifts. Therefore, these results are robust against dust modelling uncertainties. 
\item We fit the combined and dust-attenuated rest-frame UV luminosity functions with Schechter functions to quantify their shapes in more detail. This analysis revealed a rest-frame UV luminosity function faint-end slope, which is rather steep compared to observationally inferred values especially towards lower redshifts. For instance, at $z\sim 2$ we predict a best-fit slope around $\sim -1.6$ to $\sim -1.8$ for all three dust models. This values is steeper than typical observationally inferred values and also steeper than most other theoretical predictions. The observations usually suggest a shallower faint end slope around $\sim -1.3$. This result is consistent with the slight overproduction of faint galaxies discussed above. 
\end{itemize}

In addition to the rest-frame UV luminosity functions, we have also determined apparent galaxy luminosity functions for the NIRCam wide filters F070W, F090W, F115W, F150W, F200W, F277W, F356W and F444W based on dust Model C. We described all these apparent luminosity functions through Schechter fits and discuss the best-fit Schechter parameters redshift evolution.
\begin{itemize}[leftmargin=*]
\item Similarly to the rest-frame UV luminosity function, our predictions for the faint-end slope of {\it JWST} NIRCam bands are also steeper than those of other theoretical predictions. 
\item  Based on the derived best-fit Schechter functions, we then calculate concrete observational estimates for {\it JWST} NIRCam observations. Specifically, we provide predictions for the expected number of galaxies per unit redshift observed in the NIRCam F200W band within the ${2.2\times2.2 \,{\rm arcmin}^{2}}$ field of view of the {\it JWST} NIRCam instrument. Our predictions are based on two exposure times $T_{\rm exp}=10^4{\rm s}$ and $T_{\rm exp}=10^5{\rm s}$ and two different signal-to-noise ratios of $5$ and $10$. For example, at $z=8\pm 0.5$ we predict about $\sim 200$ galaxies
to be detected with a signal-to-noise ratio of $5$ for an exposure time of $10^5{\rm s}$. This number drops to $\sim 40$ for an exposure time of $10^4{\rm s}$. For $z=4\pm 0.5$ we predict more than $\sim 1000$ galaxies to be detected with a signal-to-noise ratio of $10$ and the longer exposure time. These results are consistent with previous estimates.   
\item In addition we have also calculated number counts for {\it JWST} NIRCam high-$z$ dropout
surveys. Specifically, we have provided predictions for two extragalactic surveys
that are currently planned with {\it JWST}. The first is
a large ($∼720$ hours) observational program, the {\it JWST}
Advanced Deep Extragalactic Survey (JADES). The second is the Cosmic
Evolution Early Release Science Survey (CEERS). We have used the galaxy apparent UV luminosities to make predictions for galaxy number counts for these future {\it JWST} surveys. 
For example, at $z=8$, we find that JADES-D can detect $\sim 3$ galaxies per unit redshift at the bright end up to $m_{\rm UV}\sim 27\mmag$, while having a decent detection limit $m_{\rm UV}\sim29.7\mmag$ at faint end. $235$ galaxies per unit redshift can be detected in total. JADES-M has wider sky coverage and can detect $\sim5$ times larger number of galaxies at the bright end. The total number of galaxies detected is $238$ comparable to that of JADES-D.
\end{itemize}

Theoretically, the galaxy luminosity function has to be derived through stellar population synthesis and dust modelling based on simulation predictions. The galaxy stellar mass function, on the other hand, is the fundamental property that is directly predicted by the galaxy formation model. Observationally, galaxy stellar mass functions can be derived through a convolution of the rest-frame UV galaxy luminosity with mass-to-light ratio relations. 
We have therefore explored the stellar mass to UV magnitude relation, i.e. the mass-to-light ratio relation. This relation provides another important test of our galaxy formation, and dust attenuation model. 
\begin{itemize}[leftmargin=*]
\item We find that our mass-to-light ratio relations are consistent with observationally inferred ratios except for brighter galaxies, where we tend to predict more massive galaxies at a given rest-frame UV luminosity. 
This discrepancy towards the bright end is potentially related to the fact that observations
are based on UV selected galaxies and have a potential bias missing
dusty star-forming galaxies and quiescent galaxies. This can lead to
a shift in the relation towards UV brighter galaxies at a given stellar
mass. We have studied in detail the scatter in the $M_{\ast}$ $M_{\rm UV}$ relation. This scatter increases
from the faint end towards the bright end.
\item The mass-to-light ratio relation is tighter towards higher redshifts due to
younger stellar populations with dominant UV emission. 
\item Towards lower redshifts, the scatter in the mass-to-light ratio relation increases since the stellar populations
become older, which weakens the correlation between stellar mass
and UV luminosity causing a visibly larger scatter. 
\item  We have employed the derived mass-to-light ratio relation to reconstruct the galaxy stellar mass function from the IllustrisTNG rest-frame UV luminosity functions. As expected, we find that this reconstruction of the galaxy stellar mass function only yields reliable results if the scatter in the mass-to-light ratio relation is taken into account to populate the massive end of the galaxy stellar mass function.  
\end{itemize}

Finally, we have explored how well star formation rates of IllustrisTNG correlate with observational star formation rate indicators. Spectral indices like the H$\alpha$ line luminosity are among the most important star formation rate indicators. We have therefore also extracted H$\alpha$ line luminosities for TNG50 galaxies.
\begin{itemize}[leftmargin=*]
\item Our predictions correctly recover the observed relation between H$\alpha$ emission line luminosities and galaxy star formation rates at $z=2$ demonstrating a tight relation between star formation rates and H$\alpha$ luminosity. 
\item We also find a marginally consistent H$\alpha$ luminosity function at $z=2$ in reasonable agreement with observational data. This provides another verification of our dust modelling, and the dust calibration procedure. 
\item In addition we also provided a prediction for the H$\alpha$ luminosity function at $z=4$  that can be confronted against upcoming observations.
\end{itemize}

We conclude that the combination of the three IllustrisTNG simulations, TNG50, TNG100, and TNG300, allows detailed predictions for the high redshift galaxy population up to $z \sim 10$.  The derived galaxy number densities enable the derivation of the expected number of detectable galaxies with {\it JWST} in multiple bands assuming different survey characteristics leading to important forecasts for the first observations with {\it JWST}. A significant uncertainty stems from the unknown dust abundances in galaxies, which have to be inferred through a dust calibration procedure. Here novel more self-consistent dust models~\citep[e.g.][]{McKinnon2016,McKinnon2017,McKinnon2018} would be required to overcome these limitations. 

\section*{Acknowledgements}

MV acknowledges support through an MIT RSC award, a Kavli Research Investment Fund, NASA ATP grant NNX17AG29G, and NSF grants AST-1814053 and AST-1814259. FM is supported by the Program ``Rita Levi Montalcini'' of the Italian MIUR.
The primary IllustrisTNG simulations were realised with compute time granted by the Gauss Centre for Supercomputing (GCS): TNG50 under GCS Large-Scale Project GCS-DWAR, and TNG100 and TNG300 under GCS-ILLU on the GCS share of the supercomputer Hazel Hen at the High Performance Computing Center Stuttgart (HLRS). GCS is the alliance of the three national supercomputing centres HLRS (Universit\"at Stuttgart), JSC (Forschungszentrum J\"ulich), and LRZ (Bayerische Akademie der Wissenschaften), funded by the German Federal Ministry of Education and Research (BMBF) and the German State Ministries for Research of
Baden-W\"urttemberg (MWK), Bayern (StMWFK) and Nordrhein-Westfalen (MIWF). Additional simulations were carried out on the Draco and Cobra supercomputers at the Max Planck Computing and Data Facility (MPCDF). Ancillary and test runs of the project
were also run on the Stampede supercomputer at TACC/XSEDE (allocation
AST140063), at the Hydra and Draco supercomputers at the Max Planck Computing
and Data Facility, and on the MIT/Harvard computing facilities supported by FAS
and MIT MKI.

\bibliographystyle{mnras}
%\bibliography{bibdesk} % if your bibtex file is called bibdesk.bib

% Don't change these lines
\bsp	% typesetting comment
\label{lastpage}
\end{document}